\documentclass[preprint,showpacs,preprintnumbers,amsmath,amssymb,superscriptaddress,10pt,graphicx]{revtex4}

\usepackage{graphicx}
\usepackage{dcolumn}
\usepackage{bm}
\usepackage{amssymb}

\begin{document}

\title{On the road to discovery of relic gravitational waves: The $TE$ and
$BB$ correlations in the cosmic microwave background radiation}

\author{W.~Zhao}
\email{Wen.Zhao@astro.cf.ac.uk} \affiliation{School of Physics
and Astronomy, Cardiff University, Cardiff, CF24 3AA, United
Kingdom} \affiliation{Wales Institute of Mathematical and
Computational Sciences, United Kingdom} \affiliation{Department of
Physics, Zhejiang University of Technology, Hangzhou, 310014,
P.~R.~China}
\author{D.~Baskaran}
\email{Deepak.Baskaran@astro.cf.ac.uk} \affiliation{School of
Physics and Astronomy, Cardiff University, Cardiff, CF24 3AA,
United Kingdom} \affiliation{Wales Institute of Mathematical and
Computational Sciences, United Kingdom}
\author{L.~P.~Grishchuk}
\email{Leonid.Grishchuk@astro.cf.ac.uk} \affiliation{School of
Physics and Astronomy, Cardiff University, Cardiff, CF24 3AA,
United Kingdom} \affiliation{Sternberg Astronomical Institute,
Moscow State University, Moscow 119899, Russia}

%\date{\today}

%%%%%%%%%%%%%%%%%%%%%%%%%%%%%%%%%%%%%%%%%%%%%%%%%%%%%%%%%%%%%%%%%%%%%%%%%%%%%%%%%%%
%%%%%%%%%%%%%%%%%%%%%%%%%%%%%%%%%%  ABSTRACT  %%%%%%%%%%%%%%%%%%%%%%%%%%%%%%%%%%%%%%%%%%
%%%%%%%%%%%%%%%%%%%%%%%%%%%%%%%%%%%%%%%%%%%%%%%%%%%%%%%%%%%%%%%%%%%%%%%%%%%%%%%%%%%

\begin{abstract}
The detection of primordial gravitational waves is one of the
biggest challenges of the present time. The existing 
(Wilkinson Microwave Anisotropy Probe)
observations are helpful on the road to this goal, and the
forthcoming experiments (Planck) are likely to complete this
mission. We show that the 5-year Wilkinson Microwave 
Anisotropy Probe $TE$ data contains a hint of
the presence of gravitational wave contribution. In
terms of the parameter $R$, which gives the ratio of contributions
from gravitational waves and density perturbations to the
temperature quadrupole, the best-fit model produced $R=0.24$.
Because of large residual noises, the uncertainty of this
determination is still large, and it easily includes the $R=0$
hypothesis. However, the uncertainty will be strongly reduced in
the forthcoming observations which are more sensitive. We numerically
simulated the Planck data and concluded that the relic
gravitational waves with $R=0.24$ will be present at a better than
3$\sigma$ level in the $TE$ observational channel, and at a better
than 2$\sigma$ level in the `realistic' $BB$ channel. The
balloon-borne and ground-based observations may provide a healthy
competition to Planck in some parts of the lower-$\ell$ spectrum.
\end{abstract}

%%%%%%%%%%%%%%%%%%%%%%%%%%%%%%%%%%%%%%%%%%%%%%%%%%%%%%%%%%%%%%%%%%%%%%%%%%%%%%%%%%%
%%%%%%%%%%%%%%%%%%%%%%%%%%%%%%%%%%%%%%%%%%%%%%%%%%%%%%%%%%%%%%%%%%%%%%%%%%%%%%%%%%%

\pacs{98.70.Vc, 04.30.-w, 98.80.Cq}

\maketitle

%%%%%%%%%%%%%%%%%%%%%%%%%%%%%%%%%%%%%%%%%%%%%%%%%%%%%%%%%%%%%%%%%%%%%%%%%%%%%%%%%%%
%%%%%%%%%%%%%%%%%%%%%%%%%%%%%%%%%%  SECTION 1   %%%%%%%%%%%%%%%%%%%%%%%%%%%%%%%%%%%%%%%%%%
%%%%%%%%%%%%%%%%%%%%%%%%%%%%%%%%%%%%%%%%%%%%%%%%%%%%%%%%%%%%%%%%%%%%%%%%%%%%%%%%%%%

\section{Introduction \label{intro}}

The relic gravitational waves must have been generated by strong
variable gravitational field of the very early Universe
\cite{grishchuk1974}. We got used to the notion that the
generation of gravitational waves requires some sort of a
time-dependent quadrupole moment in the energy distribution of
matter and fields. But, by definition, there is no such
asymmetries in the distribution of matter and fields comprising a
homogeneous isotropic universe. So, how the relic gravitational
waves could come into existence? In the context of classical
gravitational waves, the word `amplification' is probably a more
accurate explanation than the word `generation'. Indeed, the
pre-existing classical gravitational waves could be amplified in
the course of time, because the nonlinear character of gravitation
makes them parametrically coupled to the variable gravitational
field of the isotropic homogeneous universe (gravitational pump
field). The coupling is strong and effective if and when the
variations of the pump field are so fast that the inverse
time-scale of variations becomes comparable with the frequency of
the wave. This is when the amplitude of the wave increases over
and above the limits of the adiabatic law.

Moreover, the phenomenon of relic gravitational waves is more general than
this classical picture; it does not require the previously generated waves
and does not rely on their presence. The reason is that in
the quantum world there always exist, even in the absence of classical waves,
the inevitable ground (vacuum) state fluctuations of the corresponding field.
One can think of these zero-point quantum oscillations as those that are being
amplified, as a result of their interaction with the gravitational pump field.
Speaking more accurately, the quantum-mechanical Schrodinger evolution
transforms the initial no-particle (vacuum) state of gravitational waves into a
multi-particle (strongly squeezed vacuum) state. In this sense, relic
gravitational waves have been generated from `nothing' by the external pump
field. This process is called the superadiabatic (parametric) amplification.
For a recent review of the subject, see \cite{0707}.

Under certain extra conditions, the cosmological density perturbations
({\it dp}), which combine specific perturbations of the
gravitational field and matter fields, can also be generated by
the same mechanism of superadiabatic (parametric) amplification.
This takes place in addition to the generation of gravitational
waves ({\it gw}). The theoretical ingredients describing the
independently quantized `tensor' ({\it t}, gravitational waves)
and `scalar' ({\it s}, density perturbations) degrees of freedom
of the perturbed early Universe are almost identical. This results
in approximately equal amplitudes of the long-wavelength metric
perturbations representing gravitational waves on one side and
density perturbations on the other side. Therefore, assuming that
the observed anisotropies in cosmic microwave background radiation (CMB) are
indeed caused by cosmological perturbations of quantum-mechanical
origin, one expects that the contributions of density
perturbations and gravitational waves to the lower-order CMB
multipoles are of the same order of magnitude (see \cite{0707} and
references there).

The content of the paper is as follows. In Sec.~\ref{section1} we explain the properties
of the gravitational field perturbations $h_{ij}(\eta,{\bf x})$, and summarize the 
definitions and notations that will be used in the paper. We shall be working with the
expansion of the field $h_{ij}$ over spatial Fourier harmonics $e^{\pm i {\bf n}\cdot {\bf x}}$, 
and with the associated Fourier coefficients $\stackrel{s}{c_{\bf n}}$ and 
$\stackrel{s}{c_{\bf n}}^{\dagger}$. In the rigorous quantum-mechanical version of the theory,
$\stackrel{s}{c_{\bf n}}$ and $\stackrel{s}{c_{\bf n}}^{\dagger}$ are the annihilation and creation 
operators acting on the quantum states. But, to simplify calculations, we are using a `classical' 
version of the theory, whereby $\stackrel{s}{c_{\bf n}}$ and $\stackrel{s}{c_{\bf n}}^{\dagger}$
are treated as classical random complex numbers $\stackrel{s}{c_{\bf n}}$ and $\stackrel{s}{c_{\bf n}}^*$.

In Sec.~\ref{section2} we explain in
detail how the statistical properties of the underlying cosmological
perturbations, encoded in the probability density functions (pdf's) of complex random variables
$\stackrel{s}{c}_{\bf n}$, $\stackrel{s~*}{c_{\bf n}}$, translate into the
statistical properties of the CMB
multipole coefficients $a_{\ell m}^X$ ($X=T,E,B$). The statistics of the
CMB anisotropies, often referred to in the form of a loose and poetic concept
of `cosmic variance', is an important ingredient of the whole problem of
deriving cosmological conclusions from the CMB data. We formulate the
pdf's for individual quantities $a_{\ell m}^X$ and
for the full set $\{a_{\ell m}^X\}$ of these quantities. The main emphasis is
on the correlation coefficient $\rho_\ell$ in the joint pdf for $T$ and $E$
components of the CMB.

In Sec.~\ref{section3} we formulate the pdf's for the best unbiased
estimators of the auto-correlation and cross-correlation power spectra.
Again, the main emphasis is on the estimator $D_{\ell}^{TE}$ of the $TE$
power spectrum. This is because the gravitational waves have a distinctive
signature in the region of lower $\ell$'s: the mean value of $D_{\ell}^{TE}$
must be positive for density perturbations and negative for gravitational
waves \cite{deepak}. The individual outcomes may have signs opposite to the
sign of the mean value. This is why we evaluate the probabilities of finding
positive (negative) outcomes for $D_{\ell}^{TE}$ in the situation when the
expected value of this variable is negative (positive).

The actual numerical amount of the $TE$ cross-correlation at every $\ell$ is
governed by the numerical value of the correlation coefficient $\rho_{\ell}$.
In Sec.~\ref{s3} we derive $\rho_{\ell}$ for specific cosmological
models with various amounts of density perturbations and gravitational waves.
The parameters of the background cosmological model are always taken from
the 5-year Wilkinson Microwave Anisotropy Probe (WMAP) best-fit $\Lambda$CDM cosmology \cite{5map_cosmology}, but the
perturbation parameters are allowed to vary. The derived coefficients
$\rho_{\ell}$ make it possible to build the $\ell$-dependent confidence
intervals surrounding the mean value of the estimator $D_{\ell}^{TE}$.

The inherent uncertainty of the expected CMB signal, ultimately explicable
in terms of the quantum-mechanical origin of cosmological perturbations, is
exacerbated by the `real life' effects, such as instrumental noises,
foreground emissions, incomplete sky coverage, etc. In Sec.~\ref{sa4} we use
the parameters of the WMAP and Planck missions to show how these effects
broaden the size of the confidence intervals. This allows us to put to the
test the existing WMAP $TE$ data and make predictions for the Planck mission.

The WMAP5 best-fit analysis suggests that the observed CMB anisotropies have no
contribution from gravitational waves \cite{5map_cosmology}. We call this
conclusion the null hypothesis $H_0$. In Sec.~\ref{s4} we test the null
hypothesis using the WMAP5 $TE$ data. Although the evidence for rejection of
$H_0$ is weak, within 1$\sigma$, the data points demonstrate the tendency to
lie below the $H_0$ expectation curve, which is actually required by the
presence of relic gravitational waves.

Since the existence of relic gravitational waves is a necessity dictated by
general relativity and quantum mechanics, we include gravitational waves in
the $TE$ data analysis as a compulsory ingredient (Sec.~\ref{s5}). We consider
a class of models which contain gravitational waves and are consistent with
the available $TT$, $EE$ and $BB$ data. Although our physical picture is 
guided by gravitational waves of quantum-mechanical origin \cite{0707}, 
the analysis is general and applies to gravitational waves of any origin.
One has to note, though, that classical gravitational waves generated at later stages of cosmological 
evolution have wavelengths much shorter than the Hubble radius and therefore they do not affect the lower-order CMB multipoles.
We build and study the likelihood
function for $R$, where $R$ is the ratio of $\it gw$ and $\it dp$
contributions to the temperature quadrupole $C_{\ell=2}^{TT}$. The maximum of
the likelihood function turns out to be at $R=0.24$. This number also
determines other perturbation parameters in our class of models. Because of
the large WMAP noises, the confidence interval for $R$ is quite broad, so that
the WMAP team's suggestion $R=0$ is only 1$\sigma$ away from our maximum
likelihood value $R=0.24$. Nevertheless, we consider our best-fit model with
$R=0.24$ as a benchmark model, which we believe is to be confirmed by the
more sensitive Planck mission. We numerically simulate the future Planck data
and show that the relic gravitational waves are to be detected by the $TE$
method at a better than 3$\sigma$ level.

The prospects of the Planck mission are discussed at some length
in the last Sec.~\ref{s6}. We compare the $TE$ method with the more
often mentioned $BB$ method. The $B$-mode of polarization does not
contain the contribution from density perturbations but is
inherently weak and prone to various systematic effects. We
distinguish the `optimistic' and `realistic' cases in detection of
gravitational waves through the $BB$ correlation function. It
appears that the `realistic' $BB$ case offers a little better than
2$\sigma$ detection of relic gravitational waves in the $R=0.24$
model. We also show that the $BB$ method mostly relies on the
reionization era, whereas the $TE$ correlation function is mostly
sensitive to the era of recombination.

The Conclusions summarize our findings and emphasize the necessity
of concentrated efforts in the race for discovery of relic
gravitational waves.

%%%%%%%%%%%%%%%%%%%%%%%%%%%%%%%%%%%%%%%%%%%%%%%%%%%%%%%%%%%%%%%%%%%%%%%%%%%%%%%%%%%%
%%%%%%%%%%%%%%%%%%%%%%%%%%%%%%%%%%%  SECTION 2   %%%%%%%%%%%%%%%%%%%%%%%%%%%%%%%%%%%%%%%%%%
%%%%%%%%%%%%%%%%%%%%%%%%%%%%%%%%%%%%%%%%%%%%%%%%%%%%%%%%%%%%%%%%%%%%%%%%%%%%%%%%%%%%

\section{Gravitational field perturbations \label{section1}}

Let us recall that the gravitational field of a
slightly perturbed flat Friedmann-Lemaitre-Robertson-Walker (FLRW) universe is described by
\begin{eqnarray}
 ds^2=-c^2dt^2+a^2(t)(\delta_{ij}+h_{ij})dx^idx^j 
 =a^2(\eta)[-d\eta^2+(\delta_{ij}+h_{ij})dx^idx^j],
\nonumber
\label{metric}
\end{eqnarray}
where $t$-time and $\eta$-time are related by $c d t=a d \eta$.
The metric perturbation field $h_{ij}(\eta,{\bf x})$ can be expanded over
spatial Fourier harmonics $e^{\pm i{\bf n}\cdot{\bf x}}$:
\begin{eqnarray}
\label{h}
 h_{ij}(\eta,{\bf
 x})=\frac{\mathcal{C}}{(2\pi)^{3/2}}\int_{-\infty}^{+\infty}\frac{d^3{\bf
 n}}{\sqrt{2n}} \sum_{s=1,2}\left[\stackrel{s}{p}_{ij}({\bf n})
 \stackrel{s}{h}_{n}(\eta)e^{ i{\bf n}\cdot{\bf x}}
\stackrel{s}{c}_{\bf n}+ \stackrel{s~*}{p_{ij}}({\bf n})
\stackrel{s~*}{h_{n}}(\eta)e^{-i{\bf n}\cdot{\bf x}}
\stackrel{s~\dag}{c_{\bf n}} \right],
\end{eqnarray}
where ${\bf n}$ is the dimensionless time-independent wavevector. The
dimensionless wavenumber $n$ is $n=(\delta_{ij}n^in^j)^{1/2}$.

The dimensionless $n$ is related with the dimensionful $k$ by $k= {n}/{2l_H}$,
where $l_H=c/H_0$ is today's Hubble radius. The value $n_H=4\pi$ labels the
wave whose today's wavelength is equal to the `half of the size of the
Universe', that is, equal to today's Hubble radius. The use of $n$,
rather than $k$, is very convenient as it guides the evaluation of the CMB
multipoles: cosmological perturbations with wavenumber $n$ are mostly
responsible for the CMB temperature anisotropies at the multipole
$\ell \approx n$.

The general form of Eq.~(\ref{h}) is common for metric
perturbations $h_{ij}(\eta,{\bf x})$ characterizing gravitational
waves and density perturbations, but  ${\it gw}$ and ${\it dp}$ differ in the explicit
content of the two polarization tensors $\stackrel{s}{p}_{ij}({\bf
n})$ ($s=1,2$). In the {\it gw} case the field $h_{ij}$ consists
of two `transverse-traceless' components, whereas in the {\it dp}
case it consists of the `scalar' and `longitudinal-longitudinal'
components (see, for example, \cite{deepak}). The {\it gw} mode
functions $\stackrel{s}{h}_{n}(\eta)$ satisfy the perturbed
Einstein equations linearized around an FLRW background. In the
{\it dp} case the gravitational mode functions
$\stackrel{s}{h}_{n}(\eta)$ are necessarily accompanied by matter
mode functions. For many models of matter, the full set of
dynamical equations for density perturbations leaves independent
only one out of two polarization components $s=1,2$.

In the rigorous quantum-mechanical version of the theory, the quantities
$h_{ij}$ are quantum-mechanical operators acting on quantum states. The
$\stackrel{s}{c}_{\bf n}$ and $\stackrel{s~\dag}{c_{\bf n}}$
are the annihilation and creation operators, satisfying the commutation
relations $[\stackrel{s'}{c}_{\bf n},~\stackrel{s~\dag}{c_{\bf
 n'}}]=\delta_{s's}\delta^{(3)}({\bf n}-{\bf n'})$.
In the case of density perturbations, the same operators
$\stackrel{s}{c}_{\bf n}$ and $\stackrel{s~\dag}{c_{\bf n}}$ are
inherited, through the linearized Einstein equations, by the
matter fields. According to the physical formulation of the
problem, the initial quantum state of the quantized perturbations
is the ground state $|0\rangle$ of the corresponding Hamiltonian.
The ground state satisfies the condition $\stackrel{s}{c}_{\bf
n}|0\rangle=0$. For gravitational waves, the normalization
constant $\mathcal{C}$ is $\mathcal{C}=\sqrt{16\pi}l_{\rm Pl}$,
whereas for density perturbations $\mathcal{C}=\sqrt{24\pi}l_{\rm
Pl}$, where $l_{\rm Pl}=\left(G\hbar/c^3\right)^{1/2}$ is the
Planck length. Obviously, if the Planck constant $\hbar$ is
artificially sent to zero, the initial zero-point quantum
fluctuations, as well as the field $h_{ij}$ itself, vanish.

To  simplify calculations, we will be using a `classical' version of the
theory, whereby the quantum-mechanical operators $\stackrel{s}{c}_{\bf n}$
and $\stackrel{s~\dag}{c_{\bf n}} $ are treated as classical random
complex numbers $\stackrel{s}{c}_{\bf n}$ and $\stackrel{s~*}{c_{\bf n}}$.
We are loosing some delicate quantum aspects of the rigorous theory but they
are not our main concern. The statistical properties
of $\stackrel{s}{c}_{\bf n}$, $\stackrel{s~*}{c_{\bf n}}$ define the
statistical properties of the classical field $h_{ij}$ (\ref{h}) and,
through the linear radiative transfer equations, they determine the
statistical properties of the CMB anisotropies.

Since the Schrodinger evolution of the initial vacuum state into a
squeezed vacuum state retains the Gaussianity of the state (while
developing strongly unequal variances in the amplitude and the
phase), we are justified in assuming that each individual complex
coefficient $\stackrel{s}{c}_{\bf n}$ is described by a complex pdf
\cite{distribution,book5} with a zero mean and unit variance. If necessary,
one can imagine that the space of wavevectors $\bf n$ is
discretised into a fine 3-lattice. Moreover, since the different
modes ${\bf n}$ and polarization states $s$ are not supposed to
`know about each other', we are justified in assuming that the
joint distribution for all coefficients $\{\stackrel{s}{c}_{\bf
n}\}$ (that is, for the full set including all possible values of
${\bf n}$ and $s$) is a complex multivariate Gaussian pdf
comprising a product of individual Gaussian pdf's
\cite{book5,distribution}.

The first and the second moments calculated with the joint pdf amount to
\begin{eqnarray}
\label{nrelations}
\langle
\stackrel{s}{c}_{\bf n}\rangle = \langle
\stackrel{s~*}{c_{\bf n}} \rangle = 0,
~~\langle\stackrel{s}{c}_{\bf n} \stackrel{s'~*}{c_{\bf n'}}\rangle
=\delta_{ss'}\delta^{(3)}({\bf n}-{\bf n'}),~~
\langle\stackrel{s}{c}_{\bf n}
\stackrel{s'}{c}_{\bf n'}\rangle
=\langle\stackrel{s~*}{c_{\bf n}}
\stackrel{s'~*}{c_{\bf n'}} \rangle=0.
\end{eqnarray}
Since the coefficients $\stackrel{s}{c}_{\bf n}$ obey a joint multivariate
Gaussian (normal) distribution, the higher order averages are expressible in
terms of the second moments.

It is clear from Eq.~(\ref{nrelations}) that the mean value of the field
$h_{ij}$ is zero, but the variance is non-zero and is calculated to be
\begin{eqnarray}
\nonumber
\frac{1}{2}\langle h_{ij}(\eta,{\bf x})h^{ij}(\eta,{\bf x})\rangle
= \int\limits_{0}^{\infty} \frac{dn}{n}~h^2(n,\eta).
\end{eqnarray}
The function
\begin{eqnarray}
\label{powerspectrumhdef}
h^2(n,\eta) \equiv \frac{\mathcal{C}^2}{2\pi^2}n^2
\sum_{s=1,2}|\stackrel{s}{h}_n(\eta)|^2,
\end{eqnarray}
gives the mean-square value of the gravitational field perturbations
in a logarithmic interval of the wavenumbers $n$. It is called the metric
power spectrum. The spectrum (\ref{powerspectrumhdef}) refers to gravitational
waves or density perturbations depending on whether this quantity is
calculated from {\it gw} or {\it dp} mode functions.

The CMB calculations are usually being done numerically, whereby
the initial conditions for differential equations are imposed at
times deep in the radiation-dominated era
\cite{BertschingerMa,zaldarriaga}. The wavelengths of relevant
modes ${\bf n}$ are much longer than the Hubble radius at that
time, and the primordial power spectrum $h^2(n)$ is a smooth
function of $n$. The early pump fields (scale factors) $a(\eta)$
which are power-law functions of $\eta$ generate primordial
spectra $h^2(n)$ which are power-law functions of $n$ \cite{0707}:
\begin{equation}
\label{primsp}
h^2(n)~({\it gw}) = B_t^2 n^{{\rm n}_t}, ~~~~~h^2(n)~({\it dp}) =
B_s^2 n^{{\rm n}_s -1}.
\end{equation}
The quantum-mechanical evolution of the initial vacuum states of
{\it gw} and {\it dp} fields results \cite{0707} in approximately
equal primordial amplitudes $B_t^2$, $B_s^2$, and also in the
relationship $n_t \approx n_s -1$. Although, in general, the four
numbers $B_t^2, B_s^2, n_t, n_s$ specified at the
radiation-dominated stage do not contain enough information to
impose all the starting conditions for numerical calculation of
the ensuing CMB anisotropies, some additional simplifying
assumptions \cite{BertschingerMa,zaldarriaga}, \cite{deepak} allow one to
do that.

At the end of this section it is necessary to warn the reader
that according to the claims of inflationary theory the amplitudes
of density perturbations should be arbitrarily large in the limit
of models with small values of the early Universe parameter
$\epsilon$, where $\epsilon = -{\dot H}/H^2$. Specifically, the
incorrect (inflationary) treatment of quantized scalar
perturbations has led to the proposition that the power spectrum
of scalar metric perturbations (often denoted by $\zeta$ or
$\cal{R}$) should be divergent as $1/\epsilon$, right from the very 
early time when initial conditions in the high-frequency regime of the 
perturbations were imposed. Inflationary theory starts its line of argument 
with vacuum fluctuations of the scalar field (inflaton) in de Sitter space 
time and finishes with infinitely large scalar metric perturbations in the 
same de Sitter spacetime. This predicted
divergency of density perturbations in the limit of $\epsilon=0$
($n_s =1, n_t =0$) is customarily being substituted by the claim
that it is the amount of primordial gravitational waves that is
expected to be very small. This is expressed in the form of small
values of the `tensor-to-scalar' ratio $r$:
\begin{equation}
\label{tsr}
r= 16 \epsilon = - 8 n_t.
\end{equation}
In particular, according to the inflationary Eq.~(\ref{tsr}), the interval of
a `standard' (de Sitter) inflation, i.e. $\epsilon =0$, generates a zero amount
of primordial gravitational waves, i.e. $r=0$. This is in full contradiction
with the mechanism of superadiabatic (parametric) amplification \cite{grishchuk1974}.
Certainly, we are not using Eq.~(\ref{tsr}), neither as a valid theoretical
result nor as an element of CMB data analysis. For a detailed critical
discussion of inflationary predictions and the data analysis based on
inflationary theory, see \cite{0707}.

%

%%%%%%%%%%%%%%%%%%%%%%%%%%%%%%%%%%%%%%%%%%%%%%%%%%%%%%%%%%%%%%%%%%%%%%%%%%%%%%%%%%%%
%%%%%%%%%%%%%%%%%%%%%%%%%%%%%%%%%%%  SECTION 3   %%%%%%%%%%%%%%%%%%%%%%%%%%%%%%%%%%%%%%%%%%
%%%%%%%%%%%%%%%%%%%%%%%%%%%%%%%%%%%%%%%%%%%%%%%%%%%%%%%%%%%%%%%%%%%%%%%%%%%%%%%%%%%%

\section{Anisotropies in the Cosmic Microwave Background Radiation
\label{section2}}

%%%%%%%%%%%%%%%%%%%%%%%%%%%%%%%%%%%%%%%%%%%%%%%%%%%%%%%%%%%%%%%%%%%%%%%%%%%%%%%%%%%
                                                                                                       %%%%%% SUBSECTION 3.1 %%%%%%
%%%%%%%%%%%%%%%%%%%%%%%%%%%%%%%%%%%%%%%%%%%%%%%%%%%%%%%%%%%%%%%%%%%%%%%%%%%%%%%%%%%

\subsection{Characterization of the radiation field}

The radiation field is usually characterized by four Stokes
parameters ($I,Q,U,V$), where $I$ is the total intensity of
radiation, $Q$ and $U$ describe the magnitude and direction of
linear polarization, and $V$ is the circular polarization. The
Stokes parameters can be viewed as functions of ($t,x^i,\nu,e^i$),
where $\nu$ is the photon's frequency, and $e^{i}$ is a unit
vector in the direction of observation (opposite to the photon's
propagation). In a given space-time point ($t,x^i$) and for a
fixed $\nu$, the Stokes parameters are functions of $\theta,\phi$,
where $\theta,\phi$ are coordinates on a unit sphere indicating
the direction of observation: $d\sigma^2=g_{ab}dx^adx^b=d\theta^2+\sin^2\theta d\phi^2$.
%\begin{eqnarray}
%\nonumber
%%\label{new1}
%d\sigma^2=g_{ab}dx^adx^b=d\theta^2+\sin^2\theta d\phi^2.
%\end{eqnarray}

The Stokes parameters of the radiation are components of the
polarization tensor $P_{ab}$ \cite{landau} which can be written as
\begin{eqnarray}
P_{ab}(\theta,\phi)= \frac{1}{2}\left(
\begin{array}{c}
~~I+Q~~~~~~~-(U-iV)\sin\theta\\
-(U+iV)\sin\theta~~~~~(I-Q)\sin^2\theta
\end{array}
\right).
\nonumber
\end{eqnarray}
Under arbitrary transformation of coordinates $(\theta,\phi)$, the components
of $P_{ab}(\theta,\phi)$ transform as components of a tensor, but some
combinations of $P_{ab}(\theta,\phi)$ and its derivatives remain invariant.

Two algebraic invariants are given by
 \begin{eqnarray}
 \label{new2}
 I(\theta,\phi)=g^{ab}(\theta,\phi)P_{ab}(\theta,\phi),~~
 V(\theta,\phi)=i\epsilon^{ab}(\theta,\phi)P_{ab}(\theta,\phi),
 \nonumber
 \end{eqnarray}
where $\epsilon^{ab}(\theta,\phi)$ is a completely antisymmetric
pseudotensor
 \begin{eqnarray}
 \nonumber
 \epsilon^{ab}= \left( \begin{array}{c}
 ~~~0~~~~~~~~~~~-\left(\sin\theta\right)^{-1}\\
\left(\sin\theta\right)^{-1}~~~~~~~0~~~
 \end{array}
 \right).
 \end{eqnarray}
Now one can single out the symmetric trace-free (STF) part
$P_{ab}^{\rm STF}$ of $P_{ab}$: $P_{ab}(\theta,\phi)=\frac{1}{2}I
 g_{ab}-\frac{i}{2}V\epsilon_{ab}+P_{ab}^{\rm STF}$,
%\begin{eqnarray}
%P_{ab}(\theta,\phi)=\frac{1}{2}I
% g_{ab}-\frac{i}{2}V\epsilon_{ab}+P_{ab}^{\rm STF},
% \nonumber
%\end{eqnarray}
\begin{eqnarray}
 P_{ab}^{\rm STF}= \frac{1}{2}\left(
\begin{array}{c}
~~Q~~~~~~~-U\sin\theta\\
-U\sin\theta~~~-Q\sin^2\theta
\end{array}\right).
\nonumber
\end{eqnarray}
Two other invariants require the use of second covariant derivatives of
$P_{ab}^{\rm STF}$. It can be shown that there exist only two linearly
independent invariants that can be built from second derivatives \cite{deepak}:
\begin{eqnarray}
E(\theta,\phi)=-2(P_{ab}^{\rm STF})^{;a;b},~~~B(\theta,\phi)=
-2(P_{ab}^{\rm STF})^{;b;d}\epsilon^a_{~d}~.
\nonumber
\end{eqnarray}
In the above expression ``$;$" denotes covariant differentiation on the sphere.
The quantities $I$ and $E$ are scalars, while $V$ and $B$ are pseudoscalars.

In CMB applications one usually ignores $V$, as well as the
unperturbed part of $I$, and regards the remaining quantities as
measured in the temperature units, specifically in micro-Kelvin
$({\mu}{\rm K})$. These quantities (temperature and polarization
anisotropies) can be expanded over ordinary spherical harmonics
$Y_{{\ell}m}(\theta,\phi)$:
\begin{subequations}
\begin{eqnarray}
\label{Iexpend}
 I(\theta,\phi)=\sum_{{\ell}=0}^{\infty}\sum_{m=-{\ell}}^{{\ell}} a_{{\ell}m}^T
 Y_{{\ell}m}(\theta,\phi),
 \end{eqnarray}
 \begin{eqnarray}
 E(\theta,\phi)=\sum_{{\ell}=2}^{\infty}\sum_{m=-{\ell}}^{{\ell}}
 \left[\frac{({\ell}+2)!}{({\ell}-2)!}\right]^{\frac{1}{2}}a_{{\ell}m}^E
 Y_{{\ell}m}(\theta,\phi),
~
 B(\theta,\phi)=\sum_{{\ell}=2}^{\infty}\sum_{m=-{\ell}}^{{\ell}}
 \left[\frac{({\ell}+2)!}{({\ell}-2)!}\right]^{\frac{1}{2}}a_{{\ell}m}^B
 Y_{{\ell}m}(\theta,\phi).
\nonumber \\
\label{Eexpend}
\end{eqnarray}
\end{subequations}
%The spherical harmonics satisfy the relations
% \begin{eqnarray}
% \nonumber %\label{ylm}
% \begin{array}{c}
% \int_0^{\pi}\int_0^{2\pi} Y_{\ell m}(\theta,\phi)
% Y^*_{\ell' m'}(\theta,\phi)\sin\theta~d\phi~d\theta
% = \delta_{\ell\ell'}\delta_{mm'}, \\
% Y_{\ell m}^*(\theta,\phi)=(-1)^mY_{\ell -m}(\theta,\phi).
% \end{array}
%\end{eqnarray}
The $\ell$-dependent factors in front of $a_{\ell m}^{E}$ and
$a_{\ell m}^{B}$ in (\ref{Eexpend}) have been introduced in order to work
with definitions consistent with previous literatures \cite{zaldarriaga}, \cite{KKS1997}.

Since $I$, $E$, $B$ are real functions on the sphere, the complex multipole
coefficients in (\ref{Iexpend}) and (\ref{Eexpend}) satisfy the following
conditions
\begin{eqnarray}
\label{arelation}
 a_{\ell m}^{X}=(-1)^m a_{\ell, -m}^{X*}~,~~~(X=T,E,B,~~-\ell\leq
 m\leq \ell).
 \end{eqnarray}
%The multipole coefficients of a given field $I$, $E$, $B$ are calculable as
% \begin{subequations}
% \begin{eqnarray}
% \nonumber % \label{at}
%a_{\ell m}^{T}=\int_0^{\pi}\int_0^{2\pi}Y_{\ell
% m}^*(\theta,\phi)I(\theta,\phi)\sin\theta~d\phi~d\theta,
%\end{eqnarray}
%\begin{eqnarray}
% \nonumber %\label{ae}
%  a_{\ell m}^{E}=\left[\frac{(\ell-2)!}{(\ell+2)!}\right]^{\frac{1}{2}}
%\int_0^{\pi}
%  \int_0^{2\pi}Y_{\ell m}^*(\theta,\phi)E(\theta,\phi)\sin\theta~d\phi~d\theta,
%\end{eqnarray}
%\begin{eqnarray}
% \nonumber %\label{ab}
%a_{\ell m}^{B}=\left[\frac{(\ell-2)!}{(\ell+2)!}\right]^{\frac{1}{2}}
%\int_0^{\pi}\int_0^{2\pi}
%Y_{\ell m}^*(\theta,\phi)B(\theta,\phi)\sin\theta~d\phi~d\theta~.
%\end{eqnarray}
%\end{subequations}
The set of multipole coefficients
$(a_{{\ell}m}^T,a_{{\ell}m}^E,a_{{\ell}m}^B)$ completely
characterizes the CMB field that we are studying.

%%%%%%%%%%%%%%%%%%%%%%%%%%%%%%%%%%%%%%%%%%%%%%%%%%%%%%%%%%%%%%%%%%%%%%%%%%%%%%%%%%%
                                                                                                       %%%%%% SUBSECTION 3.2 %%%%%%
%%%%%%%%%%%%%%%%%%%%%%%%%%%%%%%%%%%%%%%%%%%%%%%%%%%%%%%%%%%%%%%%%%%%%%%%%%%%%%%%%%%

 \subsection{Statistical properties of the CMB field\label{subsection2.3}}

The radiative transfer equations relate the set of random
coefficients $\stackrel{s}{c}_{\bf n}$ with the multipole
coefficients $(a_{{\ell}m}^T,a_{{\ell}m}^E,a_{{\ell}m}^B)$. As a
consequence, the multipole coefficients $a_{{\ell}m}^X$
($X=T,E,B$) inherit the randomness of $\stackrel{s}{c}_{\bf n}$.

The calculations usually begin from a single Fourier mode of
perturbations considered in a special reference frame whose
$z$-axis is parallel to the wavevector ${\bf n}$. This is how one
arrives at the multipole coefficients denoted
$a_{{\ell}m}^X(n,s)$. The quantities $a_{{\ell}m}^X(n,s)$ depend
in a complicated but calculable manner on the metric (plus matter,
in the case of density perturbations) mode functions and the
background cosmological model
\cite{Polnarev1985,a1,a2,crittenden,ana3,frewin,ana1,ana2,grishchuk2,ana4,ana4a,ana5},
\cite{deepak}, \cite{ana6}. The next step
is the generalization of this result to the set of all Fourier
components with arbitrarily directed ${\bf n}$'s. This requires
the use of the Wigner rotation coefficients
$D_{mm'}^{\ell}(\hat{\bf n})$, where $\hat{\bf n}\equiv{\bf n}/n$.
The Wigner coefficients describe the rotations of individual
special frames associated with ${\bf n}$'s to one unique frame of
the observer. The Wigner coefficients satisfy the orthogonality
relation \cite{angular}
\begin{eqnarray}
\label{angular}
 (2\ell+1)\int {D_{mm'}^{\ell}}^*({\bf\hat{n}})
 D_{MM'}^L({\bf\hat{n}})~d~\Omega_{\bf
 \hat{n}} = 4\pi\delta_{\ell L}\delta_{mM}\delta_{m'M'}.
\end{eqnarray}

A detailed analysis shows \cite{deepak} that the resulting multipole
coefficients $a_{\ell m}^X$ are related to the random coefficients
$\stackrel{s}{c}_{\bf n}$, both for {\it gw} and {\it dp}, as follows
\begin{eqnarray}
a_{{\ell}m}^X=\frac{\mathcal{C}}{(2\pi)^{3/2}}
\int^{\infty}_{-\infty}\frac{d^3{\bf n}}{\sqrt{2n}}
 \sum_{s=1,2}\sum_{m'=-{\ell}}^{\ell}\left[{a}_{{\ell}m'}^X(n,s)
 D_{m m'}^{\ell}(\hat{\bf n})\stackrel{s}{c}_{\bf n}+
(-1)^m{a}_{{\ell} m'}^{X*}(n,s)D_{-m, m'}^{{\ell}~~~*}(\hat{\bf
n})\stackrel{s~*}{c_{\bf n}}
 \right]. 
\label{aa1}
\end{eqnarray}
The linear character of this relation implies that the probability density
functions of individual multipole coefficients $a_{\ell m}^X$ are zero-mean
Gaussian pdf's, and the joint distribution of all multipole coefficients
is described by a multivariate Gaussian pdf. The effects of nonlinearities
and quantum loops of the underlying metric field are very small, and we ignore
them. The nonlinear effects in the radiative transfer equations are possible,
but apparently they are also small. The individual distributions are marginal
distributions derivable from the joint pdf. Before formulating these pdf's
explicitely, we shall use Eq.~(\ref{nrelations}) and
Eq.~(\ref{aa1}) in order to find second moments of the distributions.

%%%%%%%%%%%%%%%%%%%%%%%%%%%%%%%%%%%%%%%%%%%%%%%%%%%%%%%%%%%%%%%%%%%%%%%%%%%%%%%%%%%%
%%%%%%%%%%%%%%%%%%%%%%%%%%%%%%%%%%%%%%%%%%%%%%%%%%%%%%%%%%%%%%%%%%%%%%%%%%%%%%%%%%%%

\subsubsection{Auto-correlation and cross-correlation
functions\label{subsection2.3.1}}

It follows from Eq.~(\ref{nrelations}) and Eq.~(\ref{aa1}) that all linear
averages of $a_{\ell m}^X$ vanish,
\begin{eqnarray}
 \nonumber %\label{linaverage}
\langle a_{\ell m}^X\rangle =\langle a_{\ell
 m}^{X*}\rangle= 0.
\end{eqnarray}
The quadratic averages $\langle a_{{\ell}m}^{X}a_{{\ell'}m'}^{X'*}\rangle$ do
not vanish. They are called the correlation functions. One gets an
auto-correlation function if $X=X'$, and a cross-correlation function
if $X\neq X'$.

Let us start from the auto-correlation functions  
\cite{ana3,ana1,ana2,ana5}, \cite{deepak}, \cite{ana6}. 
Using Eqs.~(\ref{nrelations},~\ref{aa1},~\ref{angular}) 
one can show that
\begin{eqnarray}
\label{b1}
\langle a_{\ell m}^{X}a_{\ell'm'}^{X*}\rangle
= \delta_{mm'}\delta_{\ell\ell'}C_{\ell}^{XX},~~\langle
a_{\ell m}^{X}a_{\ell'm'}^{X}\rangle=\langle
 a_{\ell m}^{X*}a_{\ell'm'}^{X*}\rangle=0,~~(X=T, E, B),
\end{eqnarray}
where the $XX$ power spectrum $C_{\ell}^{XX}$ is given by \cite{deepak}
\begin{eqnarray}
\label{clxxvalue}
C_{\ell}^{XX}=\frac{\mathcal{C}^2}{2\pi^2(2\ell+1)}\int ndn
\sum_{s=1,2}\sum_{m=-\ell}^{\ell}
\left| a_{\ell m}^{X}(n,s)\right|^2.
\end{eqnarray}
Both, density perturbations and gravitational waves, produce
distinctive $C_\ell^{TT}$, $C_\ell^{EE}$ power spectra, but the
$C_\ell^{BB}$ is zero for density perturbations and $C_\ell^{BB}
\neq 0$ for gravitational waves 
\cite{zaldarriaga,KKS1997,bb1,bb2}. Specifically, all the
coefficients $a_{\ell m}^{B}(n,s)$ vanish in the {\it dp} case. In
other words, density perturbations do not produce the $B$-mode of
polarization \cite{zaldarriaga,KKS1997,bb1,bb2}, \cite{deepak}.

Let us now turn to the cross-correlation functions. In general,
there exists three cross-correlation functions, namely $TE$, $TB$
and $EB$.  However, the $TB$ and $EB$ functions vanish identically
in the {\it dp} case, as density perturbations do not produce the
$B$-mode. In the {\it gw} case, the left and right circular
polarizations of gravitational waves give contributions of
opposite signs into the $TB$ and $EB$ functions. It is very
natural to expect that the relic gravitational waves with both
polarizations had been generated in equal amounts. Then, one
concludes that in the {\it gw} case the $TB$ and $EB$ functions do
also vanish \cite{kamionkowski}, \cite{deepak}. (However, there exist 
parity-violating processes in the later Universe, see for
example \cite{kgr}.) Therefore, we shall only deal with the $TE$
cross-correlation function.

Using Eqs.~(\ref{aa1},~\ref{nrelations},~\ref{angular}), we arrive at
\begin{eqnarray}
\label{bb21}
\begin{array}{c}
\frac{1}{2}\langle
 a_{\ell m}^{T}a_{\ell'm'}^{E*}+a_{\ell
m}^{T*}a_{\ell'm'}^{E}\rangle=\delta_{mm'}\delta_{\ell\ell'}C_\ell^{TE},~~
\frac{1}{2}\langle
 a_{\ell m}^{T}a_{\ell'm'}^{E*}-a_{\ell
m}^{T*}a_{\ell'm'}^{E}\rangle=0, \\
\langle
 a_{\ell m}^{T}a_{\ell'm'}^{E}\rangle=\langle a_{\ell
m}^{T*}a_{\ell'm'}^{E*}\rangle=0,
\end{array}
\end{eqnarray}
where the $C_\ell^{TE}$ is given by
\begin{eqnarray}
\label{cltevalue}
C_{\ell}^{TE}=\frac{\mathcal{C}^2}{4\pi^2(2\ell+1)}
\int ndn\sum_{s=1,2}\sum_{m=-\ell}^{\ell}
\left[a_{\ell m}^{T}(n,s)a_{\ell m}^{E*}(n,s)+
a_{\ell m}^{T*}(n,s)a_{\ell m}^{E}(n,s)\right].
\end{eqnarray}
Note that one can derive Eqs.~(\ref{b1},~\ref{bb21}) as soon as the
averages (\ref{nrelations}) are given, independently on whether the
actual distributions are Gaussian or not. Also, it is seen from
Eqs.~(\ref{b1},~\ref{bb21}) that only the products of coefficients
with $\ell=\ell'$ and $m=m'$ are important.

Unlike the auto-correlation power spectra $C_{\ell}^{XX}$ which are strictly
positive functions, the values of the $TE$ `power spectrum' $C_\ell^{TE}$
can be positive or negative. Most importantly for our further discussion,
in the region of lower $\ell$'s the function $C_\ell^{TE}$ is positive for
density perturbations and negative for gravitational waves \cite{deepak}.
It can also be shown, using Eqs.~(\ref{cltevalue},~\ref{clxxvalue}), that
\begin{eqnarray}
\label{ert}
(C_{\ell}^{TE})^2\leq C_{\ell}^{TT}C_{\ell}^{EE},
\end{eqnarray}
which is the analog of the Cauchy-Bunyakovsky-Schwarz inequality.

Since density perturbations and gravitational waves are independent fields,
they contribute to the total CMB power spectra additively:
\begin{eqnarray}
\label{TotalCell_dp+gw}
C_{\ell}^{XX'} = C_{\ell}^{XX'}(dp) + C_{\ell}^{XX'}(gw).
\end{eqnarray}

%%%%%%%%%%%%%%%%%%%%%%%%%%%%%%%%%%%%%%%%%%%%%%%%%%%%%%%%%%%%%%%%%%%%%%%%%%%%%%%%%%%
%%%%%%%%%%%%%%%%%%%%%%%%%%%%%%%%%%%%%%%%%%%%%%%%%%%%%%%%%%%%%%%%%%%%%%%%%%%%%%%%%%%

\subsubsection{Probability density functions for individual multipole
coefficients\label{s2.3.2}}

The Gaussian nature of $\stackrel{s}{c}_{\bf n}$, together with
Eq.~(\ref{aa1}), translate into Gaussian pdf's for individual
multipole coefficients $a_{\ell m}^X$.

It is convenient to start from distributions for
real and imaginary parts of $a_{\ell m}^X$,
\begin{eqnarray}
a_{\ell m}^{X}=a_{\ell m}^{X(r)}+ i  a_{\ell m}^{X(i)}.
 \nonumber %\label{acomplex}
\end{eqnarray}
It follows from Eq.~(\ref{arelation}) that
\begin{eqnarray}
a_{\ell 0}^{X(i)}=0, ~~~a_{\ell m}^{X(r)} = (-1)^ma_{\ell, -m}^{X(r)},
~~~{\rm and~}
a_{\ell m}^{X(i)} = (-1)^{m+1}a_{\ell, -m}^{X(i)}.
\nonumber
\end{eqnarray}
Thus, the set of multipole coefficients $a_{\ell m}^X$, where
$\ell \geq m \geq -\ell$, is fully equivalent
to the set of real quantities $a_{\ell 0}^{X(r)}, a_{\ell m}^{X(r)},
a_{\ell m}^{X(i)},~ {\rm where}~ {\ell} \geq m \geq 1$. We will be using
index $c$ to denote all quantities as $a_{\ell m}^{X(c)}$ where
$c = r ~{\rm or} ~i$ if $m \geq 1$, and $c=r$ if $m=0$.

Each individual coefficient $a_{\ell m}^{X(c)}$ satisfies a zero-mean
normal distribution
\begin{eqnarray}
  \nonumber %\label{falmc}
 f(a_{\ell m}^{X(c)})=
 \frac{1}{\sqrt{2\pi}\sigma_{\ell m}^{X(c)}}
 e^{-(a_{\ell m}^{X(c)})^2/2(\sigma_{\ell m}^{X(c)})^2}.
\end{eqnarray}
The variances, i.e. the squares of standard deviations
$\sigma_{\ell m}^{X(c)}$, are the second moments already analyzed
in the previous subsection. Comparing with Eq.~(\ref{b1}) one finds
\begin{eqnarray}
 \nonumber %\label{deltalmc}
\begin{array}{c}(\sigma_{\ell 0}^{X(r)})^2=\langle
a_{\ell 0}^{X(r)} a_{\ell 0}^{X(r)}\rangle=C_{\ell}^{XX}, \\
(\sigma_{\ell m}^{X(r)})^2=\langle a_{\ell m}^{X(r)}
a_{\ell m}^{X(r)}\rangle = \frac{1}{2} C_{\ell}^{XX},~~
(\sigma_{\ell m}^{X(i)})^2=\langle a_{\ell m}^{X(i)}a_{\ell m}^{X(i)} \rangle
=\frac{1}{2}C_{\ell}^{XX},~~(\ell \geq m \geq 1) \\
\langle a_{\ell m}^{X(r)} a_{\ell m}^{X(i)}\rangle=0.
\end{array}
\end{eqnarray}
Denoting $\sigma_\ell^X = \sqrt{C_{\ell}^{XX}}$, one can write:
$\sigma_{\ell 0}^X=\sigma_{\ell}^X$ and $\sigma_{\ell
m}^{X(r)}=\sigma_{\ell m}^{X(i)}=\sigma_{\ell}^X/\sqrt{2}$. In
other words, the standard deviations of the $(2\ell +1)$ random
variables $\frac{a^{X(r)}_{\ell 0}}{\sqrt{2}}, a_{\ell m}^{X(r)},
a_{\ell m}^{X(i)}$, ${\rm where}~\ell \geq m\geq1$, are all equal
and amount to one and the same number
$\sigma_{\ell}^{X}/{\sqrt{2}}$.

The equivalent statements follow from the pdf for complex multipole
coefficients $a_{\ell m}^X$ $(m\ge1)$. Each individual $a_{\ell m}^X$ obeys a
complex zero-mean normal distribution \cite{distribution,book5}
with the standard deviation $\sigma_{\ell}^X$:
\begin{eqnarray}
\label{lgauss} f(a_{\ell m}^X)=\frac{1}{\pi(\sigma_{\ell}^{X})^2}
e^{-(a_{\ell m}^Xa_{\ell m}^{X*})/(\sigma_{\ell}^X)^2}.
\end{eqnarray}

%%%%%%%%%%%%%%%%%%%%%%%%%%%%%%%%%%%%%%%%%%%%%%%%%%%%%%%%%%%%%%%%%%%%%%%%%%%%%%%%%%%
%%%%%%%%%%%%%%%%%%%%%%%%%%%%%%%%%%%%%%%%%%%%%%%%%%%%%%%%%%%%%%%%%%%%%%%%%%%%%%%%%%%

\subsubsection{Joint probability density function for multipole
coefficients\label{s2.3.3}}

Although the joint pdf for the set of all coefficients $\stackrel{s}{c}_{\bf n}$
is a factorizable Gaussian distribution, the radiative transfer equations make
similar factorization impossible for some subsets of all multipole
coefficients $a_{\ell m}^X$. Specifically, $a_{\ell m}^T$ and  $a_{\ell' m'}^E$
`do know about each other', when $\ell = \ell'$, $m=m'$. Their joint pdf is not
factorizable and their correlations do not vanish, as was made clear in
Eq.~(\ref{bb21}).

Let us denote sets of coefficients by curly brackets. The full set of all
multipole coefficients $\{a_{\ell m}^X\}$ is
\begin{eqnarray}
\nonumber
 \{a_{\ell m}^X\} \equiv (a_{\ell m}^{X}|~~X=T,E,B;
 ~\ell=2,3,\cdot\cdot\cdot;~m=-\ell,\cdot\cdot\cdot,\ell).
\end{eqnarray}
In this set we did not include the potentially nonzero coefficients $a_{00}^T$,
$a_{1m}^T$, and hence, from now on, we restrict our analysis to multipoles
$\ell\geq2$.

Each member of the set $\{a_{\ell m}^X\}$ is a linear combination
(\ref{aa1}) of complex coefficients $\stackrel{s}{c}_{\bf n}$
whose joint distribution is a zero-mean normal pdf. This
guarantees that the joint distribution for the set $\{a_{\ell
m}^X\}$ is a zero-mean multivariate normal distribution
\cite{distribution}. Such distributions are completely
characterized by their second moments. For the physical reasons
explained in Sec.~\ref{subsection2.3.1} we assumed that
there is no $TB$ and $EB$ second moments, so that the $B$
components `do not know' about $T$ and $E$ components. As for the
non-vanishing second moments, they were calculated in
Eq.~(\ref{b1}) and Eq.~(\ref{bb21}).

Combining the above statements into a formula, we can write the joint pdf
for the full set $\{a_{\ell m}^X\}$:
\begin{eqnarray}
 \nonumber %\label{totaljoint2}
f(\{a_{\ell m}^T, a_{\ell m}^E, a_{\ell m}^B \})=
f(\{a_{\ell m}^T, a_{\ell m}^E\}) f(\{a_{\ell m}^B\}).
\end{eqnarray}
Here,
\begin{eqnarray}
 \nonumber %\label{totaljoint3}
f(\{a_{\ell m}^B\})=
\prod_{\ell=2}^{\infty}\prod_{m=-\ell}^{\ell} f(a_{\ell m}^{B}),
\end{eqnarray}
where each of the functions $f(a_{\ell m}^{B})$ (for fixed $\ell$ and $m$)
is a single-variable normal distribution (\ref{lgauss}) with $X=B$. As for
the $T, E$ part of the joint pdf, we are interested in the subset of all
$a_{\ell m}^T, a_{\ell' m'}^E$, in which $\ell = \ell', m=m'$. This part
of the joint pdf is the product of the functions
$f(a_{\ell m}^{T},a_{\ell m}^{E})$, for each fixed pair of $\ell$
and $m$, which we are set to formulate.

The joint distribution for complex $a_{\ell m}^T$, $a_{\ell m}^E$
($m\neq0$) is a bivariate normal distribution which can be written
as \cite{distribution}
 \begin{eqnarray}
 f(a_{\ell m}^T,a_{\ell m}^E)=\frac{1}{\pi^2(\sigma_\ell^T
 \sigma_\ell^E)^2(1-\rho_\ell^2)}\exp\left\{-\frac{1}{1-\rho_{\ell}^2}
\left[\frac{{a_{\ell m}^T}a_{\ell m}^{T*}}{(\sigma_\ell^T)^2}+
\frac{{a_{\ell m}^E}a_{\ell
m}^{E*}}{(\sigma_\ell^E)^2}-\frac{\rho_{\ell}({a_{\ell m}^Ta_{\ell
m}^{E*}}+{a_{\ell m}^{T*}a_{\ell m}^{E}})}
{\sigma_\ell^T\sigma_\ell^E}\right]\right\}. \nonumber \\
\label{jointal0}
\end{eqnarray}
In the case $m=0$, the joint pdf reduces to a bivariate normal
distribution for real quantities $a_{\ell 0}^T$, $a_{\ell 0}^E$,
 \begin{eqnarray}
 f(a_{\ell 0}^T,a_{\ell 0}^E)=\frac{1}{2\pi\sigma_\ell^T
 \sigma_\ell^E\sqrt{1-\rho_\ell^2}}\exp\left\{-\frac{1}{2(1-\rho_{\ell}^2)}
\left[\frac{({a_{\ell 0}^T})^2}{(\sigma_\ell^T)^2}+
\frac{({a_{\ell
0}^E})^2}{(\sigma_\ell^E)^2}-\frac{2\rho_{\ell}{a_{\ell
0}^Ta_{\ell 0}^{E}}}
{\sigma_\ell^T\sigma_\ell^E}\right]\right\}. \nonumber \\
\label{jointal1}
\end{eqnarray}

The $\sigma_{\ell}^T$ and $\sigma_{\ell}^E$ in
Eq.~(\ref{jointal0}) and Eq.~(\ref{jointal1}) are the standard
deviations of $a_{\ell m}^T$ and $a_{\ell m}^E$, respectively. The
$\rho_{\ell}$ in Eq.~(\ref{jointal0}) is the correlation
coefficient, defined by
\begin{eqnarray}
\rho_{\ell}=\frac{\frac{1}{2}\langle a_{\ell m}^{T}a_{\ell m}^{E*} +
a_{\ell m}^{T*}a_{\ell m}^{E}\rangle}
{\sqrt{\langle a_{\ell m}^{T}a_{\ell m}^{T*}\rangle\langle a_{\ell m}^{E}
a_{\ell m}^{E*}\rangle}}=
\frac{C_{\ell}^{TE}}{ \sqrt{ C_{\ell}^{TT}C_{\ell}^{EE} } }.
\label{rholdefine}
\end{eqnarray}
Like the standard deviation $\sigma_{\ell}^X$, which, for a given $\ell$, is
one and the same for all $m$'s, the correlation coefficient $\rho_\ell$ is
also one and the same for all $m$'s, $\ell \geq m \geq -\ell$. For a bivariate
normal distribution the correlation coefficient obeys the condition
$|\rho_{\ell}|\leq1$, which essentially is Eq.~(\ref{ert}). Concrete numerical
values of $\rho_\ell$'s depend on the type of underlying perturbations and
cosmological background model.

The complex pdf (\ref{jointal0}) with $m \neq 0$ is equivalent to
the product of joint pdf's for real and imaginary parts of
$(a_{\ell m}^T,a_{\ell m}^E)$, i.e.~for $(a_{\ell
m}^{T(r)},a_{\ell m}^{E(r)})$ and $(a_{\ell m}^{T(i)},a_{\ell
m}^{E(i)})$ with $\ell \geq m \geq 1$ \cite{distribution}.
Introducing the label $c=r~{\rm or}~i$ one can write the common
formula
\begin{eqnarray}
f(a_{\ell m}^{T(c)},a_{\ell m}^{E(c)})=
 \frac{1}{ \pi\sigma_\ell^T \sigma_\ell^E\sqrt{1-\rho_\ell^2}}
 \exp\left\{-\frac{1}{(1-\rho_{\ell}^2)}
\left[ \frac{(a_{\ell m}^{T(c)})^2}{(\sigma_\ell^T)^2}+
\frac{(a_{\ell m}^{E(c)})^2}{(\sigma_\ell^E)^2}
-\frac{2\rho_{\ell} a_{\ell m}^{T(c)}a_{\ell m}^{E(c)} }
{\sigma_\ell^T\sigma_\ell^E}\right]\right\}.
\nonumber \\
\label{jointal0c}
\end{eqnarray}

The quantities
$\{\sigma_\ell^{T},\sigma_\ell^{E},\sigma_\ell^{B},\rho_\ell\}$
participating in and defining the pdf's are related with the CMB
power spectra calculated in Sec.~\ref{subsection2.3}:
\begin{eqnarray}
\label{relation}
 (\sigma_\ell^{T})^2=C_\ell^{TT},~~
 (\sigma_\ell^{E})^2=C_\ell^{EE},~~
 (\sigma_\ell^{B})^2=C_\ell^{BB},~~
 \rho_\ell\sigma_\ell^{T}\sigma_\ell^E=C_\ell^{TE}.
\end{eqnarray}
We will use these relations in our further discussion.

%%%%%%%%%%%%%%%%%%%%%%%%%%%%%%%%%%%%%%%%%%%%%%%%%%%%%%%%%%%%%%%%%%%%%%%%%%%%%%%%%%%
%%%%%%%%%%%%%%%%%%%%%%%%%%%%%%%%%%  SECTION 4   %%%%%%%%%%%%%%%%%%%%%%%%%%%%%%%%%%%%%%%%%%
%%%%%%%%%%%%%%%%%%%%%%%%%%%%%%%%%%%%%%%%%%%%%%%%%%%%%%%%%%%%%%%%%%%%%%%%%%%%%%%%%%%

\section{Estimators of the CMB power spectra and their probabilty density
functions\label{section3}}

The quantum-mechanical origin of cosmological perturbations, which
we represented by randomness of $\stackrel{s}{c}_{\bf n}$, does
not allow one to predict a unique CMB map, i.e. one concrete set
of multipole coefficients $a_{\ell m}^X$. The most that theory
allows is to formulate probability density functions, and
we have done this. The pdf's are fully characterized by the set of
numbers
$\{\sigma_\ell^{T},\sigma_\ell^{E},\sigma_\ell^{B},\rho_\ell\}$
or, equivalently, by the power spectra (\ref{relation}). Theory
can predict these parameters, while observations can check whether
these parameters are indeed as predicted.

To find a parameter from observations, one normally uses an appropriate random
variable (statistic) called an estimator. The estimator is unbiased if its mean
value is equal to the parameter, and is called the best if its variance is as small as
possible. A concrete value of the estimator calculated on the actually
observed data (in our case, on the set of actually observed multipole
coefficients $a_{\ell m}^X$) gives an estimate of the parameter. We will
denote by symbol $D^{XX'}_{\ell}$ the best unbiased estimators of the power
spectra $C^{XX'}_{\ell}$ and will consider separately the auto-correlation,
$X=X'$, and cross-correlation, $XX'=TE$, estimators.

%%%%%%%%%%%%%%%%%%%%%%%%%%%%%%%%%%%%%%%%%%%%%%%%%%%%%%%%%%%%%%%%%%%%%%%%%%%%%%%%%%%
%%%%%%%%%%%%%%%%%%%%%%%%%%%%%%%%%%%%%%%%%%%%%%%%%%%%%%%%%%%%%%%%%%%%%%%%%%%%%%%%%%%

\subsection{Estimators of the auto-correlation power spectra \label{Section3a}}

It is clear from Eq.~(\ref{b1}) that there exists plenty of unbiased
estimators of $C^{XX}_{\ell}$. Each of $2\ell +1$ estimators (random variables)
$\left(a_{{\ell}m}^{X}a_{{\ell}m}^{X*}\right)$ with $\ell \geq m \geq -\ell$
calculated on the observed multipoles provides an unbiased estimate of
$C^{XX}_{\ell}$. However, there exists only one best unbiased estimator.
It is given by
\begin{eqnarray}
 \nonumber %\label{dlxx}
 D_{\ell}^{XX}=\frac{1}{2\ell+1}
 \sum_{m=-\ell}^{\ell}\left(a_{{\ell}m}^{X}a_{{\ell}m}^{X*}\right).
\end{eqnarray}
The expectation value of this estimator is, as required,
\begin{eqnarray}
\label{35}
 \langle D_{\ell}^{XX}\rangle= \frac{1}{2\ell+1}
 \sum_{m=-\ell}^{\ell}\langle a_{{\ell}m}^{X}a_{{\ell}m}^{X*}\rangle
 =(\sigma_{\ell}^X)^2=C_{\ell}^{XX}.
\end{eqnarray}
The standard deviation is $\Delta{D_{\ell}^{XX}}\equiv
 \sqrt{\langle (D_{\ell}^{XX})^2\rangle-\langle
 D_{\ell}^{XX}\rangle^2}$,
%\[
% \Delta{D_{\ell}^{XX}}\equiv
% \sqrt{\langle (D_{\ell}^{XX})^2\rangle-\langle
% D_{\ell}^{XX}\rangle^2},
%\]
amounts to \cite{knox}, \cite{zaldarriaga}, \cite{estimator}
\begin{eqnarray}
\label{DeltaDXX}
 \Delta{D_{\ell}^{XX}}=(\sigma_{\ell}^X)^2
 \sqrt{\frac{2}{2{\ell}+1}}.
\end{eqnarray}
This is the smallest, but still a non-zero, variance among all possible
quadratic estimators \cite{estimator}. (Formula (\ref{DeltaDXX}) provides one
of interpretations of the term `cosmic variance'.)

We have access to only one realization of the inherently random
CMB field, that is, we can deal with only one observed set of
multipole coefficients $a_{\ell m}^X$. Under certain conditions,
the access to only one realization of the stochastic process is
sufficient for extraction the true parameters of the process, with
probability arbitrarily close to 1, from the data. When this is
possible the stochastic process is called ergodic. However, there
is no ergodic processes on a 2-sphere (our sky) \cite{estimator}.
We will always be facing some uncertainty surrounding the
parameters extracted from the observed CMB field.

The variance of an estimator at each $\ell$ is just a number, 
and is not sufficient for proper handling of the data. The full available
information is contained in the pdf of the estimator. In order to derive
the pdf of $D_{\ell}^{XX}$, ($XX= TT, EE, BB$), we will first rewrite it
as
\begin{eqnarray}
 \nonumber %\label{red}
D_{\ell}^{XX}
=\frac{2}{2\ell+1}\left((\frac{a_{\ell 0}^{X(r)}}{\sqrt{2}})^2
+\sum_{m=1}^{\ell} [(a_{\ell m}^{X(r)})^2
+(a_{\ell m}^{X(i)})^2]\right), ~~~(m\geq1).
\end{eqnarray}
It is clear that the estimator $D_{\ell}^{XX}$ is just the sum of squares of
$(2\ell+1)$ independent normal variables. Thus, for every fixed $\ell$,
the estimator obeys a $\chi^2$ distribution with $2\ell +1$ degrees
of freedom \cite{distribution}.

In what follows, Sec.~\ref{s3.4}, we will be dealing with the
problem of observations on a cut sky. The cut sky limitation
effectively means that, for a given $\ell$, one has access not to
all $2\ell +1$ degrees of freedom, but only to a smaller number
$n$, $n= (2\ell+1)f_{\rm sky}$, where $f_{\rm sky}$ is the
observed fraction of the sky \cite{cut1,cut2,cutpdf}. 
In preparation to this problem, we
are writing down the pdf for $D_{\ell}^{XX}$ with $n$ degrees of
freedom, $D_{\ell}^{XX}(n)$, where $n$ is not necessarily equal to
$2\ell+1$. Denoting $V\equiv
nD_{{\ell}}^{XX}(n)/(\sigma_\ell^{X})^2$, one writes
\cite{distribution}
\begin{eqnarray}
\label{f1}
 f(D_{\ell}^{XX}(n))=\frac{nV^{n/2-1}e^{-V/2}}{2^{n/2}\Gamma(n/2)
 (\sigma_\ell^{X})^2},
\end{eqnarray}
where $\Gamma$ is the $Gamma$-function. Certainly, when $n=2\ell +1$, the
mean value and the standard deviation of the variable $D_{\ell}^{XX}$
derivable from the pdf (\ref{f1}), coincide with the already written expressions
(\ref{35}) and (\ref{DeltaDXX}).

%%%%%%%%%%%%%%%%%%%%%%%%%%%%%%%%%%%%%%%%%%%%%%%%%%%%%%%%%%%%%%%%%%%%%%%%%%%%%%%%%%%
%%%%%%%%%%%%%%%%%%%%%%%%%%%%%%%%%%%%%%%%%%%%%%%%%%%%%%%%%%%%%%%%%%%%%%%%%%%%%%%%%%%

\subsection{Estimators of the cross-correlation $TE$ power spectrum
\label{sectionIIIB}}

The $TE$ cross-correlation is in the focus of this paper, so we
shall explore the $TE$ estimator in depth, without sparing
details. We know that the $C_{\ell}^{TE}$ correlation function
should be positive, if caused by density perturbations, from
$\ell=2$ and up to $\ell \approx 53$; and  it should be negative,
if caused by gravitational waves, from $\ell=2$ and up to $\ell
\approx 150$ \cite{deepak}, a further study of the
$TE$ correlation was conducted in \cite{polna1}. But $C_{\ell}^{TE}$ is only the mean
value of the correlations. In a particular realization of the
random field, i.e. for the actually observed $a_{\ell m}^X$, the
$D_{\ell}^{TE}$ outcomes can be negative, despite the fact that
the theoretical $C_{\ell}^{TE}$ is positive, and vice versa.
Suppose negative correlations are observed in a set of multipoles.
Is it necessarily a signature of gravitational waves, or it may be
a statistical fluke of density perturbations alone? Suppose
positive correlations are observed in a set of multipoles. Are
they `positive enough' to be explained by density perturbations
alone, or they require an admixture of gravitational waves? This
is the kind of fundamental questions of the $TE$ approach that we
need to answer \cite{0707}, and they can only be answered on the
basis of the probability density functions for the estimators.

It is clear from the previous discussion and Eq.~(\ref{bb21}) that the best
unbiased estimator $D_\ell^{TE}$ is given by
\begin{eqnarray}
 D_\ell^{TE}\equiv\frac{1}{2(2\ell+1)}\sum_{m=-\ell}^{\ell}
 \left(a_{{\ell}m}^{T}a_{{\ell}m}^{E*}+a_{{\ell}m}^{T*}a_{{\ell}m}^{E}\right).
 \nonumber %\label{TEestimator}
\end{eqnarray}
It can also be written as
\begin{eqnarray}
\label{dltev}
D_{\ell}^{TE}=\frac{1}{2\ell+1}\left(v_{\ell 0}^{(r)}+\sum_{m=1}^{\ell}
(v_{\ell m}^{(r)}+v_{\ell m}^{(i)})\right)
\end{eqnarray}
where
\begin{eqnarray}\label{definev}
v_{\ell 0}^{(r)}\equiv a_{\ell 0}^{T(r)} a_{\ell
0}^{E(r)},~~v_{\ell m}^{(r)}\equiv 2a_{\ell m}^{T(r)}a_{\ell
m}^{E(r)},~~v_{\ell m}^{(i)}\equiv 2a_{\ell m}^{T(i)}a_{\ell
m}^{E(i)},~~(m\geq1).
\end{eqnarray}
As before, we denote quantities (\ref{definev}) by $v_{\ell m}^{(c)}$,
where $c=r ~{\rm or}~ i$, $m \geq 0$. Obviously, as the coefficients
$a_{\ell 0}^{T}$, $a_{\ell 0}^{E}$ are real, $v_{\ell 0}^{(i)}=0$.

Each of the quantities $v_{\ell m}^{(c)}$ is an unbiased (but not the best)
estimator of $C_{\ell}^{TE}$, as follows from the averages based on the joint
pdf (\ref{jointal0c}):
\begin{eqnarray}
 \nonumber %
\langle v_{\ell m}^{(c)} \rangle=
\rho_{\ell}\sigma_{\ell}^T\sigma_{\ell}^E
=C_{\ell}^{TE},~~
\Delta  v_{\ell m}^{(c)}=
\sqrt{1+\rho^2_\ell}\sigma_{\ell}^T\sigma_{\ell}^E=
\sqrt{C_{\ell}^{TT}C_{\ell}^{EE}+(C_{\ell}^{TE})^2}.
\end{eqnarray}
In order to analyze the pdf of the estimator $D_\ell^{TE}$ in (\ref{dltev}), it is
instructive to study in the pdf's for quantities $v_{\ell m}^{(c)}$. 
In Appendix \ref{appendixA} we present a detailed discussion
of these pdfs and some of their relevant properties.

Based on the above analysis and the results from Appendix \ref{appendixA},
we are now in the position to derive the pdf for the
best unbiased estimator $D_{\ell}^{TE}$. Firstly, the mean value
and the standard deviation of $D_{\ell}^{TE}$ are known from the joint $TE$
pdf:
\begin{equation}
\label{meanw}
 \langle D_{\ell}^{TE}\rangle=\frac{1}{2(2\ell+1)}\sum_{m=-\ell}^{\ell}
\langle a_{{\ell}m}^{T}a_{{\ell}m}^{E*}+a_{{\ell}m}^{T*}a_{{\ell}m}^{E}\rangle
 =\rho_{\ell}\sigma_{\ell}^T\sigma_{\ell}^E = C_{\ell}^{TE},
\end{equation}
\begin{equation}
\label{derivationw}
 \Delta D_{\ell}^{TE}\equiv\sqrt{\langle (D_{\ell}^{TE})^2\rangle
 -\langle D_{\ell}^{TE}\rangle^2}
 =\sigma_{\ell}^T\sigma_{\ell}^E\sqrt{\frac{\rho_{\ell}^2+1}{2{\ell}+1}}
 =\sqrt{\frac{{C_\ell^{TT}C_\ell^{EE}+(C_\ell^{TE})^2}}{{2\ell+1}}}
=\frac{\Delta v_{\ell m}^{(c)}}{\sqrt{2\ell+1}}.
\end{equation}
Clearly, $\Delta D_{\ell}^{TE}$ is smaller than $\Delta v_{\ell m}^{(c)}$ by
a factor $\sqrt{2\ell+1}$. The minimum of $\Delta D_{\ell}^{TE}$ is
at $\rho_{\ell}=0$ and the maximum is at $|\rho_{\ell}|=1$. The maximum is
only $\sqrt{2}$ times greater than the minimum. For growing $\ell$ and fixed
$\sigma_{\ell}^T$, $\sigma_{\ell}^E$, $\rho_{\ell}$, the standard deviation
is decreasing as $1/\sqrt{2\ell+1}$.

To derive the pdf $f(D_{\ell}^{TE})$ one notes (see Eq.~(\ref{dltev}) that the
variable $D_{\ell}^{TE}$ is essentially the sum of $2\ell + 1$ products of
normally distributed variables from two classes, $T$ and $E$, with standard
deviations $\sigma_{\ell}^T$ and $\sigma_{\ell}^E$, respectively. Since we will
be interested in the situation where the number $n$ of accessible degrees of
freedom is smaller than $2\ell +1$, we have to make the derivation more
general. It is instructive to start from a somewhat abstract problem.

Let the three variables $x,y,z$ be defined by
 \begin{equation}\label{xyz}
 x\equiv X_1^2+X_2^2+\cdot\cdot\cdot+X_n^2,
 ~ ~y\equiv Y_1^2+Y_2^2+\cdot\cdot\cdot+Y_n^2,
 ~~ z\equiv X_1Y_1+X_2Y_2+\cdot\cdot\cdot+X_nY_n,~
 \end{equation}
where $X_i$ and $Y_i$ ($1\leq i\leq n$) are real variables.
We assume that within each class, $X$ and $Y$, the variables are statistically
independent and each $X_i$ and $Y_i$ obeys a zero-mean normal distribution
with standard deviation $\sigma^X$ and $\sigma^Y$, respectively. We also
assume that each  pair $X_i, Y_i$ obeys a zero-mean bivariate normal
distribution with correlation coefficient $\rho$. Then, the joint pdf
$f(x,y,z)$ is a Wishart distribution with $n$ degrees of freedom
\cite{distribution}:
\begin{eqnarray}
 \nonumber %\label{wishart}
f(x,y,z)&=&
\left(\frac{1}{4(1-\rho^2)(\sigma^X\sigma^Y)^2}\right)^{n/2}
\frac{(xy-z^2)^{\frac{n-3}{2}}}{\pi^{1/2}
\Gamma(\frac{n}{2})\Gamma(\frac{n-1}{2})} \\ \nonumber
&\times&\exp\left\{-\frac{1}{2(1-\rho^2)}\left(
\frac{x}{(\sigma^{X})^2}+\frac{y}{(\sigma^{Y})^2}
-\frac{2\rho z}{\sigma^{X}\sigma^Y}
\right)\right\} .
\end{eqnarray}
From this distribution, by integrating over variables $x$ and $y$, one
derives the pdf $f(z)$:
\begin{eqnarray}
\label{fz}
 f(z)=\sqrt{\frac{|z|^{n-1}}{2^{n-1}\pi(\sigma^X\sigma^Y)^{n+1}(1-\rho^2)}}
\frac{1}{\Gamma(\frac{n}{2})}\exp
\left\{\frac{\rho}{1-\rho^2}\frac{z}{\sigma^X\sigma^Y}\right\}
 K_{\frac{n-1}{2}}\left(\frac{|z|}{(1-\rho^2)\sigma^X\sigma^Y}\right),
\end{eqnarray}
where $K_{\frac{n-1}{2}}$ is the $(\frac{n-1}{2})$-order modified Bessel
function. In this derivation we have used (2.382.2) and (3.471.9) from
Ref.~\cite{book1}. In the special case $n=1$, and denoting the pair
$X_1, Y_1$ by $a_{\ell 0}^T, a_{\ell 0}^E$ or $\sqrt{2}a_{\ell m}^{T(c)},
\sqrt{2}a_{\ell m}^{E(c)}$, (for fixed $m\geq 1$, $c=r ~{\rm or}~ i$), one
finds that the pdf (\ref{fz}) returns to the pdf (\ref{pdfvl0}) for each
$v_{\ell m}^{(c)}$.

The pdf for the estimator $D_{\ell}^{TE}(n)$ with $n$ degrees of freedom,
where $n$ is not necessarily equal to $2\ell+1$, follows now from Eq.~(\ref{fz}).
Comparing Eqs.~(\ref{dltev},~\ref{xyz}) we write $D_{\ell}^{TE}(n)=z/n$ and,
then, the pdf for $D_{\ell}^{TE}(n)$ (omitting argument $n$):
 \begin{eqnarray}
 f(D_{\ell}^{TE})&=&\left(\frac{|D^{TE}_{\ell}|}{2}\right)^{\frac{n-1}{2}}
 \left(\frac{n}{\sigma_{\ell}^T\sigma_{\ell}^E}\right)^{\frac{n+1}{2}}
 \sqrt{\frac{1}{\pi(1-\rho_{\ell}^2)}}\frac{1}{\Gamma(\frac{n}{2})}  \nonumber\\ \label{pdfwl}
 &\times&\exp
\left\{\frac{\rho_{\ell}}{1-\rho_{\ell}^2}\frac{nD^{TE}_{\ell}}{\sigma_{\ell}^T
\sigma_{\ell}^E}\right\}
 K_{\frac{n-1}{2}}\left(\frac{n|D^{TE}_{\ell}|}{(1-\rho_{\ell}^2)\sigma_{\ell}^T
 \sigma_{\ell}^E}\right).
 \end{eqnarray}
Certainly, Eqs.~(\ref{meanw},~\ref{derivationw}) are consequences of this pdf,
when $n =2\ell+1$. In Appendix \ref{appendixB} we discuss some useful properties of
this pdf. These properties are used in our consequent analysis, in particular for determining
the confidence intervals of $D^{TE}_{\ell}$ for specific cosmological models.

%%%%%%%%%%%%%%%%%%%%%%%%%%%%%%%%%%%%%%%%%%%%%%%%%%%%%%%%%%%%%%%%%%%%%%%%%%%%%%%%%%%
%%%%%%%%%%%%%%%%%%%%%%%%%%%%%%%%%%  SECTION 4   %%%%%%%%%%%%%%%%%%%%%%%%%%%%%%%%%%%%%%%%%%
%%%%%%%%%%%%%%%%%%%%%%%%%%%%%%%%%%%%%%%%%%%%%%%%%%%%%%%%%%%%%%%%%%%%%%%%%%%%%%%%%%%

\section{Correlation coefficient $\rho_\ell$ and estimator $D^{TE}_{\ell}$
in specific models\label{s3}}

The sign and value of the $TE$ correlations, as well as their statistical
distribution, crucially depend on the sign and value of the correlation
coefficient $\rho_\ell$. As we know, at lower
$\ell$'s, $\rho_\ell$ should be negative for gravitational waves and positive
for density perturbations. However, specific numerical values of $\rho_\ell$
are different in different models. It is difficult to find $\rho_\ell$
analytically, so we are using numerical calculations to find it
numerically.

%%%%%%%%%%%%%%%%%%%%%%%%%%%%%%%%%%%%%%%%%%%%%%%%%%%%%%%%%%%%%%%%%%%%%%%%%%%%%%%%%%%
                                                                                                       %%%%%% SUBSECTION 3.1 %%%%%%
%%%%%%%%%%%%%%%%%%%%%%%%%%%%%%%%%%%%%%%%%%%%%%%%%%%%%%%%%%%%%%%%%%%%%%%%%%%%%%%%%%%

\subsection{Numerical evaluation of the correlation
coefficient $\rho_{\ell}$\label{s3.1}}

Assuming that the correlation functions $C_{\ell}^{TT},
C_{\ell}^{EE}, C_{\ell}^{TE}$ are known from numerical modelling,
the $\rho_\ell$ can be found from the defining expressions
(\ref{rholdefine}), (\ref{TotalCell_dp+gw}):
\begin{eqnarray}
\rho_\ell = \frac{C_{\ell}^{TE}}{\sqrt{C_{\ell}^{TT}C_{\ell}^{EE}}} =
\frac{C_{\ell}^{TE}(dp) + C_{\ell}^{TE}(gw)}{\sqrt{
\left(C_{\ell}^{TT}(dp)+C_{\ell}^{TT}(gw)\right)
\left(C_{\ell}^{EE}(dp)+C_{\ell}^{EE}(gw)\right)}}.
\label{rhol_Cls}
\end{eqnarray}

In our numerical calculations we work with a
single cosmological background model, which is the `best-fit'
$\Lambda$CDM cosmology based on the 5-year WMAP observations and other
data sets \cite{5map_cosmology}. The parameters of the background model are
\begin{eqnarray}
\begin{array}{c}
\Omega_b h^2 = 0.02265 \pm 0.00059, ~~~
\Omega_c h^2 = 0.1143 \pm 0.0034, \\
\Omega_{\Lambda}  = 0.721\pm 0.015, ~~~ \tau_{reion} = 0.084 \pm
0.016.
\end{array}
\label{WMAPBestFit}
\end{eqnarray}
The associated value of the Hubble parameter is $h =0.701 \pm 0.013$.
In calculations, we are using the central values of these parameters.

As for the perturbations, the WMAP5 `best-fit' analysis suggests
no contribution from gravitational waves, whereas density
perturbations are characterized by the spectral index $n_s$ and
the amplitude $\Delta_{\cal {R}}^2$ \cite{5map_cosmology,5map_para}:
\begin{eqnarray}
\label{WMAPBestFitDP}
n_s = 0.960^{+0.014}_{-0.013},~~~
\Delta_{ \cal{R} }^2 = ( 2.457^{+0.092}_{-0.093} ) \times 10^{-9}, ~~~
\left( {\rm at}~k = 0.002{\rm Mpc}^{-1} \right).
\end{eqnarray}
The WMAP5 `best-fit' parameters (\ref{WMAPBestFit}), (\ref{WMAPBestFitDP})
imply the temperature quadrupole
\begin{equation}
\label{bestfitC2}
\frac{\ell(\ell+1)C_{\ell=2}^{TT}}{2\pi} = 1160 ~({\mu}{\rm K}^2).
\end{equation}

Being guided by the quantum-mechanical theory of cosmological
perturbations (see Sec.~\ref{intro}), we include gravitational
waves in our analysis from the very beginning, as a necessary
ingredient of CMB anisotropies at lower $\ell$'s. We allow the
{\it dp} parameters (\ref{WMAPBestFitDP}) to vary and strive to
achieve the total theoretical $TT$, $EE$, $TE$ spectra as close as
possible to the observed data. Gravitational waves are
parameterized by the primordial spectral index $n_t$, where
$n_t=n_s -1$ (unless some other $n_t$ is chosen for purposes of
illustration), and the amplitude parameter $R$, where $R$ is the
ratio of the {\it gw} and {\it dp} contributions to the
temperature quadrupole:
\begin{eqnarray}
\label{defineR}
 R \equiv \frac{C_{\ell=2}^{TT}(gw)}{C_{\ell=2}^{TT}(dp)}.
\end{eqnarray}
To avoid any ambiguities in definitions and derived upper limits,
we do not use the `tensor-to-scalar ratio' parameter $r$ (see \cite{0707}).
For very rough comparison of results, one can keep in mind the approximate
relationship $r\simeq2R$, which follows from the study \cite{turner}. In
power spectra calculations we rely on the CMBFast code \cite{numerical1}
(Alternatively one can use the CAMB \cite{numerical2} which gives consistent results).
The correlation coefficient $\rho_{\ell}$ is then found
from Eq.~(\ref{rhol_Cls}).

%%%%%%%%%%%%%%%%%%%%%%%%%%%%%%%%%%%%
\begin{figure}
\begin{center}
\includegraphics[width=12cm,height=10cm]{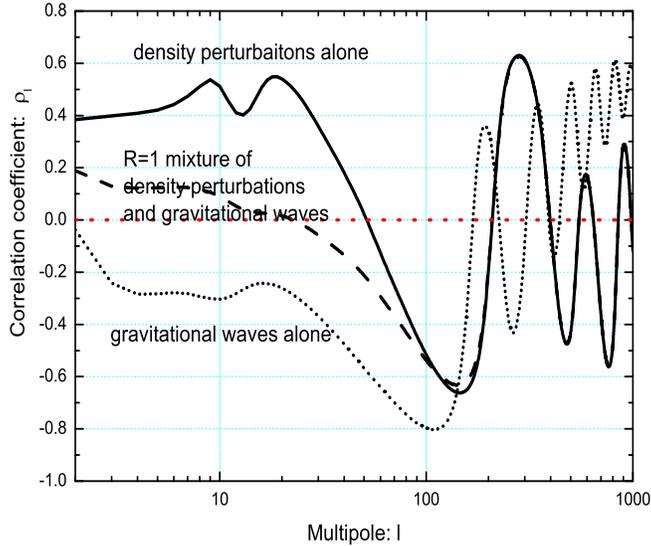}
\end{center}
\caption{The correlation coefficient $\rho_{\ell}$ as a function
of $\ell$ in models with different amounts
of gravitational waves. The solid line denotes the best-fit
$\Lambda$CDM model with density perturbation alone,
the dashed line is for the $R=1$ mixture of {\it dp} and {\it gw}, and the
dotted line is for gravitational waves alone.}
\label{section2_fig1}
\end{figure}
%%%%%%%%%%%%%%%%%%%%%%%%%%%%%%%%%%%%

In the way of illustration, we plot in Fig.~\ref{section2_fig1} the
$\rho_{\ell}$ for three representative sets of perturbations on
the background of $\Lambda$CDM cosmology (\ref{WMAPBestFit}). The
solid curve shows $\rho_{\ell}$ for density perturbations alone
with best-fit parameters (\ref{WMAPBestFitDP}). The dotted curve
shows $\rho_{\ell}$ when gravitational waves alone are present
with $n_t=0$ and the amplitude normalization such that the
temperature quadrupole is equal to the best-fit WMAP5 prediction
(\ref{bestfitC2}). The dashed curve shows $\rho_{\ell}$ arising in
the $R=1$ mixture of density perturbations ($n_s=0.96$) and
gravitational waves ($n_t=0$) with total quadrupole equal to the
WMAP5 best-fit value (\ref{bestfitC2}).

As expected \cite{deepak} and seen in Fig.~\ref{section2_fig1},
$\rho_{\ell}$ is positive for {\it dp} and negative for {\it gw}
at lower multipoles, $\ell\lesssim 50$. In the intermediate case
$R=1$, $\rho_{\ell}$ is small $|\rho_{\ell}|<0.2$, as the {\it dp}
and {\it gw} contributions largely cancel each other. At higher
multipoles, $\ell>100$, the $R=1$ curve approaches the {\it dp}
curve, as the role of gravitational waves diminishes in comparison
with density perturbations \cite{h,deepak}. The oscillatory
features at very low multipoles $\ell \sim 10 - 20 $ (clearly seen
in the solid curve) are real. They are a reflection of effects of
$\Omega_{\Lambda}$ in combination with the reionization era
characterized by the optical depth $\tau_{reion}$.

Knowing $\rho_{\ell}$ as a function of $\ell$ one can build and analyse
pdf's for the $D^{TE}_\ell$ estimator in specific models. As we show below,
the sign difference in $\rho_{\ell}$'s for density perturbations and
gravitational waves is crucial for telling one sort of perturbations from
another.

%%%%%%%%%%%%%%%%%%%%%%%%%%%%%%%%%%%%%%%%%%%%%%%%%%%%%%%%%%%%%%%%%%%%%%%%%%%%%%%%%%%
                                                                                                       %%%%%% SUBSECTION 3.2 %%%%%%
%%%%%%%%%%%%%%%%%%%%%%%%%%%%%%%%%%%%%%%%%%%%%%%%%%%%%%%%%%%%%%%%%%%%%%%%%%%%%%%%%%%

\subsection{Probability of negative values of the estimator
$D_{\ell}^{TE}$\label{s3.2}}

\begin{figure}
\begin{center}
\includegraphics[width=12cm,height=9cm]{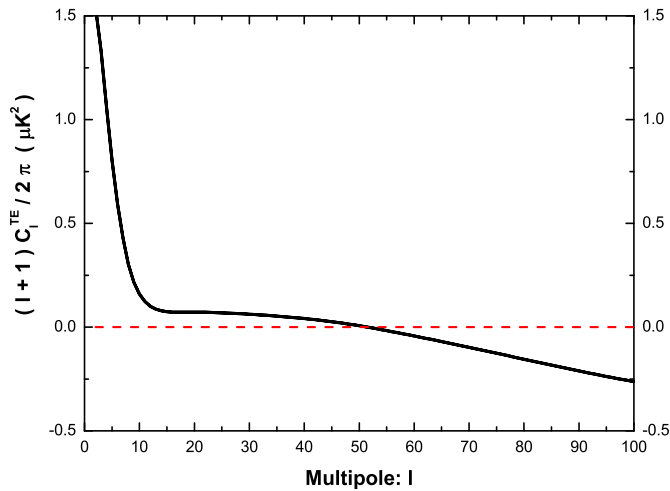}
\end{center}
\caption{The $(\ell +1)C_\ell^{TE}/2\pi$ power spectrum in units
of $({\mu}{\rm K}^2)$ for the best-fit WMAP5 model with no gravitational
waves.} \label{CellTE}
\end{figure}

In the center of our attention is the WMAP5 best-fit model (\ref{WMAPBestFit}),
(\ref{WMAPBestFitDP}) with no gravitational waves, $R=0$. The mean value of
the estimator $D_{\ell}^{TE}$, i.e. the $TE$ power spectrum, is shown in
Fig.~\ref{CellTE}. To assess the consistency of this graph with the actually
observed data points \cite{5map_cosmology,lambda} (see also
\cite{map_statement}), many of which are negative
at $\ell\lesssim50$, we start from the pdf for this model. We first ignore
noises and systematic effects, but will take them into account in
Sec.~\ref{sa4}.

%%%%%%%%%%%%%%%%%%%%%%%%%%%%%%%%%%%%
%%%%%%%%%%%%%%%%%%%%%%%%%%%%%%%%%%%%

\begin{figure}
\begin{center}
\includegraphics[width=16cm,height=12cm]{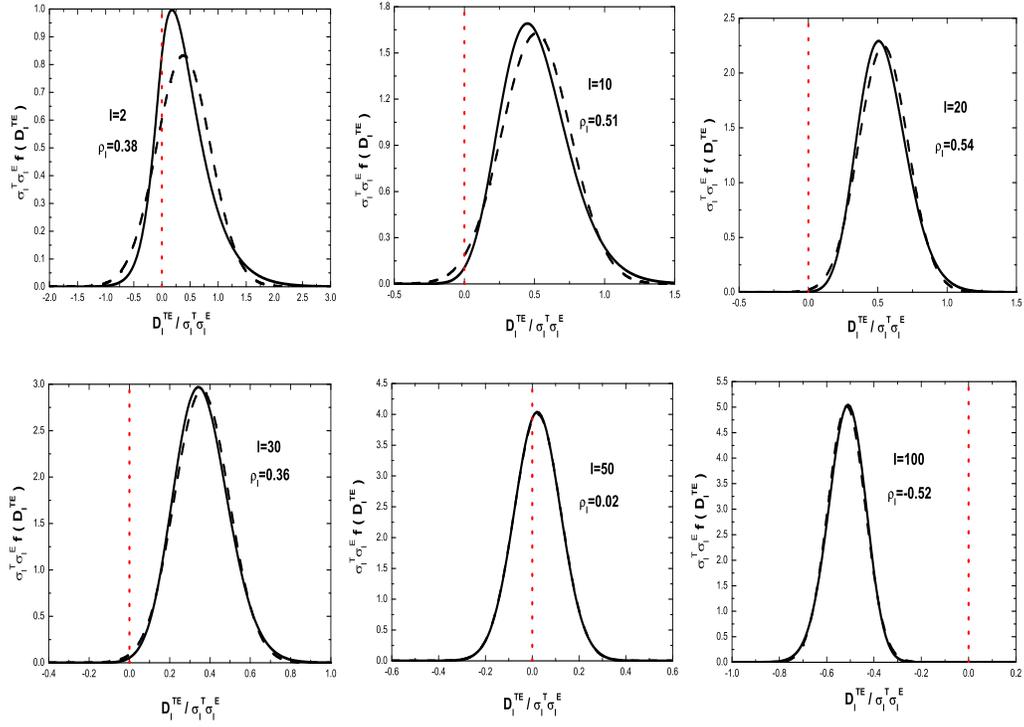}
\end{center}
\caption{Probability density functions $f(D^{TE}_{\ell})$ at
${\ell}=2,10,20,30,50,100$ for the WMAP5 best-fit model (\ref{WMAPBestFit}),
(\ref{WMAPBestFitDP}). The dashed curves show the Gaussian approximation to
the exact pdf's.}
\label{section3_fig4}
\end{figure}

Our derivation is based on Eq.~(\ref{pdfwl}), with $n=2\ell+1$, and
$\rho_{\ell}$, as plotted by solid line in Fig.~\ref{section2_fig1}.
In Fig.~\ref{section3_fig4} we show $f(D_{\ell}^{TE})$ for a series of
multipoles, $\ell=2,10,20,30,50,100$. When $\ell$ increases, maxima of
individual pdf's move from positive to negative values, crossing zero at
$\ell \approx 53$, in agreement with Fig.~\ref{CellTE}. The spread of
individual pdf's becomes narrower with increasing $\ell$ (note increasing
horizontal scale on different panels),
in agreement with the decreasing standard deviation $\Delta{D^{TE}_\ell}
\propto \sigma_{\ell}^T\sigma_{\ell}^E/\sqrt{(2\ell+1)}$, as reflected in
(\ref{derivationw}). We also plot by dashed curves the
Gaussian approximation to the exact pdf's,
\begin{eqnarray}
\label{GaussianApproximation}
 f_{G}(D_{\ell}^{TE})= \frac{1}{\sqrt{2\pi}\Delta
 D_{\ell}^{TE}}\exp\left[-\frac{(D_{\ell}^{TE}-C_{\ell}^{TE})^2}{2(\Delta
 {D_{\ell}^{TE}})^2}\right].
\end{eqnarray}
As could be anticipated, the Gaussian approximation becomes progressively
better with the increasing $\ell$.

We shall now draw various confidence intervals surrounding the mean value
at each $\ell$. The confidence regions are determined by the shortest
interval between the upper $(D^{TE}_{\ell})_U$ and the lower
$(D^{TE}_{\ell})_L$ boundaries enclosing a given surface area under the
pdf, as discussed in Appendix~\ref{appendixA}.

%%%%%%%%%%%%%%%%%%%%%%%%%%%%%%%%%%%%
%%%%%%%%%%%%%%%%%%%%%%%%%%%%%%%%%%%%

\begin{figure}
\begin{center}
\includegraphics[width=16cm,height=7cm]{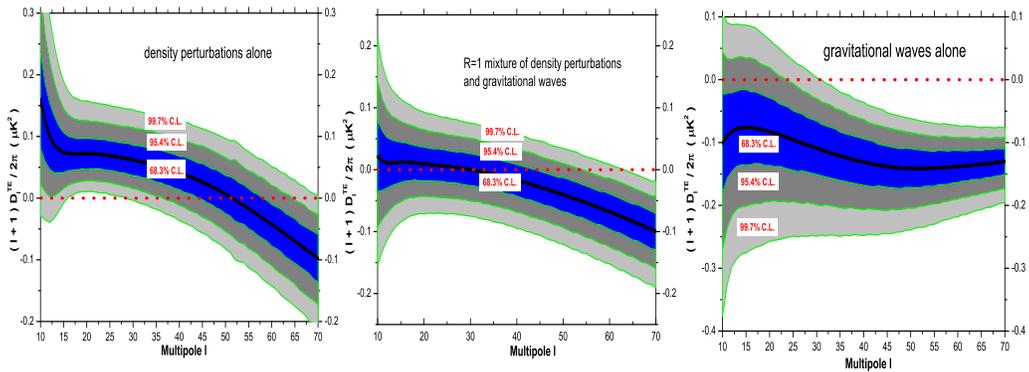}
\end{center}
\caption{The $68.3\%$, $95.4\%$, $99.7\%$ confidence intervals of
the variable $D_{\ell}^{TE}$ for three representative models
shown in Fig.~\ref{section2_fig1}. In all three panels, the solid black curves
are the mean values of the statistic $(\ell+1)D^{TE}_{\ell}/2\pi$, i.e.
the power spectra $(\ell+1)C^{TE}_{\ell}/2\pi$.}
\label{section3_fig5&6}
\end{figure}

In Fig.~\ref{section3_fig5&6}, we draw confidence intervals for all three
models considered in Fig.~\ref{section2_fig1}. Namely, the best-fit model
with density perturbations alone (left panel), the $R=1$ mixture of density
perturbations and gravitational waves (middle panel), and a model with
gravitational waves alone (right panel). It is seen on the left panel that the
$99.7\%$ confidence region is above the zero line for
$15\lesssim{\ell}\lesssim29$. The $95.4\%$ confidence region is above the
zero line for ${\ell}<37$, while the $68.3\%$ confidence region
lies in the positive territory for all ${\ell}<45$. In contrast, the right
panel for gravitational waves alone shows that all plotted confidence
regions are well below the zero line for ${\ell}\gtrsim25$.

%%%%%%%%%%%%%%%%%%%%%%%%%%%%%%%%%%%%
%%%%%%%%%%%%%%%%%%%%%%%%%%%%%%%%%%%%

\begin{figure}
\begin{center}
\includegraphics[width=12cm,height=10cm]{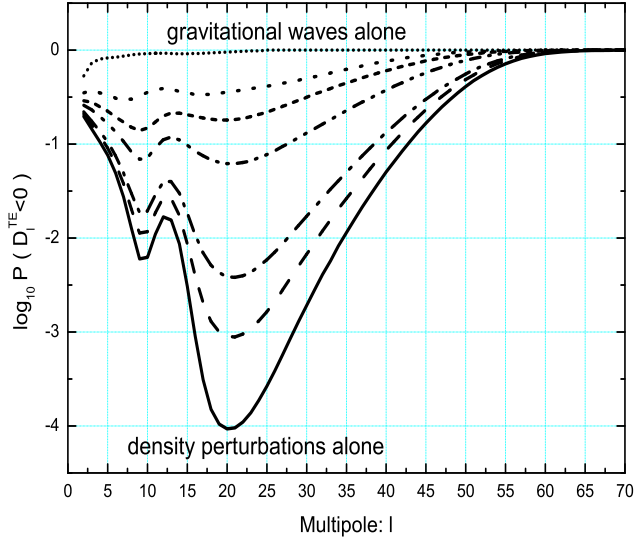}
\end{center}
\caption{The probability of finding negative $TE$ data points in
different models. The curves are ordered from the bottom to the
top by increasing $R$, $R=0.0,0.05,0.1,0.3,0.5,1.0$ with {\it gw}
alone at the very top.} \label{section3_fig8}
\end{figure}

Finally, in Fig.~\ref{section3_fig8}, we show the probability
$P(D^{TE}_{\ell}<0)$ of finding negative data points in models with
varying amounts of density perturbations and gravitational waves.
In all combinations of {\it dp} and {\it gw}, the spectral indices are
$n_s=0.96$, $n_t=0$ and the total quadrupole is equal to the best-fit
value (\ref{bestfitC2}). The black solid curve is for the WMAP5
best-fit model with no gravitational
waves, $R=0$. It is seen from the graph that in the case of density
perturbations alone, the probability $P(D^{TE}_{\ell}<0)$ is pretty small,
especially in the range of multipoles $10<\ell<30$. The lowest point is at
$\ell=20$, with $P(D^{TE}_{\ell=20}<0)=9.33\times10^{-5}$. Obviously, when
the amount of gravitational waves increases, the $P(D^{TE}_{\ell}<0)$
approaches 1.

%%%%%%%%%%%%%%%%%%%%%%%%%%%%%%%%%%%%
%%%%%%%%%%%%%%%%%%%%%%%%%%%%%%%%%%%%

%%%%%%%%%%%%%%%%%%%%%%%%%%%%%%%%%%%%%%%%%%%%%%%%%%%%%%%%%%%%%%%%%%%%%%%%%%%%%%%%%%%
%%%%%%%%%%%%%%%%%%%%%%%%%%%%%%%%%%  SECTION 5   %%%%%%%%%%%%%%%%%%%%%%%%%%%%%%%%%%%%%%%%%%
%%%%%%%%%%%%%%%%%%%%%%%%%%%%%%%%%%%%%%%%%%%%%%%%%%%%%%%%%%%%%%%%%%%%%%%%%%%%%%%%%%%

\section{Noises and other sources of increased uncertainty \label{sa4}}

The spread of the probability density functions discussed so far is the
unavoidable consequence of the inherent randomness of the sought after signal
$a_{{\ell}m}^{X}$. Additional randomness, and additional uncertainty
in the parameter estimation, is brought in by the process of measurement.
The output (observed) readings are contaminated by instrumental noises and
various foreground radiations, such as the synchrotron radiation, thermal
emission from dust, unresolved extragalactic sources, etc.
\cite{knox,map-window, planck,map1-te,5map_tt,5map_noise, lewis2}.
It is common to treat the residual systematic causes as `effective noises'
alongside with the `genuine' instrumental noises. In addition, the incomplete
sky coverage and the sky map cuts \cite{cut1,cut2,wishart2,cutpdf} introduce
extra loss of information in comparison to what is available on the full
sky. It is common to treat this complication by a proper reduction of
the potentially available $2\ell +1$ degrees of freedom at each $\ell$.

We shall now see how these factors expand the confidence intervals of various
estimators, and in particular the $D_{\ell}^{TE}$. Only by taking all these
factors into account, we will be in the position to test the `null' (no
gravitational waves) hypothesis (\ref{WMAPBestFit}), (\ref{WMAPBestFitDP}),
and make predictions for the forthcoming experiments.

%%%%%%%%%%%%%%%%%%%%%%%%%%%%%%%%%%%%%%%%%%%%%%%%%%%%%%%%%%%%%%%%%%%%%%%%%%%%%%%%%%%
                                                                                                       %%%%%% SUBSECTION 5.1 %%%%%%
%%%%%%%%%%%%%%%%%%%%%%%%%%%%%%%%%%%%%%%%%%%%%%%%%%%%%%%%%%%%%%%%%%%%%%%%%%%%%%%%%%%

\subsection{Effects of noises\label{s3.3}}

In general, the multipole coefficients ${a_{{\ell}m}^{X}}(o)$
observed in a CMB measurement consist of a signal convolved
with the beam window function, $a_{{\ell}m}^{X}(s)W_{\ell}$, and a
collective noise contribution $a_{{\ell}m}^{X}(n)$:
\begin{eqnarray}
{a_{{\ell}m}^{X}}(o)=a_{{\ell}m}^{X}(s)W_{\ell}+a_{{\ell}m}^{X}(n).
\nonumber
\end{eqnarray}
For WMAP and Planck satellites, the window function $W_{\ell}$ is close to 1
at $\ell\lesssim100$ \cite{map-window,planck}, so we set $W_{\ell}=1$.

As usual \cite{knox,map_gauss,wmap_verd,wishart2}, we assume that the noise
terms $a_{\ell m}^{X}(n)$ and the signal terms $a_{\ell m}^{X}(s)$ are
statistically independent of each other. Also, any two members in the set
$\{a_{\ell m}^{X}(n)|X=T,E,B; \ell=2,3,\cdot\cdot\cdot;
m=-\ell,\cdot\cdot\cdot,\ell\}$ are statistically independent, while each
member $a_{\ell m}^{X}(n)$ obeys a normal distribution with zero mean and
standard deviation $\sigma_{\ell}^{X}(n)$,
\begin{eqnarray}
\langle a_{\ell m}^{X}(n) a_{\ell m}^{X*}(n)\rangle=(\sigma_{\ell}^X(n))^2
=N_{\ell}^{XX}.
\nonumber
\end{eqnarray}
It is presumed that the noise power spectra $N_{\ell}^{XX}$ are known from
independent evaluations. Obviously, the $TE$ noise power spectrum vanishes,
$N_{\ell}^{TE}=0$, due to the absence of correlations between
$a_{\ell m}^{T}(n)$ and $a_{\ell m}^{E}(n)$.

%%%%%%%%%%%%%%%%%%%%%%%%%%%%%%%%%%%%%%%%%%%%%%%%%%%%%%%%%%%%%%%%%%%%%%%%%%%%%%%%%%%
%%%%%%%%%%%%%%%%%%%%%%%%%%%%%%%%%%%%%%%%%%%%%%%%%%%%%%%%%%%%%%%%%%%%%%%%%%%%%%%%%%%

\subsubsection {Probability density functions for multipole coefficients
$a_{\ell m}^X(o)$ and estimators $D_{\ell}^{XX'}(o)$}

Since ${a_{{\ell}m}^{X}}(s)$ and ${a_{{\ell}m}^{X}}(n)$ are statistically
independent and obey zero-mean normal distributions, their sum
${a_{{\ell}m}^{X}}(o)$ obeys a zero-mean normal distribution with the
variance $(\sigma_{\ell}^X(o))^2$, which is the sum of the signal and noise
variances,
\begin{eqnarray}
 \nonumber %
 (\sigma_{\ell}^X(o))^2=
 \langle {a_{{\ell}m}^{X}}(o) {a_{{\ell}m}^{X*}}(o)\rangle=
 \langle {a_{{\ell}m}^{X}}(s) {a_{{\ell}m}^{X*}}(s)\rangle+
 \langle {a_{{\ell}m}^{X}}(n)
 {a_{{\ell}m}^{X*}}(n)\rangle=C_{\ell}^{XX}+N_{\ell}^{XX}.
\end{eqnarray}
The joint pdf $f(a_{\ell m}^T(o), a_{\ell m}^E(o))$ is a zero-mean bivariate
normal distribution with the correlation coefficient $\rho_{\ell}(o)$,
\begin{eqnarray}
\label{rholdefine3}
 \rho_{\ell}(o)=\frac{\frac{1}{2}\langle {a_{{\ell}m}^{T}}(o)
 {a_{{\ell}m}^{E*}}(o)+{a_{{\ell}m}^{T*}}(o)
 {a_{{\ell}m}^{E}}(o)\rangle}{\sqrt{\langle {a_{{\ell}m}^{T}}(o)
 {a_{{\ell}m}^{T*}}(o)\rangle \langle {a_{{\ell}m}^{E}}(o)
 {a_{{\ell}m}^{E*}}(o)\rangle}}=\frac{C_{\ell}^{TE}}{\sqrt{(C_{\ell}^{TT}+N_
\ell^{TT})(C_{\ell}^{EE}+N_\ell^{EE})}}.
\end{eqnarray}
Comparing Eq.~(\ref{rholdefine3}) with Eq.~(\ref{rholdefine}) one finds that
the noises make the output correlation coefficient $\rho_{\ell}(o)$ smaller
than the no-noise correlation coefficient $\rho_{\ell}$.

The unbiased auto-correlation estimators in the presence of noise are defined
by
\begin{eqnarray}
 \nonumber %\label{dlxxo}
 {D_{\ell}^{XX}}(o)=\frac{1}{2\ell+1}\sum_{m=-\ell}^{\ell}(a_{\ell
 m}^{X}(o)a_{\ell m}^{X*}(o))-N_{\ell}^{XX}.
\end{eqnarray}
The expectation value and the standard deviation amount to
\begin{eqnarray}
\label{deltadlxxo}
\begin{array}{c}
 \langle D_{\ell}^{XX}(o) \rangle=(\sigma_{\ell}^{X}(o))^2-
N_{\ell}^{XX}=C_{\ell}^{XX},~~
\\
\\
\Delta D_{\ell}^{XX}(o)=\sqrt{\frac{2}{2\ell+1}}(\sigma_{\ell}^X(o))^2=
\sqrt{\frac{2}{2\ell+1}}(C_{\ell}^{XX}+N_{\ell}^{XX}).
\end{array}
\end{eqnarray}
Comparing with Eq.~(\ref{DeltaDXX}) one finds that the noises make
the standard deviation $\Delta D_{\ell}^{XX}(o)$ larger than the
no-noise value $\Delta D_{\ell}^{XX}$.

The pdf of the estimator $D_{\ell}^{XX}(o)$ with $n$ degrees of freedom can
be derived in a way similar to that in Sec.~\ref{Section3a},
\begin{eqnarray}
 \nonumber %
f(D_{\ell}^{XX}(o))=\frac{n(V(o))^{n/2-1}e^{-V(o)/2}}{2^{n/2}\Gamma(n/2)
(\sigma_{\ell}^{X}(o))^2},
\end{eqnarray}
where $V(o)\equiv {n(D_{\ell}^{XX}(o)+N_{\ell}^{XX})}/
{(\sigma_{\ell}^X(o))^2}$.
%\begin{eqnarray}
%V(o)\equiv {n(D_{\ell}^{XX}(o)+N_{\ell}^{XX})}/
%{(\sigma_{\ell}^X(o))^2}.
%\nonumber
%\end{eqnarray}

Finally, we have to consider the unbiased cross-correlation estimator
$D_{\ell}^{TE}(o)$ which is of prime importance for us,
\begin{eqnarray}
 \nonumber %\label{dlteo}
 {D_{\ell}^{TE}}(o)=\frac{1}{2(2\ell+1)}\sum_{m=-\ell}^{\ell}(a_{\ell
 m}^{T}(o)a_{\ell m}^{E*}(o)+a_{\ell
 m}^{T*}(o)a_{\ell m}^{E}(o)).
\end{eqnarray}
The expectation value and the standard deviation amount to
\begin{eqnarray}
\label{deltadltro}
\begin{array}{c}
 \langle D_{\ell}^{TE}(o) \rangle=\rho_{\ell}(o)\sigma_{\ell}^{T}(o)
\sigma_{\ell}^{E}(o)=C_{\ell}^{TE},~~
\\
\\
\Delta D_{\ell}^{TE}(o)=\sqrt{\frac{(\rho_{\ell}(o))^2+1}{2\ell+1}}
\sigma_{\ell}^T(o)\sigma_{\ell}^E(o)=\sqrt{\frac{(C_{\ell}^{TT}+N_{\ell}^{TT})
(C_{\ell}^{EE}
+N_{\ell}^{EE})+(C_{\ell}^{TE})^2}{2\ell+1}}.
\end{array}
\end{eqnarray}
Again, the noises make the standard deviation swell in comparison with
the no-noise case (\ref{derivationw}).

The pdf for the estimator $D_{\ell}^{TE}(o)$ with $n$ degree of
freedom is similar to that in Sec.~\ref{sectionIIIB},
\begin{eqnarray}
 f(D_{\ell}^{TE}(o))&=&\left(\frac{|{{D}^{TE}_{\ell}(o)}|}{2}\right)
 ^{\frac{n-1}{2}}\left(\frac{n}{{\sigma_{\ell}^T(o)}{\sigma_{\ell}^E(o)}}
\right)^{\frac{n+1}{2}}
 \sqrt{\frac{1}{\pi(1-(\rho_{\ell}(o))^2)}}\frac{1}{\Gamma\left(\frac{n}{2}
\right)}
  \nonumber \\
 \label{newpdfwl}
 &\times&
 \exp
\left\{\frac{{\rho_{\ell}(o)}}{1-(\rho_{\ell}(o))^2}\frac{nD^{TE}_{\ell}(o)}
{{\sigma_{\ell}^T(o)}{\sigma_{\ell}^E(o)}}\right\}
 K_{\frac{n-1}{2}}\left[\frac{n|D^{TE}_{\ell}(o)|}
 {(1-(\rho_{\ell}(o))^2){\sigma_{\ell}^T(o)}{\sigma_{\ell}^E(o)}}\right].
\end{eqnarray}

To summarize, the pdf (\ref{newpdfwl}), which includes noises,
retains the functional form of the no-noise pdf (\ref{pdfwl}), but
with the anticipated substitutions. Namely, $D_{\ell}^{TE}$,
$\sigma_{\ell}^{T}$, $\sigma_{\ell}^E$, $\rho_{\ell}$ are being
replaced with $D_{\ell}^{TE}(o)$, $\sigma_{\ell}^{T}(o)$,
$\sigma_{\ell}^E(o)$, $\rho_{\ell}(o)$, respectively.

%%%%%%%%%%%%%%%%%%%%%%%%%%%%%%%%%%%%%%%%%%%%%%%%%%%%%%%%%%%%%%%%%%%%%%%%%%%%%%%%%%%
%%%%%%%%%%%%%%%%%%%%%%%%%%%%%%%%%%%%%%%%%%%%%%%%%%%%%%%%%%%%%%%%%%%%%%%%%%%%%%%%%%%

\subsubsection {Numerical values of the noise power spectra \label{sa4a}}

The symbolic, so far, noises $N_{\ell}^{XX}$ have specific numerical values in
concrete experiments. We use evaluations quoted in the WMAP and Planck
literature (we adopt the experimental parameters but not the fundamental physics
in these references) and apply them to the $TE$ estimators in the WMAP5
best-fit model (\ref{WMAPBestFit}), (\ref{WMAPBestFitDP}).

\begin{figure}[t]
\centerline{\includegraphics[width=12cm,height=9cm]{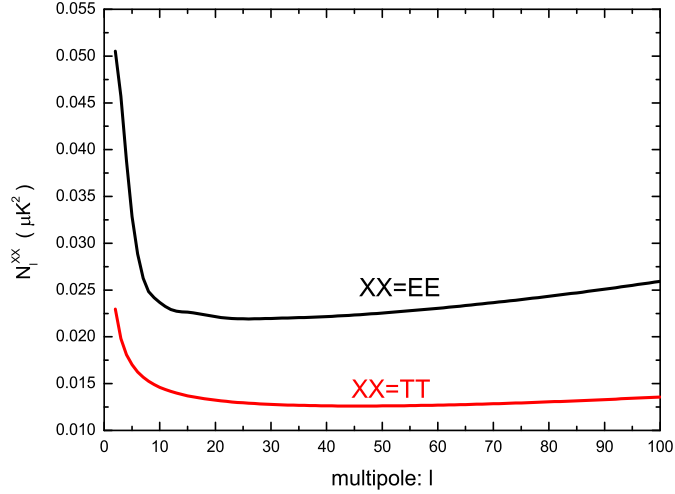}}
\caption{The estimation of the WMAP5 noises $N_{\ell}^{TT}$ and
$N_{\ell}^{EE}$.} \label{NttNee}
\end{figure}

The evaluation of the 5-year WMAP noises, which include the
instrumental noise, point sources noise and some systematic errors
\cite{wmap_verd}, follows from
Ref.~\cite{map1-te,5map_tt,5map_noise,lambda}. Starting with
numerical values for $\Delta D_{\ell}^{TT}$, $\Delta
D_{\ell}^{TE}$ cited in \cite{lambda}, and using Eq.(10) from
\cite{map1-te} and Eq.(1) from \cite{5map_tt}, we derive the noise
terms $N_{\ell}^{TT}$, $N_{\ell}^{EE}$ and plot them in
Fig.~\ref{NttNee}. We use these graphs when we refer to the WMAP5
noises. The noise power spectra have a weak multipole dependence.
For example, when $\ell=10$, one has $N_{\ell}^{TT} = 1.46\times
10^{-2} {\mu}{\rm K}^2$, $N_{\ell}^{EE} = 2.42\times 10^{-2} {\mu}{\rm K}^2$;
while for $\ell=100$, one has $N_{\ell}^{TT} = 1.36\times 10^{-2}
{\mu}{\rm K}^2$, $N_{\ell}^{EE} = 2.59\times 10^{-2} {\mu}{\rm K}^2$. (Actually,
it seems to us that these numbers somewhat overestimate the true
WMAP5 noises, as will be discussed in Sec.~\ref{s4}.)

%%%%%%%%%%%%%%%%%%%%%%%%%%%%%%%%%
%%%%%%%%%%%%%%%%%%%%%%%%%%%%%%%%%

For the Planck experiment, we use the expected instrumental noise in the
$143$GHz frequency channel. This channel is supposed to have low
foreground levels and the smallest noise. The quoted noise power spectra
have no $\ell$-dependence \cite{planck}:
\begin{eqnarray}
N_{\ell}^{TT}=1.53\times 10^{-4} \mu
{\rm K}^2,~~~N_{\ell}^{EE}=N_{\ell}^{BB}=5.58\times 10^{-4} {\mu}{\rm K}^2.
\label{Plancknoise}
\end{eqnarray}

%%%%%%%%%%%%%%%%%%%%%%%%%%%%%%%%%%%%
\begin{figure}
\begin{center}
\includegraphics[width=16cm,height=8cm]{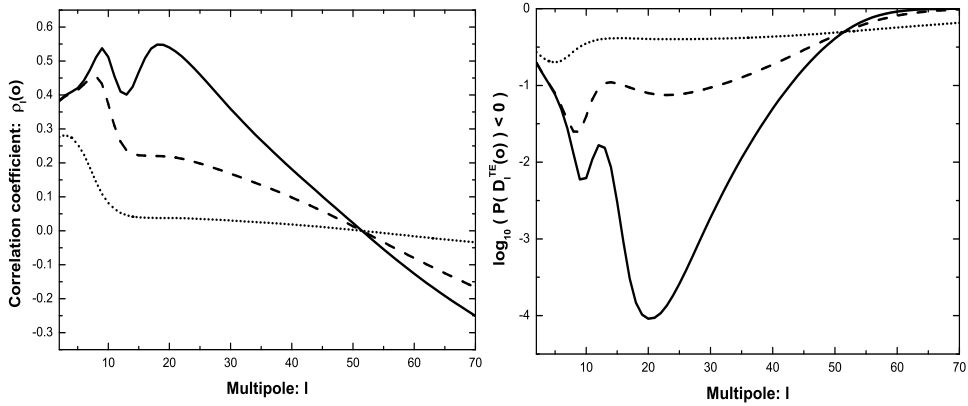}
\end{center}
\caption{The correlation coefficient ${\rho_\ell}(o)$ (left panel) and the
probability $\log_{10}(D_{\ell}^{TE}(o)<0)$ (right panel) in the WMAP5
best-fit model. In both panels the solid, dashed and dotted lines mark the
no-noise, Planck and WMAP5 curves, respectively.}
\label{section3_fig3.1}
\end{figure}
%%%%%%%%%%%%%%%%%%%%%%%%%%%%%%%%%%%%

It is demonstrated in Eq.~(\ref{rholdefine3}) that noises degrade the
correlation coefficient ${\rho_\ell}(o)$. In the left panel of
Fig.~\ref{section3_fig3.1} we plot ${\rho_\ell}(o)$ as a function of $\ell$ for
WMAP5 and Planck experiments in comparison with the no-noise graph from
Fig.~\ref{section2_fig1}. In the right panel of Fig.~\ref{section3_fig3.1}
we show $P({D}_{\ell}^{TE}(o)<0)$ for WMAP5 and Planck
experiments in comparison with the no-noise graph
from Fig.~\ref{section3_fig8}. Clearly, large WMAP5 noises make the probability
$P({D}_{\ell}^{TE}(o)<0)$ dangerously close to $0.5$.

\begin{figure}
\begin{center}
\includegraphics[width=18cm,height=8cm]{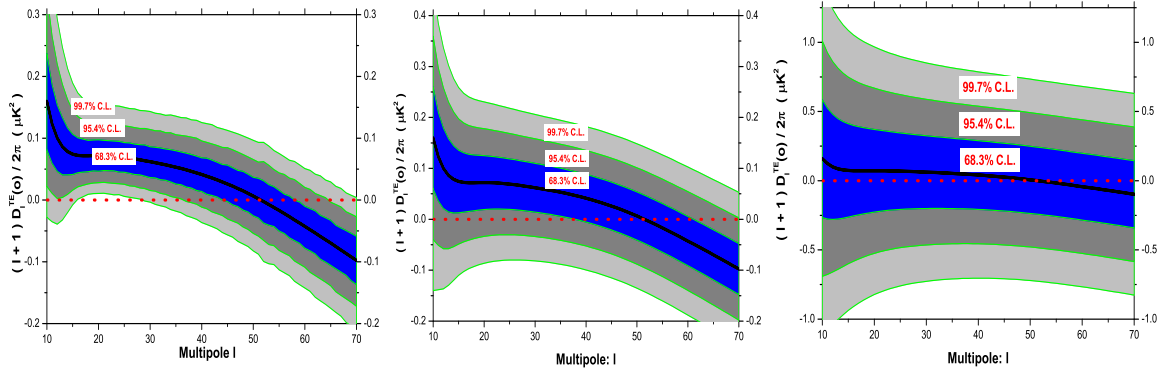}
\end{center}
\caption{The $68.3\%$, $95.4\%$ and $99.7\%$ confidence intervals
in the WMAP5 best-fit model. Panels from left to right: no-noise, Planck,
WMAP5. Note changes in scaling on the vertical axis.}
\label{section3_fig3.2}
\end{figure}

In Fig.~\ref{section3_fig3.2}, we show the $68.3\%$, $95.4\%$, $99.7\%$
confidence intervals of the $D_{\ell}^{TE}(o)$ estimator for Planck and WMAP5
experiments, starting from the no-noise case in the left panel (the same as
the left panel in Fig.~\ref{section3_fig5&6}). Obviously, with increasing noise
the confidence intervals become progressively wider.

%%%%%%%%%%%%%%%%%%%%%%%%%%%%%%%%%%%%
%%%%%%%%%%%%%%%%%%%%%%%%%%%%%%%%%%%%

%%%%%%%%%%%%%%%%%%%%%%%%%%%%%%%%%%%%%%%%%%%%%%%%%%%%%%%%%%%%%%%%%%%%%%%%%%%%%%%%%%%
                                                                                                       %%%%%% SUBSECTION 5.2 %%%%%%
%%%%%%%%%%%%%%%%%%%%%%%%%%%%%%%%%%%%%%%%%%%%%%%%%%%%%%%%%%%%%%%%%%%%%%%%%%%%%%%%%%%

\subsection{The effects of the cut sky\label{s3.4}}

Some additional stretching of confidence intervals should be attributed
to the incomplete sky coverage. Although the shape of the sky cut is, in
principle, important for accurate calculations \cite{wishart2}, we adopt a
simplified approach wherein the available number of degrees of freedom is
reduced to $n=(2\ell+1) f_{\rm sky}$, where $f_{\rm sky}$ is the observed
portion of the sky. We use the extra symbol $c$ to denote pdf's in this
cut sky approximation. Referring to Eqs.~(\ref{pdfwl}),~(\ref{newpdfwl})
we write
\begin{eqnarray}
\label{cutpdfwl}
f_{\rm c}({D}_{\ell}^{XX'}(o))=\left.f({D}_{\ell}^{XX'}(o))
\right|_{n=(2\ell+1) f_{\rm sky}}.
\end{eqnarray}

The mean value of the $XX'$ estimator calculated with
$f_{\rm c}({D}_{\ell}^{XX'}(o))$ is the same as in the case of the full
sky coverage, but the standard deviation increases,
\begin{eqnarray}
\begin{array}{c}
 \langle
{{D}^{XX}_{\ell}(o)}\rangle=C_{l}^{XX},~~
\langle {{D}^{TE}_{\ell}(o)}\rangle=C_{l}^{TE};
\\
\\
\Delta {{D}^{XX}_{\ell}(o)}=\sqrt{\frac{2}{(2\ell+1)
f_{\rm sky}}}(\sigma_{\ell}^{X}(o))^2,~~
\Delta {{D}^{TE}_{\ell}(o)}=\sqrt{\frac{(\rho_{\ell}(o))^2+1}{(2\ell+1)
f_{\rm sky}}}\sigma_{\ell}^{T}(o)\sigma_{\ell}^{E}(o).
\end{array}
\label{cd5}
 \end{eqnarray}
Comparing Eq.~(\ref{cd5}) with Eqs.~(\ref{deltadlxxo},~\ref{deltadltro})
we see that the standard deviation increases by the further factor
$1/\sqrt{f_{\rm sky}}$. In particular, the confidence intervals in
Fig.~\ref{section3_fig3.2} should be stretched in the vertical direction by
this factor.

The WMAP Collaboration \cite{5map_cut} and  the `Planck Blue Book'
\cite{planck} quote the cut sky factor
\begin{eqnarray}
\label{Planckcov}
f_{\rm sky}({\rm WMAP})=0.85,~~f_{\rm sky}({\rm Planck})=0.65,
\end{eqnarray}
respectively. The former one leads to the $1/\sqrt{f_{\rm sky}}\approx1.08$ increase in the
uncertainty of the estimators, and the latter one leads to $1/\sqrt{f_{\rm sky}}\approx1.24$.

In our further analysis we use these stretched confidence intervals, but it
may turn out that the quoted noises and $f_{\rm sky}$ are overly pessimistic,
especially for Planck, so the real level of uncertainty may prove to be
smaller.

%%%%%%%%%%%%%%%%%%%%%%%%%%%%%%%%%%%%%%%%%%%%%%%%%%%%%%%%%%%%%%%%%%%%%%%%%%%%%%%%%%%
%%%%%%%%%%%%%%%%%%%%%%%%%%%%%%%%%%  SECTION 6   %%%%%%%%%%%%%%%%%%%%%%%%%%%%%%%%%%%%%%%%%%
%%%%%%%%%%%%%%%%%%%%%%%%%%%%%%%%%%%%%%%%%%%%%%%%%%%%%%%%%%%%%%%%%%%%%%%%%%%%%%%%%%%

\section{Tests of the null hypothesis: WMAP5 best-fit model with no
gravitational waves\label{s4}}

%%%%%%%%%%%%%%%%%%%%%%%%%%%%%%%%%%%%
\begin{figure}
\begin{center}
\includegraphics[width=12cm,height=8cm]{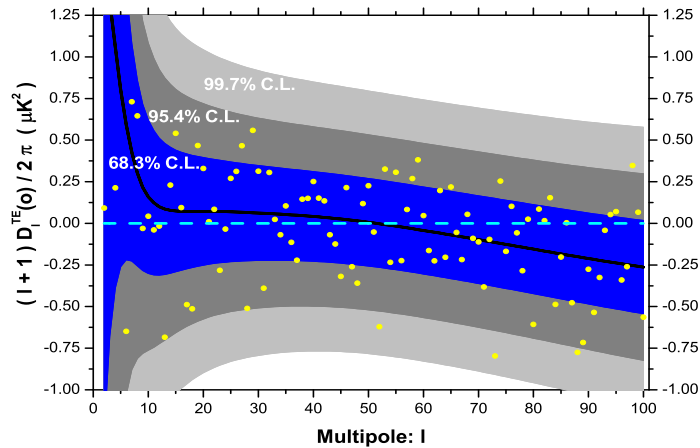}
\end{center}
\caption{The WMAP5 unbinned $TE$ data points on the background of confidence
intervals which include WMAP noises from Fig.~\ref{NttNee} and
$f_{\rm sky}=0.85$. The black solid line is the expectation value according to
the $H_0$ hypothesis.}
\label{section4_fig1}
\end{figure}
%%%%%%%%%%%%%%%%%%%%%%%%%%%%%%%%%%%%

The developed theory can now be confronted with real and simulated data.
In Fig.~\ref{section4_fig1}, based on information from
Ref.~\cite{5map_cosmology,lambda}, we show the WMAP5 unbinned $TE$ data points
on the background of the derived in Sec.~\ref{sa4} $TE$ confidence
intervals. Despite the messiness of the picture, the visual impression is such
that the curve of the `center of gravity' of the data points lies somewhat
below the black solid curve expected of density perturbations alone. This
tendency is of course required by the presence of gravitational waves, as we
discussed in Sec.~\ref{s3}. The visual impression should be
quantified by means of the `null hypothesis testing'. The null hypothesis
under investigation, denoted $H_0$, is the WMAP5 best-fit model
(\ref{WMAPBestFit}), (\ref{WMAPBestFitDP}) with no gravitational waves. We
probe $H_0$ with the mean value (M) and the variance (K) tests.

%%%%%%%%%%%%%%%%%%%%%%%%%%%%%%%%%%%%%%%%%%%%%%%%%%%%%%%%%%%%%%%%%%%%%%%%%%%%%%%%%%%
                                                                                                       %%%%%% SUBSECTION 6.1 %%%%%%
%%%%%%%%%%%%%%%%%%%%%%%%%%%%%%%%%%%%%%%%%%%%%%%%%%%%%%%%%%%%%%%%%%%%%%%%%%%%%%%%%%%

\subsection{The mean value test \label{meanvtest}}

We introduce the statistic $M$
\begin{eqnarray}
\label{mtest}
M \equiv\sum_{\ell}\frac{{D}_{\ell}^{TE}(o)-C_{\ell}^{TE}}{{\Delta
{D}_{\ell}^{TE}(o)}}.
\end{eqnarray}
To test observations, one takes ${D}_{\ell}^{TE}(o)$ as the actually observed
values of the estimator. The expectation amount $C_{\ell}^{TE}$ and the
standard deviation $\Delta{D}_{\ell}^{TE}(o)$, which weighs the scatter of
the data points in Eq.~(\ref{mtest}), are calculated from Eq.~(\ref{cd5}) assuming the validity of
the hypothesis $H_0$. The negative (positive) value of $M$ quantifies the
propensity of the data points to lie below (above) the theoretical
prediction. For unbinned data, the summation in Eq.~(\ref{mtest}) is
over $\ell=2,3,\cdot\cdot\cdot,100$.

\begin{figure}
\begin{center}
\includegraphics[width=12cm,height=8cm]{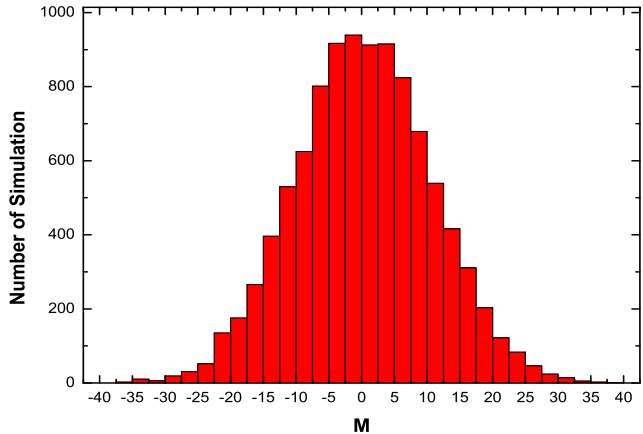}
\end{center}
\caption{The outcome values of the test quantity $M$ in simulated
samples. In the actually observed realization, $m=-8.69$.}
\label{section4_fig5}
\end{figure}

In confirmation of the visual impression, the value of $M$ on the actually
observed $TE$ data points turns out to be negative and equal to
\begin{eqnarray}
\label{m1}
 m=-8.69.
\end{eqnarray}
To assess the case for rejection of $H_0$ one needs to know the pdf for the
variable $M$. Using pdf (\ref{cutpdfwl}) we generate $10^4$ random
samples of the estimator ${D}_{\ell}^{TE}(o)$, where each sample includes all
$\ell=2,3,\cdot\cdot\cdot,100$. We then calculate $M$ for each sample.
The distribution of these $M$ (effectively, the pdf for the variable
$M$) is shown in Fig.~\ref{section4_fig5}.

The significance level $\alpha$ of the test, or in other words the
probability of incorrect rejection of $H_0$, is
\begin{eqnarray}
\alpha =P \left(\left.\frac{}{}|M|\geq |m|~ \right| H_0\right).
 \nonumber %\label{alphaM}
\end{eqnarray}
Calculating $\alpha$ from the pdf in Fig.~\ref{section4_fig5}, and with $m$
from Eq.~(\ref{m1}), we find
\begin{eqnarray}
\label{alpha1}
\alpha=40.63\% .
\end{eqnarray}
Although the case for rejection of the hypothesis $H_0$ is weak,
within 1$\sigma$, Eqs.~(\ref{m1},~\ref{alpha1}) serve as an indication that
one may try to find a better hypothesis.

%%%%%%%%%%%%%%%%%%%%%%%%%%%%%%%%%%%%
%%%%%%%%%%%%%%%%%%%%%%%%%%%%%%%%%%%%

\begin{figure}
\begin{center}
\includegraphics[width=12cm,height=8cm]{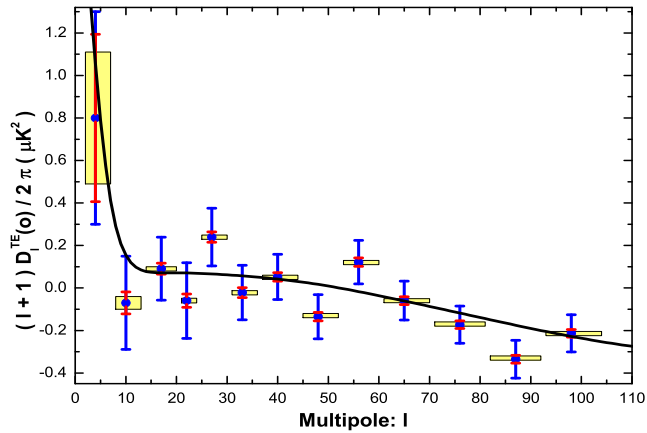}
\end{center}
\caption{In this figure, based on 5-year WMAP observations
\cite{5map_cosmology,lambda}, we show binned $TE$ data points
$(\ell+1){D}_{\ell}^{TE}(o)/2\pi$ surrounded by various 1$\sigma$
uncertainties: WMAP5 (blue, largest), Planck (red, intermediate),
no-noise (yellow, smallest).}
\label{section4_fig2}
\end{figure}

To check the stability of this conclusion, we
expanded the $M$ test. First, we carried out the same procedure with the
3-year WMAP data and found similar results. Second, we assumed the full
sky coverage $f_{\rm sky}=1.0$ with WMAP5 unbinned data and arrived at
a slightly stronger evidence for rejection of $H_0$, $m=-9.43,\alpha=32.91\%$.
%\begin{eqnarray}
% \nonumber %
%m=-9.43,~~~\alpha=32.91\%.
%\end{eqnarray}
Third, we applied the $M$ test to the WMAP5 data binned at thirteen multipoles
$\ell =4, 10, 17, 22, 27, 33, 40, 48, 56, 65, 76, 87, 98$
\cite{5map_cosmology,lambda}, as shown in Fig.~\ref{section4_fig2}. We made
simplifying theoretical assumptions \cite{page1,bin2,bin3} according to which
the estimator at every binned multipole obeys a Gaussian distribution with
the mean value $C_{\ell}^{TE}$ taken at $\ell$ representing the bin, and with
the standard deviation properly averaged over the bin and reduced by the size
of the bin. Binned multipoles are supposed to be statistically independent.
In Fig.~\ref{section4_fig2}, the binned WMAP5 $TE$ data
$(\ell+1){D}_{\ell}^{TE}(o)/2\pi$ are shown by (blue) dots, and the 1$\sigma$
intervals, which include WMAP5 noises from Fig.~\ref{NttNee} and
$f_{\rm sky}=0.85$, are shown by the largest (blue) bars. Following a somewhat incorrect
practice, we transferred the 1$\sigma$ confidence intervals from the
theoretical pdf to the data points themselves. The size of
the bins is indicated by the width of the (yellow) rectangular regions. The black solid line
is the expectation value according to $H_0$ hypothesis. Under the simplifying
assumptions made, the quantity $M$ satisfies a zero-mean Gaussian distribution
with variance $n=13$, determined by the number $n=13$ of the observed (binned)
data points. The analogue of Eq.~(\ref{mtest}), with summation over
thirteen binned multipoles, resulted in $m=-3.72,\alpha=30.21\%$.
% \begin{eqnarray}
%  \nonumber %\label{m2}
% m=-3.72,~~~\alpha=30.21\%.
% \end{eqnarray}
The additional assumption $f_{\rm sky}=1$ reduces $\alpha$ to $\alpha=26.52\%$.

The crucial importance of efforts to reduce noises in future
experiments is illustrated by the following exercise. We assumed
that the Planck mission would observe exactly the same $TE$ data
points as WMAP5 did, but which should now be analyzed against the
Planck's smaller noises. We first calculated the confidence
intervals of the unbinned data using the Planck noises
(\ref{Plancknoise}) and $f_{\rm sky}=0.65$, and then we have
binned the data using the same procedure that was used in the WMAP
analysis \cite{5map_cosmology,lambda}. The resulting 1$\sigma$
intervals are shown by the (red) intermediate-size bars in Fig.~\ref{section4_fig2}. With
these smaller uncertainties, the $M$ test gives $m=-15.67,\alpha=1.39\times10^{-5}$.
% \begin{eqnarray}
%  \nonumber %\label{M3}
% m=-15.67,~~~\alpha=1.39\times10^{-5}.
% \end{eqnarray}
The further assumption $f_{\rm sky}=1$ makes $\alpha$ even smaller,
$\alpha=7.08\times10^{-8}$. In other words, in these circumstances, the
null hypothesis would be rejected at a higher than 4$\sigma$ level
($\alpha\simeq 7\times 10^{-5}$).

Finally, we have looked at what would happen in the ideal, no-noise, case,
that is, when the standard deviation is at its minimum level
$\Delta{D}_{\ell}^{TE}$ and $f_{\rm sky}=1$. The 1$\sigma$ intervals are shown
by the height of the (yellow) rectangular regions in Fig.~\ref{section4_fig2}. The $M$ test
results in $m=-27.17,\alpha = 5.19\times 10^{-14}$.
%\begin{eqnarray}
%\nonumber %
% m=-27.17,~~~ \alpha = 5.19\times 10^{-14}.
% \end{eqnarray}
Thus, in the ideal case, the WMAP5 $TE$ observations would have
rejected the WMAP5 $H_0$ hypothesis at a higher than 7$\sigma$
level. The results of all $M$ tests are summarized in Table
\ref{5mtable}.

%%%%%%%%%%%%%%%%%%%%%%%%%%%%%%%%%%%%
%%%%%%%%%%%%%%%%%%%%%%%%%%%%%%%%%%%%

%%%%%%%%%%%%%%%%%%%%%%%%%%%%%%%%%%%%
\begin{table}
\caption{Significance level $\alpha$ in the mean value tests based on WMAP5
data.}
\begin{center}
\label{5mtable}
\begin{tabular}{|c|c|c|c|}
  \hline
  data & noises & $\alpha$ (cut sky) & $\alpha$ (full sky) \\
  \hline
  unbinned data & WMAP5 noises  & $40.63\%$  & $32.91\%$  \\
  binned data & WMAP5 noises & $30.21\%$  & $26.52\%$  \\
  binned data & Planck noises  & $1.39\times10^{-5}$  &  $7.08\times10^{-8}$ \\
  binned data & no noise  & ------ &  $5.19\times 10^{-14}$ \\
  \hline
\end{tabular}
\end{center}
\end{table}
%%%%%%%%%%%%%%%%%%%%%%%%%%%%%%%%%%%%

%%%%%%%%%%%%%%%%%%%%%%%%%%%%%%%%%%%%%%%%%%%%%%%%%%%%%%%%%%%%%%%%%%%%%%%%%%%%%%%%%%%
                                                                                                       %%%%%% SUBSECTION 6.2 %%%%%%
%%%%%%%%%%%%%%%%%%%%%%%%%%%%%%%%%%%%%%%%%%%%%%%%%%%%%%%%%%%%%%%%%%%%%%%%%%%%%%%%%%%

\subsection{The variance test}

\begin{figure}
\begin{center}
\includegraphics[width=12cm,height=8cm]{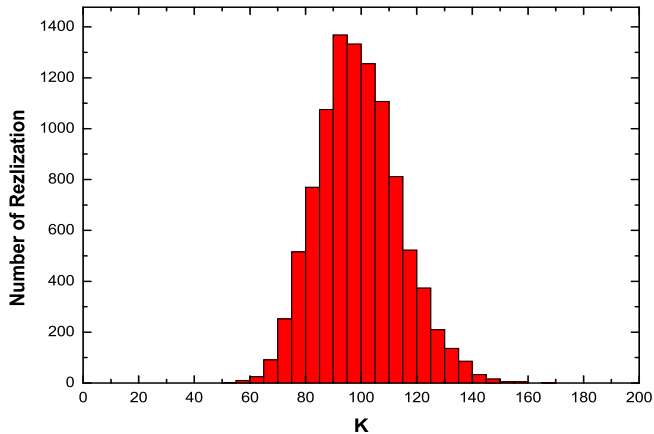}
\end{center}
\caption{The distribution of the test quantity $K$ from the
simulated samples. In the observed realization, $k=85.38$.}
\label{section4_fig3}
\end{figure}

The scatter of the observed data points is probed by the $K$ statistic
\begin{eqnarray}
 \nonumber %\label{ktest}
 K\equiv\sum_{\ell}\frac{({D}_{\ell}^{TE}(o)-C_{\ell}^{TE})^2}{(\Delta
 {D}_{\ell}^{TE}(o))^2}.
\end{eqnarray}
The significance level $\alpha$ of the $K$ test is defined by
\begin{eqnarray}
\alpha=P\left(\left. \frac{}{} K\geq k~\right| H_0\right).
 \nonumber %\label{alphaK}
\end{eqnarray}
First, we apply the $K$ test to the WMAP5 $TE$ unbinned data,
shown on Fig. \ref{section4_fig1}. The pdf of the test statistic
$K$, built by the simulation method described in the previous
subsection, is shown in Fig.~\ref{section4_fig3}. The value of the
$K$ statistic and the corresponding significance level are as
follows $k = 85.38,\alpha=82.63\%$.
%\begin{eqnarray}
% \nonumber %
%k = 85.38,~~~\alpha=82.63\% .
%\end{eqnarray}
Under the assumption $f_{\rm sky}=1$, $\alpha$ reduces to $\alpha=44.38\%$.
Thus, with WMAP5 noises, the variance test does not provide enough evidence for
rejection of the null hypothesis. Nevertheless, the somewhat lower than
expected value of $k$ points toward a possible overestimation of the WMAP5
noises.

%%%%%%%%%%%%%%%%%%%%%%%%%%%%%%%%%%%%
%%%%%%%%%%%%%%%%%%%%%%%%%%%%%%%%%%%%

We have also applied the $K$ test to the binned 5-year WMAP data. We have made
the same simplifying Gaussian assumptions that were discussed in
Sec.~\ref{meanvtest}. Under these assumptions, the quantity $K$ obeys a
$\chi_{n}^2$ distribution with $n=13$ degrees of freedom \cite{book7}.
The peak of this function is at $K\approx n$. Using this pdf, we
find $k=11.48$ and $\alpha=57.06\%$. The full sky assumption reduces $\alpha$
to $\alpha=40.96\%$. Again, with WMAP5 noises, the case for rejection of $H_0$
is weak.

The value of $\alpha$ dramatically decreases when the WMAP5 data are
accompanied by the Planck noises. In this case, $k \approx 211$ and
$\alpha \approx 10^{-37}$. The $H_0$ hypothesis would have been rejected with
a huge margin. The $\alpha$ is even smaller in the no-noise case
with $f_{\rm sky}=1$. The summary of $K$ tests is shown in Table \ref{5ktable}.

It is important to note that, with the Gaussian approximations made for the pdf 
of the estimator ${D}_{\ell}^{TE}(o)$, the variance test is equivalent to 
the test, where the calculations are being done directly with 
the weighted values of the pdf at individual data points: 
\begin{eqnarray} 
K = -2 \ln \left( \frac{{\prod_{\ell}f({D}_{\ell}^{TE}(o))}}{{\prod_{\ell}
f(C_{\ell}^{TE})}}\right) = - 2 \ln \prod_{\ell} \frac{f({D}_{\ell}^{TE}(o))}
{f(C_{\ell}^{TE})}, 
\end{eqnarray} 
where $f({D}_{\ell}^{TE}(o))$ is the value of the pdf at the observed data 
point ${D}_{\ell}^{TE}(o)$, and $f(C_{\ell}^{TE})$ is the value of the pdf 
at the expected point ${D}_{\ell}^{TE}(o)=C_{\ell}^{TE}$.

%%%%%%%%%%%%%%%%%%%%%%%%%%%%%%%%%%%%
\begin{table}
\caption{The significance level $\alpha$ of $K$ tests of WMAP5 data}
\begin{center}
\label{5ktable}
\begin{tabular}{|c|c|c|c|}
  \hline
  data & noises & $\alpha$ (cut sky) & $\alpha$ (full sky) \\
  \hline
  unbinned data &  WMAP noises  & $82.63\%$  &  $44.38\%$ \\
  binned data & WMAP noises & $57.06\%$  & $40.96\%$  \\
  binned data & Planck noises  & $\approx 10^{-37}$  &  $< 10^{-37}$ \\
  binned data & no noise  & ------ &  $\ll 10^{-37}$ \\
  \hline
\end{tabular}
\end{center}
\end{table}
%%%%%%%%%%%%%%%%%%%%%%%%%%%%%%%%%%%%

%%%%%%%%%%%%%%%%%%%%%%%%%%%%%%%%%%%%%%%%%%%%%%%%%%%%%%%%%%%%%%%%%%%%%%%%%%%%%%%%%%%
%%%%%%%%%%%%%%%%%%%%%%%%%%%%%%%%%%  SECTION 7   %%%%%%%%%%%%%%%%%%%%%%%%%%%%%%%%%%%%%%%%%%
%%%%%%%%%%%%%%%%%%%%%%%%%%%%%%%%%%%%%%%%%%%%%%%%%%%%%%%%%%%%%%%%%%%%%%%%%%%%%%%%%%%

\section{Likelihood analysis of the quadrupole ratio $R$\label{s5}}

The existence of relic gravitational waves is a necessity dictated by general
relativity and quantum mechanics, and not an inflationary `bonus' which, as
follows from inflationary theory, most likely should not be awarded, see
Sec.~\ref{section1}. We do not use our theoretical position as a technical `prior'
in data analysis, but we certainly are willing to reconsider the WMAP5
$TE$ data from this point of view. We also intend to make predictions for
future experiments. The immediate goal is to evaluate the quadrupole parameter
$R$, Eq.~(\ref{defineR}), from the existing observations. The mild indications
in the WMAP5 $TE$ data favoring the rejection of the $H_0$ hypothesis $(R=0)$
have been discussed above.

%%%%%%%%%%%%%%%%%%%%%%%%%%%%%%%%%%%%%%%%%%%%%%%%%%%%%%%%%%%%%%%%%%%%%%%%%%%%%%%%%%%
                                                                                                       %%%%%% SUBSECTION 6.1 %%%%%%
%%%%%%%%%%%%%%%%%%%%%%%%%%%%%%%%%%%%%%%%%%%%%%%%%%%%%%%%%%%%%%%%%%%%%%%%%%%%%%%%%%%

\subsection{Expanding the parameter space}

The inclusion of gravitational waves with two new free parameters $B_t^2$
and $n_t$ (see Eq.~(\ref{primsp})) would require, strictly speaking, a new
full-scale likelihood analysis of all the data. Certainly, the addition of
gravitational waves must be consistent with all the measured correlation
functions and upper limits, Eq.~(\ref{TotalCell_dp+gw}). In what follows, we
simplify this approach, while retaining the main points of what we want to
demonstrate. First, we keep intact the background cosmological parameters, as
quoted in Eq.~(\ref{WMAPBestFit}). Second, we bound the four perturbation
parameters $B_s^2$, $n_s$, $B_t^2$, $n_t$ by three constraints, thus leaving
independent only one parameter. We choose this independent parameter to be
the quadrupole ratio $R$. The three mentioned constraints are as follows:
 \begin{eqnarray}
 \label{nsR}
n_t=n_s-1,~~~ 
\ell(\ell+1)C_{\ell=10}^{TT}/2\pi=840{\mu}{\rm K}^2,~~~
n_s(R)=n_s(0)+0.35R-0.07R^2.
 \end{eqnarray}
The first constraint is purely theoretical, it reflects the origin
of cosmological perturbations \cite{h}. The second constraint
plays the role of an overall normalization, it requires the
perturbation parameters to satisfy the demand that the joint, $dp$
plus $gw$, temperature anisotropy at $\ell=10$ were fixed at the
well-measured level. The third constraint, where $n_s(0)=0.960$
\cite{5map_cosmology}, is empirical. The function $n_s(R)$ was
designed to be such that the best studied correlation function
$C_{\ell}^{TT}$ were as close as possible to the actually observed
data for a wide range of possible $R \geq 0$. In contrast to other
analyses, we are not using the inflationary `consistency relation'
$n_t=-r/8$, which affects the results regarding gravitational
waves (see comments in \cite{0707} on the self-contradictory
nature of the data analysis based on inflationary formulas).

\begin{figure}
\begin{center}
\includegraphics[width=8cm]{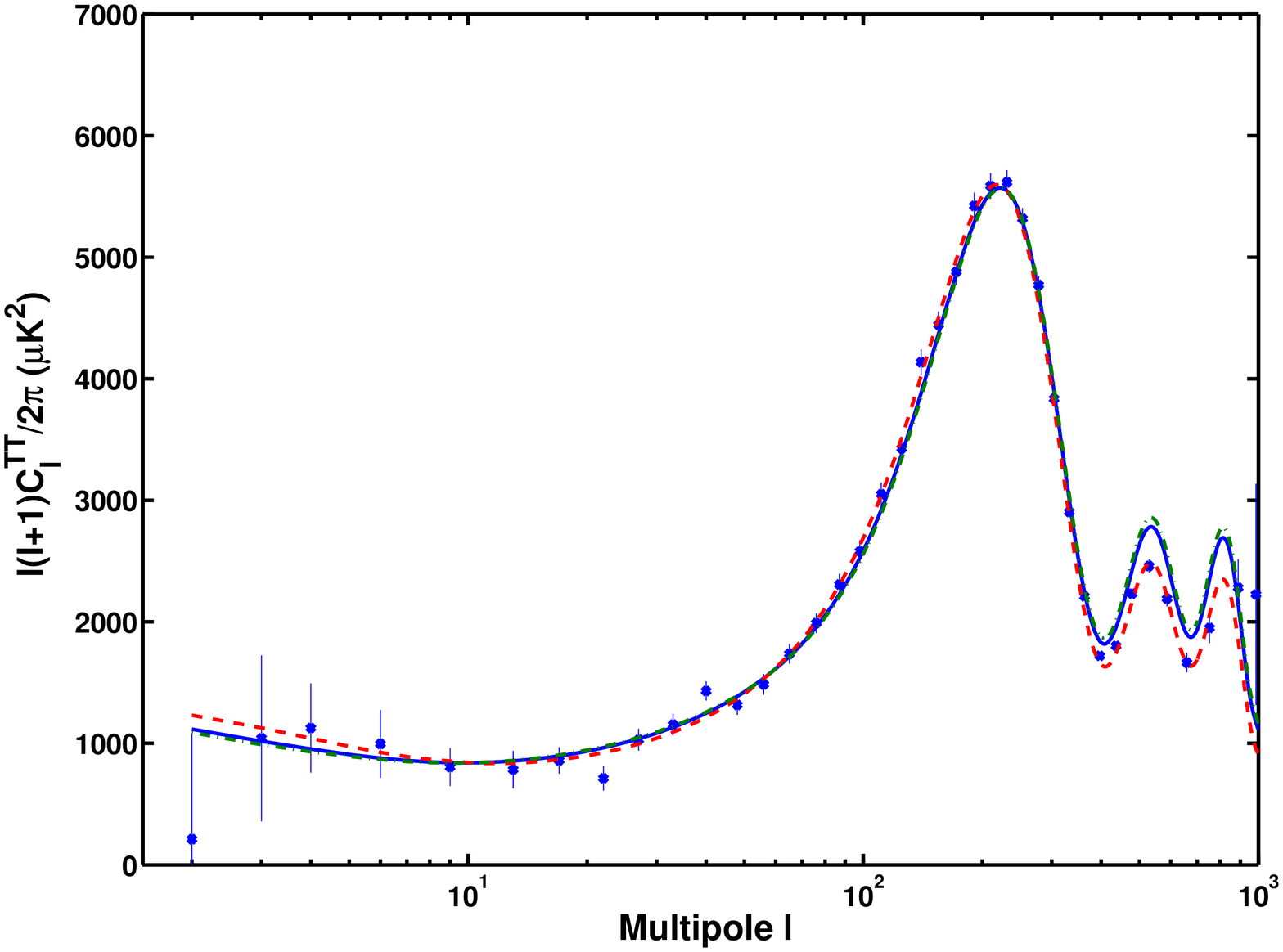}
\includegraphics[width=8cm]{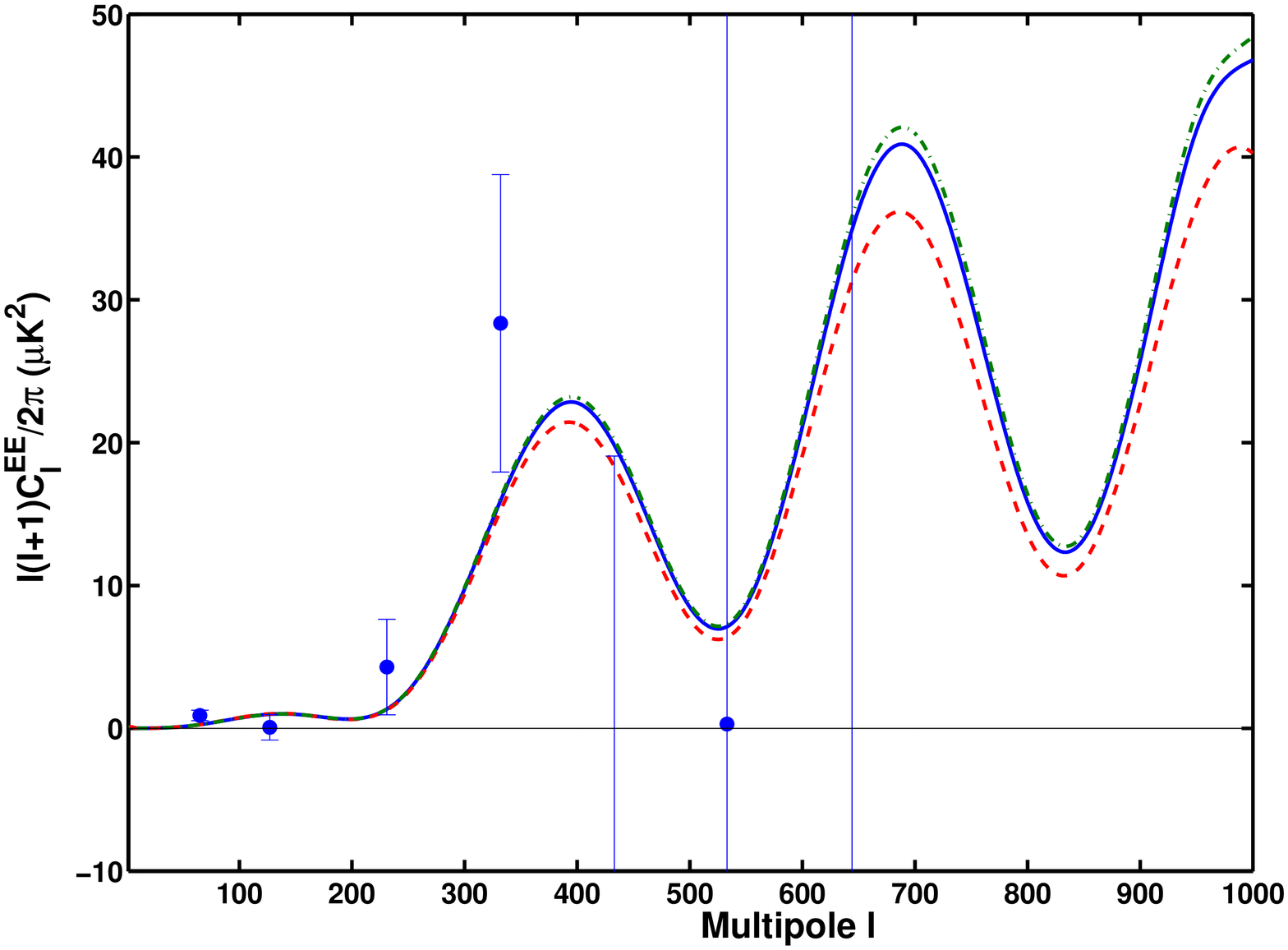}
\end{center}
\caption{The $TT$ (left) and $EE$ (right) power spectra in models with
varying amounts of gravitational waves. The dashed (red) line is for the
$H_0$ hypothesis, $R=0$. The solid (blue) line shows the model with $R=0.24$,
and the dash-dotted (green) line is for the model with $R=0.3$.}
\label{section5_fig0}
\end{figure}

In Fig.~\ref{section5_fig0} we illustrate, for selected values of
$R$, the consistency of our class of models with the $TT$ and $EE$
observations. As usual, some corrections at very high $\ell$'s
could be achieved, if desired, by the assumption of the `running'
spectral index ${\rm n}_s$, but we did not resort to this
possibility.  It is also important to emphasize that the $BB$
spectrum in our class of models is well below the existing upper
limits \cite{5map_tt}. Indeed, the WMAP limit is
$\ell(\ell+1)C_{\ell=2-6}^{BB}/2\pi<0.15{\mu}{\rm K}^2$ ($95\%$C.L.).
This corresponds to $r < 20$ \cite{5map_tt}, which is roughly $R <
10$, whereas we will be mostly dealing with $R <1$.

%%%%%%%%%%%%%%%%%%%%%%%%%%%%%%%%%%%%
%%%%%%%%%%%%%%%%%%%%%%%%%%%%%%%%%%%%

\subsection{Likelihood function for $R$ from the WMAP $TE$ data \label{s5.2}}

The likelihood function for a parameter is a pdf for a random variable, where
the parameter is considered unknown, whereas the value of the random variable
is considered known from experiment (see for example \cite{knox,mcmc,wishart2}).
Since the variables $D_{\ell}^{TE}(o)$ with differing $\ell$'s are independent,
their joint pdf is a product of individual pdf's $f_{\rm c}(D_{\ell}^{TE}(o))$
over $\ell$. Therefore, the likelihood function for $R$ is
\begin{eqnarray}
\label{like0}
 {\mathcal{L}} (R)
= C \prod_{\ell} f_{\rm c}(D_{\ell}^{TE}(o)),
\end{eqnarray}
where $f_{\rm c}(D_{\ell}^{TE}(o))$ is given by
Eqs.~(\ref{cutpdfwl},~\ref{newpdfwl}), the quantities
$D_{\ell}^{TE}(o)$ are taken from observations, and $C$ is a
normalization constant to be fixed later.

\begin{figure}
\begin{center}
\includegraphics[width=12cm,height=10cm]{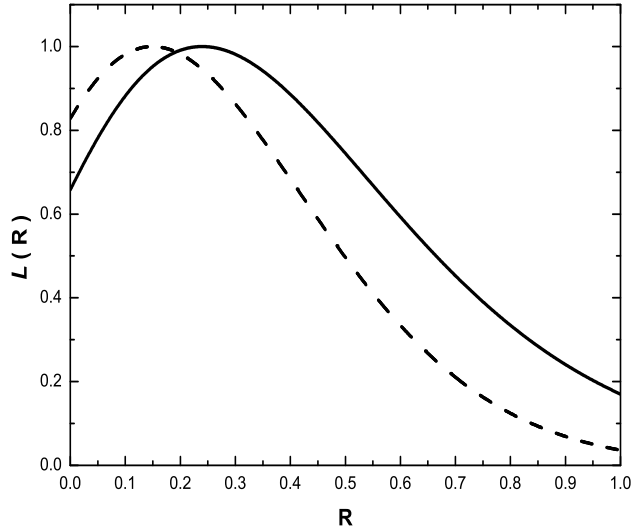} \qquad
\end{center}
\caption{The likelihood function for $R$. The solid line uses
the unbinned 5-year WMAP data, whereas the dashed line is for
the unbinned 3-year WMAP data.}
\label{section5_fig1}
\end{figure}

Moving by small steps $\Delta R =0.01$ from $R=0$ we build ${\mathcal{L}}(R)$
numerically. The $D_{\ell}^{TE}(o)$ participating in Eq.~(\ref{like0}) are the
unbinned data shown in Fig.~\ref{section4_fig1}, the other quantities are
determined by Eqs.~(\ref{WMAPBestFit},~\ref{nsR}), Fig.~\ref{NttNee} and
$f_{\rm sky}=0.85$. The result of this calculation is shown by a solid line
in Fig.~\ref{section5_fig1}.

\begin{figure}
\begin{center}
\includegraphics[width=12cm,height=10cm]{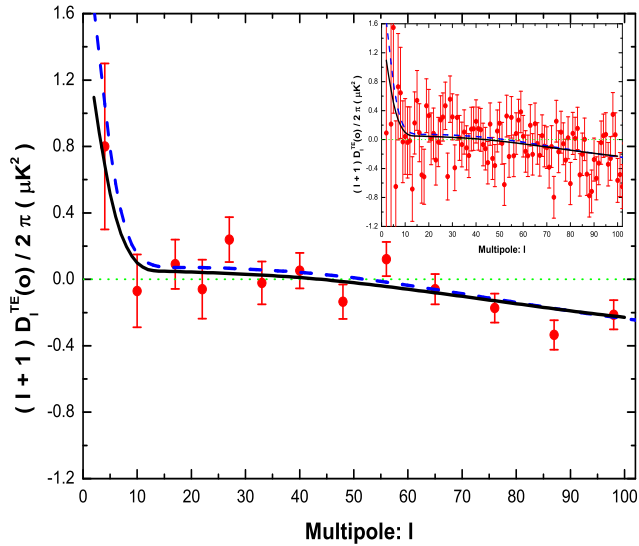}
\end{center}
\caption{The solid line is our best fit to the WMAP5 binned $TE$ data,
in comparison with the $H_0$ hypothesis (dashed line). The depicted error bars
are taken from the $R=0$ model. In the insert, we plot the same graph for
unbinned data.}
\label{section5_fig4}
\end{figure}

The maximum of ${\mathcal{L}}(R)$ is at $R=0.240$. The constant $C$ was chosen
to make ${\mathcal{L}}(R)=1$ at maximum, as illustrated in
Fig.~\ref{section5_fig1}. Taking into account $R=0.240$ and Eq.~(\ref{nsR}), the full set of our
best-fit perturbation parameters is defined by
\begin{eqnarray}
\label{OurBestFit}
\ell(\ell+1)C_{\ell=10}^{TT}/2\pi=840{\mu}{\rm K}^2, ~~~R=0.240, ~~n_s=1.040, ~~
n_t=0.040.
\end{eqnarray}
This model has a blue(ish) primordial power spectrum and
accommodates a significant amount of relic gravitational waves.
The $TE$ spectrum for this model is shown by a solid line in
Fig.~\ref{section5_fig4}. The main difference with the $TE$ spectrum of the
WMAP5 best-fit model (dashed line) is at multipoles $\ell<50$. This could be
anticipated, as the role of gravitational waves is essential only at
$\ell \lesssim 100$. The $TT$ and $EE$ spectra for our model ($R=0.240$) are
plotted in Fig.~\ref{section5_fig0}.

%%%%%%%%%%%%%%%%%%%%%%%%%%%%%%%%%%%%
%%%%%%%%%%%%%%%%%%%%%%%%%%%%%%%%%%%%

The difference with WMAP conclusions is indicative but not significant,
owing to large WMAP noises. Indeed, the confidence interval around the
maximum likelihood value $R=0.240$ is broad. Using the techniques of
Appendix~\ref{appendixA} we find that the $68.3\%$ confidence interval is represented
by $R=0.240^{+0.291}_{-0.225}$. (Lower boundary cannot be less than $-0.240$
as the quantity $R$ is non-negative.) Thus, the value $R=0$ is only barely
outside the $68.3\%$ confidence interval. This result is
consistent with the null hypothesis testing in Sec.~\ref{s4}.

%%%%%%%%%%%%%%%%%%%%%%%%%%%%%%%%%%%%
%%%%%%%%%%%%%%%%%%%%%%%%%%%%%%%%%%%%

We have also calculated the likelihood function ${\mathcal{L}}(R)$
for 3-year WMAP $TE$ data. The background cosmological parameters
are taken as the WMAP3 $\Lambda$CDM model: $\Omega_{m}h^2=0.1265$,
$\Omega_{b}h^2=0.0223$, $h=0.735$, $\tau_{reion}=0.088$
\cite{map3}. In line with the WMAP3 findings we set $n_s(0)=0.951$
\cite{map3} and use Eq.~(\ref{nsR}) to relate $n_s$ and $R$. The
resulting likelihood function is shown in Fig.~\ref{section5_fig1}
by a dashed line. The maximum likelihood value, along with the
associated $68\%$ confidence interval, is
$R=0.149^{+0.247}_{-0.149}$. The best-fit spectral indices are
$n_s=1.002$ and $n_t=0.002$. Again, the $R=0$ is within the
$68.3\%$ confidence interval. Nevertheless, it is worth noting
that in transition from WMAP3 to WMAP5 data, the peak value of $R$
increased from $R=0.149$ to $R=0.240$.

%%%%%%%%%%%%%%%%%%%%%%%%%%%%%%%%%%%%%%%%%%%%%%%%%%%%%%%%%%%%%%%%%%%%%%%%%%%%%%%%%%%
                                                                                                       %%%%%% SUBSECTION 6.3 %%%%%%
%%%%%%%%%%%%%%%%%%%%%%%%%%%%%%%%%%%%%%%%%%%%%%%%%%%%%%%%%%%%%%%%%%%%%%%%%%%%%%%%%%%

\subsection{Likelihood analysis of simulated $TE$ data \label{s5.3}}

Being guided by the experience that in astrophysics very often `today's
indication is tomorrow's discovery', we treat our best-fit model
(\ref{OurBestFit}) as a benchmark model to be confirmed by Planck and other
sensitive missions. Using the parameters of this model we simulate the $TE$
data for Planck, taking into account the expected noises and $f_{\rm sky}$.
We also vary the perturbation parameters, noises and sky coverage in order
to assess what will be happening with nearby models.

The procedure for Planck is as follows. First, we choose the input value
$R_{\rm i}$ and determine other perturbation parameters from Eq.~(\ref{nsR}). The
background parameters are fixed by Eq.~(\ref{WMAPBestFit}). Second, we adopt
Planck parameters, Eqs.~(\ref{Plancknoise},~\ref{Planckcov}), and randomly
generate 10 sets of data points
$\{D_{\ell}^{TE}(o)|\ell=2,3,\cdot\cdot\cdot,100\}$, where each set includes
all indicated $\ell$'s. We use for this purpose the underlying pdf
$f_{\rm c}(D_{\ell}^{TE}(o))$, Eqs.~(\ref{cutpdfwl},~\ref{newpdfwl}).
Obviously, even if all parameters are fixed, the outcome data points
$D_{\ell}^{TE}(o)$ are supposed to be random due to the randomness of signal
and noises. Third, for every set of simulated data we build, in a manner
described in Sec.~\ref{s5.2}, the likelihood functions $\mathcal{L}(R)$ and
explore their properties.

\begin{figure}
\begin{center}
\includegraphics[width=17cm,height=9cm]{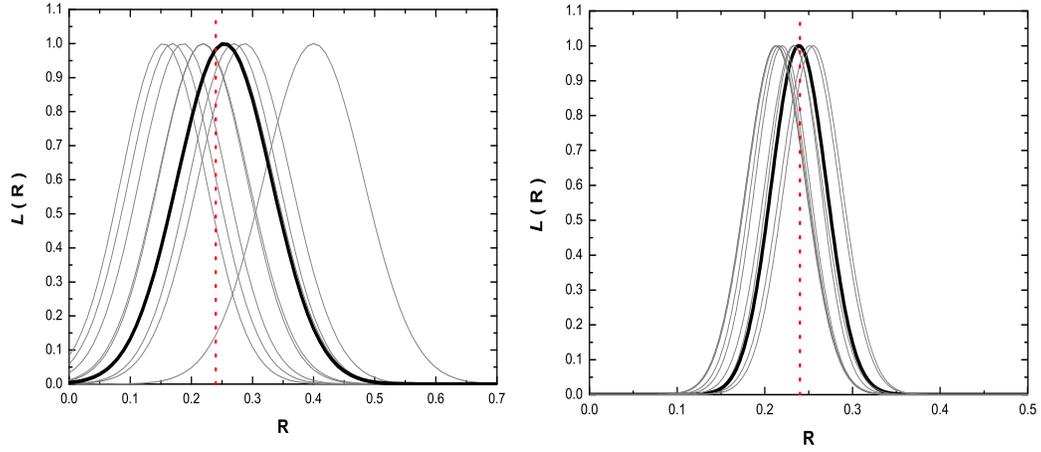}
\end{center}
\caption{The likelihood functions for $R$ with the input value $R_{\rm i}=0.240$.
The Planck conditions are used in the left panel, the ideal conditions - in
the right panel. The `typical' likelihood functions are shown by the dark
lines.}
\label{section5_fig2}
\end{figure}

Let us start from the benchmark value $R_{\rm i}=0.24$. The 10
likelihood functions built from 10 random realizations are shown
in Fig.~\ref{section5_fig2} (left plot). As expected, the maxima of
the likelihood functions concentrate around the input value
$R_{\rm i}=0.240$, and the widths of the likelihood functions are
approximately equal to each other. The arithmetical mean of the
maxima is $\overline{R_{\rm p}}=0.242$, and the arithmetical mean
width of the $68.3\%$ confidence intervals, i.e.~1$\sigma$
interval, is $\overline{\Delta R}=0.074$. The realization whose
maximum is closest to the input value $R_{\rm i}=0.240$ (in a sense, a
`typical' likelihood function) is shown by a dark line.

On the ground of these simulations we conclude that the Planck satellite will
be able to see a relatively strong evidence for relic gravitational waves,
if the $R_{\rm i}=0.240$ model is correct. In other words, Planck will be able to
exclude the value $R=0$ at more than 3$\sigma$ confidence level. We have also
done simulations for a more optimistic cut sky factor $f_{\rm sky} = 0.85$.
In this case, $\overline{\Delta R}=0.062$, thus allowing to exclude the $R=0$
assumption at 4$\sigma$ level.

The best achievable conditions for detecting the signature of
relic gravitational waves are reached in the idealized case of no
noise and full sky coverage. With the input value $R_{\rm i}=0.240$, the
10 likelihood functions for $TE$ experiment in ideal conditions
are shown in Fig.~\ref{section5_fig2} (right plot). The mean value
of the maxima is $\overline{R_{\rm p}}=0.228$, and the mean width
of the $68.3\%$ confidence intervals is $\overline{\Delta
R}=0.032$. Thus, the ideal $TE$ experiment would be able to detect
gravitational waves and exclude $R=0$ at more than 7$\sigma$
confidence level.

%%%%%%%%%%%%%%%%%%%%%%%%%%%%%%%%%%%%
%%%%%%%%%%%%%%%%%%%%%%%%%%%%%%%%%%%%

\begin{figure}
\begin{center}
\includegraphics[width=17cm,height=9cm]{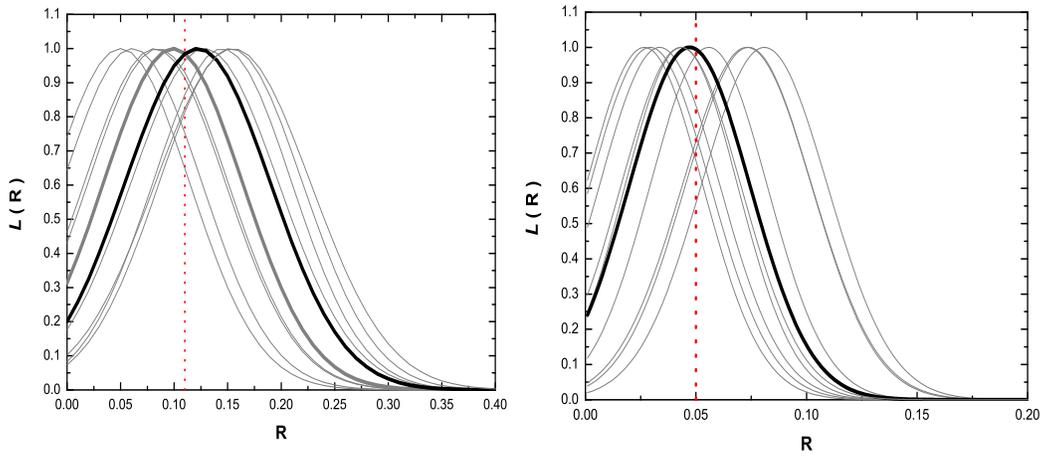}
\end{center}
\caption{The likelihood functions for the smallest $R$ detectable at a
2$\sigma$ level. The left panel uses $R_{\rm i}=0.110$ and Planck parameters.
The right panel is for an ideal experiment with $R_{\rm i}=0.050$. The `typical'
functions are shown by dark lines.}
\label{section5_fig3}
\end{figure}

We have also evaluated the $R$ detectable at a 2$\sigma$ level by
Planck or ideal experiment. In the left panel of Fig.~\ref{section5_fig3} we
show 10 likelihood functions for the input $R_{\rm i}=0.110$ and Planck
parameters. The mean value of the maxima is at $\overline{R_{\rm p}}=0.108$,
and the mean width of the 1$\sigma$ confidence intervals is
$\overline{\Delta R}=0.067$. Thus, using the $TE$ method, the Planck satellite
would be able to detect $R=0.110$ at nearly 2$\sigma$ level. The left panel of
Fig.~\ref{section5_fig3} shows the results of solving the same problem in the
ideal case. The input $R$ for the 10 likelihoods is $R_{\rm i}=0.050$. The mean
value of the peaks is $\overline{R_{\rm p}}=0.051$, and the mean
width of the 1$\sigma$ confidence intervals is $\overline{\Delta R}=0.025$.
Thus, an ideal experiment would detect $R=0.05$ at exactly 2$\sigma$ level.
Even more possibilities are considered below in the context of comparison
of $TE$ and $BB$ methods.

%%%%%%%%%%%%%%%%%%%%%%%%%%%%%%%%%%%%
%%%%%%%%%%%%%%%%%%%%%%%%%%%%%%%%%%%%

%%%%%%%%%%%%%%%%%%%%%%%%%%%%%%%%%%%%%%%%%%%%%%%%%%%%%%%%%%%%%%%%%%%%%%%%%%%%%%%%%%%
%%%%%%%%%%%%%%%%%%%%%%%%%%%%%%%%%%  SECTION 8   %%%%%%%%%%%%%%%%%%%%%%%%%%%%%%%%%%%%%%%%%%
%%%%%%%%%%%%%%%%%%%%%%%%%%%%%%%%%%%%%%%%%%%%%%%%%%%%%%%%%%%%%%%%%%%%%%%%%%%%%%%%%%%

\section{Comparison of $TE$ and $BB$ methods in search for relic gravitational
waves and prospects for the Planck mission\label{s6}}

The advantage of the usually discussed $B$-mode searches for
primordial gravitational waves is that in the $BB$ correlation
function there is no competing contribution from density
perturbations. However, the $BB$ signal is inherently weak, it is
about 50 times smaller than the {\it g.w.} contribution to the
$TE$ correlation. It is true that the nominal $BB$ noise is also
smaller than the $TE$ noise, so that the nominal `signal to noise'
ratio may be slightly in favor of the $BB$ method. But one must
take into account the large systematic effects that the $B$-mode
is prone to. They include the various contaminations
\cite{galaxy,cut2,asymmetry,lensing}, the disadvantages of dealing
with an auto-correlation ($BB$) rather than with a
cross-correlation ($TE$) function, the actual weakness of the
signal which can lead to the danger of unexpected leakages and
couplings, etc. It is only the experimenters who can properly
assess these complications.

When only the evaluation of the
Planck's instrumental $BB$ noise, Eq.~(\ref{Plancknoise}), is used
in calculations, we call it an `optimistic' $BB$ case. As for the
additional complications mentioned above, we try to treat them as
`effective' noises in the framework of our formalism. We give our
evaluation of these `effective' noises and increase the nominal
instrumental $BB$ noise to the level which we call the `realistic'
$BB$ case. We think that it is only this `realistic' $BB$ case
that can be taken as a fair comparison with the $TE$ method. The
comparison of $TE$ and $BB$ methods was also considered in a
recent paper \cite{polna1} (see also earlier papers \cite{Crittenden,Melchiorri}).

%%%%%%%%%%%%%%%%%%%%%%%%%%%%%%%%%%%%%%%%%%%%%%%%%%%%%%%%%%%%%%%%%%%%%%%%%%%%%%%%%%%
                                                                                                       %%%%%% SUBSECTION 7.1 %%%%%%
%%%%%%%%%%%%%%%%%%%%%%%%%%%%%%%%%%%%%%%%%%%%%%%%%%%%%%%%%%%%%%%%%%%%%%%%%%%%%%%%%%%

\subsection{The likelihood function for $R$ in the $BB$ method \label{s6.1}}

We start from the `optimistic' $BB$ case where the $BB$ noise is taken as
$N_{\ell}^{BB}=5.58\times10^{-4}{\mu}{\rm K}^2$, see Eq.~(\ref{Plancknoise}).
We generate the mock $BB$ data using the pdf
$f_{\rm c}({D}_{\ell}^{BB}(o))$ from Eq.~(\ref{cutpdfwl}), Eq.~(\ref{f1})
~(setting $XX=BB$), and Eq.~(\ref{Planckcov}). The procedure is exactly the same as
was described in Sec.~\ref{s5.3} in the context of the $TE$ mock data. The 10
sets of simulated data for every input $R_{\rm i}$ are then used in the likelihood
function
\begin{eqnarray}
 \nonumber % \label{lbb}\label{lbb2}
 \mathcal{L}(R) =C\prod_{\ell} f_{\rm c}({D}_{\ell}^{BB}(o)),
\end{eqnarray}
where $C$ is a normalization constant.

%%%%%%%%%%%%%%%%%%%%%%%%%%%%%%%%%%%%
\begin{figure}
\begin{center}
\includegraphics[width=17cm,height=8cm]{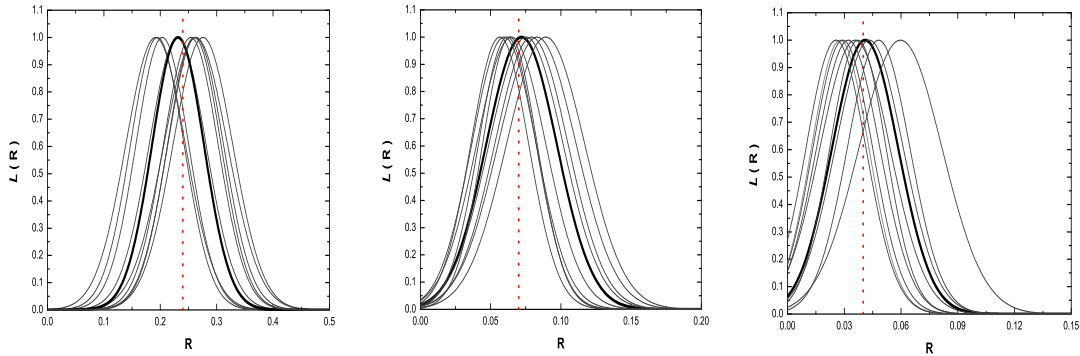}
\end{center}
\caption{Likelihood functions for $R$ from simulated $BB$
data with Planck parameters $N_{\ell}^{BB}=5.58\times 10^{-4} {\mu}{\rm K}^2$,
$f_{\rm sky}=0.65$. The input values in the panels from left to right
are $R_{\rm i} = 0.24.~0.07,~0.04$. The dark solid lines show the `typical'
likelihood functions.}\label{section6_fig1}
\end{figure}
%%%%%%%%%%%%%%%%%%%%%%%%%%%%%%%%%%%%

First, we discuss the benchmark model $R_{\rm i}=0.240$. The 10 likelihood
functions are shown in Fig.~\ref{section6_fig1} (left panel). The mean
value of the maxima is $\overline{R_{\rm p}}=0.237$. The mean width of
the $68.3\%$ confidence intervals is $\overline{\Delta R}=0.050$. This
is a bit smaller than the mean width in the $TE$ likelihood functions.
For illustration, we considered other examples as well. The 10 likelihood
functions with the input value $R_{\rm i}=0.070$ are plotted in the middle
panel of Fig.~\ref{section6_fig1}. The mean values are
$\overline{R_{\rm p}}=0.070$ and $\overline{\Delta R}=0.024$. This shows
that $R=0.070$ could be detected by the $BB$ method at nearly 3$\sigma$ level.
Finally, the likelihood functions with the input value $R_{\rm i}=0.040$
are shown in the right plot of Fig.~\ref{section6_fig1}.
The mean values are $\overline{R_{\rm p}}=0.038$, $\overline{\Delta R}=0.018$.
This $R$ is at somewhat better than 2$\sigma$ level of detection.

Introducing the quantity
\begin{eqnarray}
 \nonumber %\label{snr-define}
\frac{S}{N}\equiv \frac{R_{\rm i}}{\overline{\Delta R}}
\end{eqnarray}
as the `signal to noise' measure of detecting gravitational waves, we show
in Table \ref{tebb} the comparative performances of $TE$ and `optimistic' $BB$
methods in terms of $S/N$ for a range of input values $R_{\rm i}$. The Table
also shows the mean values $\overline{R_{\rm p}}$ of the maxima of
the likelihood functions and the mean widths $\overline{\Delta R}$ of
the $68.3\%$ confidence intervals. (The number of displayed decimal
digits is an artefact of numerical calculations, and not a demonstration of
our responsibility for the accuracy of results.)

Moving to the discussion of the `realistic' $BB$ case, we think that all the
non-instrumental `effective' noises can increase the level of the noise
variables $a_{{\ell}m}^{B}(n)$, Sec.~\ref{sa4}, by at least a factor of 2.
This means that the noise term $N_{\ell}^{BB}$ participating in our analysis,
should be raised from its level in Eq.~(\ref{Plancknoise}) by at least a
factor of 4. In other words, we are working with the `realistic' noise term
$N_{\ell}^{BB}=2.24 \times 10^{-3}{\mu}{\rm K}^2$. Repeating all the calculations
with this noise term, we come up with the `realistic' evaluation of the
Planck's $BB$ performance, as shown in Table \ref{tebb2}.

%%%%%%%%%%%%%%%%%%%%%%%%%%%%%%%%%%%%
\begin{table}
\caption{Comparison of the $TE$ and `optimistic' $BB$ methods in terms
of the signal to noise ratio $S/N$.}
\begin{center}
\label{tebb}
\begin{tabular}{|c|c|c|c|c|c|c|c|}
\hline
$R_{\rm i}$ & 0.30 & 0.24 & 0.20 & 0.15 & 0.11 & 0.07 & 0.04 \\
 \hline\hline
TE: $\overline{R_{\rm p}}$ & 0.28 & 0.24 & 0.19 & 0.13 & 0.11 &
0.07 & -------\\
TE: $\overline{\Delta R}$ & 0.08 & 0.07 & 0.07 & 0.07 & 0.07 & 0.07
& ------- \\
TE: $S/N$ & 3.90 & 3.24 & 2.82 & 2.21 & 1.64 & 1.06 & ------- \\
 \hline
BB: $\overline{R_{\rm p}}$ & 0.31 & 0.24 & 0.20 & 0.16 & 0.11 &
0.07 & 0.04 \\
 BB: $\overline{\Delta R}$ & 0.06 & 0.05 & 0.04 & 0.04 & 0.03 & 0.02
& 0.02 \\
 BB: $S/N$ & 5.36 & 4.80 & 4.55 & 3.95 & 3.44 & 2.92 & 2.22 \\
\hline
\end{tabular}
\end{center}
\end{table}
%%%%%%%%%%%%%%%%%%%%%%%%%%%%%%%%%%%%%%%%%%%%%%%%%%%%%%%

%%%%%%%%%%%%%%%%%%%%%%%%%%%%%%%%%%%%
\begin{table}
\caption{Signal to noise ratio $S/N$ in the `realistic' $BB$ case.}
\begin{center}
\label{tebb2}
\begin{tabular}{|c|c|c|c|c|c|c|}
\hline
$R_{\rm i}$ & 0.30 & 0.24 & 0.20 & 0.15 & 0.11 & 0.07 \\
\hline\hline
BB: $\overline{R_{\rm p}}$ & 0.29 & 0.26 & 0.20 & 0.15 & 0.10 &
0.06 \\
BB: $\overline{\Delta R}$ & 0.13 & 0.11 & 0.09 & 0.07 & 0.06 &
0.05 \\
BB: $S/N$ & 2.40 & 2.24 & 2.30 & 2.06 & 1.72 & 1.32 \\ \hline
\end{tabular}
\end{center}
\end{table}
%%%%%%%%%%%%%%%%%%%%%%%%%%%%%%%%%%%%

\begin{figure}
\begin{center}
\includegraphics[width=12cm,height=10cm]{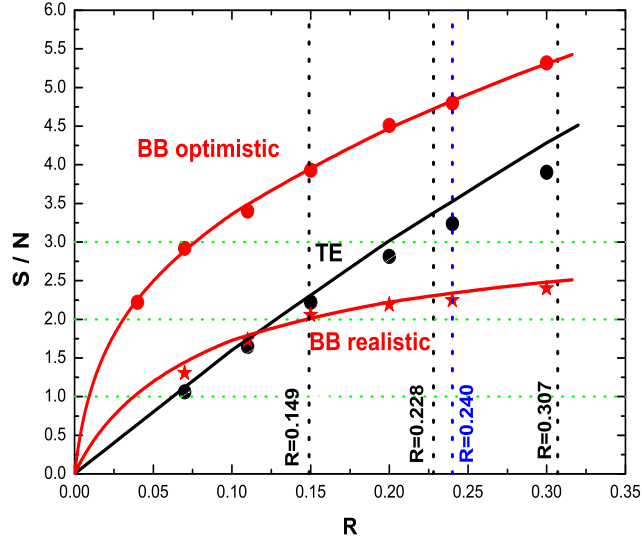}
\end{center}
\caption{The signal to noise ratio $S/N$ as a function of $R$
for the $TE$ (black) and $BB$ (red) methods.
The points show numerical results, whereas the curves are analytical
approximations.}\label{section6_fig3}
\end{figure}

The numerical results for $TE$ and $BB$ methods are also presented as
individual $S/N$ points in Fig.~\ref{section6_fig3}. The solid lines
in this figure show the analytical approximations which we are set to
discuss.

%%%%%%%%%%%%%%%%%%%%%%%%%%%%%%%%%%%%
%%%%%%%%%%%%%%%%%%%%%%%%%%%%%%%%%%%%

%%%%%%%%%%%%%%%%%%%%%%%%%%%%%%%%%%%%%%%%%%%%%%%%%%%%%%%%%%%%%%%%%%%%%%%%%%%%%%%%%%%
                                                                                                       %%%%%% SUBSECTION 7.2 %%%%%%
%%%%%%%%%%%%%%%%%%%%%%%%%%%%%%%%%%%%%%%%%%%%%%%%%%%%%%%%%%%%%%%%%%%%%%%%%%%%%%%%%%%

\subsection{Understanding the signal to noise ratio $S/N$ in $TE$ and $BB$
methods\label{s6.3.3}}

Our numerical calculations are supported by simple analytical expressions for
$S/N$ in $TE$ and $BB$ methods. Let $XX'$ denote either $TE$
or $BB$. With Gaussian approximation for individual pdfs,
Eq.~(\ref{GaussianApproximation}), and given the statistical independence
of differing $\ell$'s, the likelihood function $\mathcal{L}(R)$ reads
\begin{eqnarray}
 \nonumber %\label{gaussianapproxlikelihood}
\mathcal{L}(R) = \prod_{\ell=2}^{\ell_{max}}
\frac{1}{\sqrt{2\pi}\Delta D_{\ell}^{XX'}} ~
\exp\left[-\frac{1}{2}\left(
\frac{D_{\ell}^{XX'}-C_{\ell}^{XX'}}{\Delta D_{\ell}^{XX'}}
\right)^2\right].
\end{eqnarray}
The $R$-dependence of $\mathcal{L}(R)$ is coming from the $R$-dependence
of $C_{\ell}^{XX'}$ and $\Delta D_{\ell}^{XX'}$.

The estimate of the width $\Delta R$ of $\mathcal{L}(R)$ with the input
(maximum likelihood) value $R= R_{\rm i}$ is given by \cite{kendall}
\begin{eqnarray}
\label{estimatedeltar}
\Delta R = \left[ -\left<  \left.
\frac{\partial^2\ln \mathcal{L}(R)} {\partial R^2} \right|_{R=R_{\rm i}}
\right>\right]^{-\frac{1}{2}} ,
\end{eqnarray}
where the angle brackets denote the average over the joint pdf for
$D_{\ell}^{XX'}$. The second derivative participating in
Eq.~(\ref{estimatedeltar}) is
\begin{eqnarray}
- \frac{\partial^2\ln \mathcal{L}(R)} {\partial R^2}
&= &\sum_{\ell=2}^{\ell_{max}}
\frac{1}{{\Delta D^{XX'}_{\ell}}^2}
\left.\left\{
\left(\frac{\partial C_{\ell}^{XX'}}{\partial R}\right)^2
\right.\right.  
\left.\left.
+ \left[
-\left(D_{\ell}^{XX'}-C_{\ell}^{XX'}\right)
\frac{\partial^2 C_{\ell}^{XX'}}{\partial R^2}
\right.\right.\right. \nonumber \\
&+&\left.\left.\left.
 \frac{4\left(D_{\ell}^{XX'}-C_{\ell}^{XX'}\right)}{\Delta D_{\ell}^{XX'}}
\frac{\partial C_{\ell}^{XX'}}{\partial R}
\frac{\partial \Delta D_{\ell}^{XX'}}{\partial R}
\right.\right.\right. \nonumber \\
&+&\left.\left.\left.
\left(
\frac{3\left(D_{\ell}^{XX'}-C_{\ell}^{XX'}\right)^2
- \Delta {D_{\ell}^{XX'}}^2}{\Delta {D_{\ell}^{XX'}}^2}
\right)
\left(\frac{\partial \Delta D_{\ell}^{XX'}}{\partial R}\right)^2
\right.\right.\right. \nonumber \\
&+&\left.\left.
\left\{
-\left(D_{\ell}^{XX'}-C_{\ell}^{XX'}\right)
\frac{\partial^2 C_{\ell}^{XX'}}{\partial R^2}
\right.\right.\right.  
+\left.\left.\left.
\frac{4\left(D_{\ell}^{XX'}-C_{\ell}^{XX'}\right)}{\Delta D_{\ell}^{XX'}}
\frac{\partial C_{\ell}^{XX'}}{\partial R}
\frac{\partial \Delta D_{\ell}^{XX'}}{\partial R}
\right.\right.\right. \nonumber \\
&+&\left.\left.\left.
\left(
\frac{3\left(D_{\ell}^{XX'}-C_{\ell}^{XX'}\right)^2
- \Delta {D_{\ell}^{XX'}}^2}{\Delta {D_{\ell}^{XX'}}^2}
\right)
\left(\frac{\partial \Delta D_{\ell}^{XX'}}{\partial R}\right)^2
\right.\right.\right. \nonumber \\
&+&\left.\left.\left.
\left(
\frac{\Delta {D_{\ell}^{XX'}}^2
- \left(D_{\ell}^{XX'}-C_{\ell}^{XX'}\right)^2}{\Delta {D_{\ell}^{XX'}}}
\right)
\left(\frac{\partial^2 \Delta D_{\ell}^{XX'}}{\partial R^2}\right)
\right]
\right\}\right. .
 \nonumber %\label{estimatedeltar1a}
\end{eqnarray}
Using the averages
\begin{eqnarray}
\left< D_{\ell}^{XX'} \right> = C_{\ell}^{XX'}, ~~~
\left<\left( D_{\ell}^{XX'} - C_{\ell}^{XX'}\right)^2\right>
= {\Delta D_{\ell}^{XX'}}^2
\nonumber
\end{eqnarray}
we arrive at an intermediate expression
\begin{eqnarray}
\Delta R =  \left[ \sum_{\ell=2}^{\ell_{max}}
\frac{1}{{\Delta D^{XX'}_{\ell}}^2}\left(
\left( \frac{\partial C^{XX'}_{\ell}}{\partial R}\right)^2 +
2\left( \frac{\partial \Delta D^{XX'}_{\ell}}{\partial R}\right)^2
\right)\right]^{-\frac{1}{2}}.
\label{estimatedeltar1b}
\end{eqnarray}
Since the noises are $R$-independent and $\partial \Delta
D^{XX'}_{\ell}/\partial R$ is smaller than $\partial
C^{XX'}_{\ell}/\partial R$ by a factor $\propto
1/\sqrt{(2\ell+1)f_{\rm sky}}$, one can neglect the second term in
Eq.~(\ref{estimatedeltar1b}) bringing it to
\begin{eqnarray}
\Delta R =  \left[ \sum_{\ell=2}^{\ell_{max}}
\left( \frac{1}{{\Delta D^{XX'}_{\ell}}}
\frac{\partial C^{XX'}_{\ell}}{\partial R}\right)^2 \right]^{-\frac{1}{2}}.
\label{estimatedeltar2}
\end{eqnarray}

The general expression (\ref{estimatedeltar2}) is a well-known
result \cite{kendall} applied to CMB subjects on several occasions
(see for example \cite{Tegmark1996,Jaffe}).
Eq.~(\ref{estimatedeltar2}) allows one to write the following
analytical estimate for $S/N$:
\begin{eqnarray}
\label{analyticalsnr}
S/N  = \frac{R}{\Delta R} = \sqrt{ \sum_{\ell=2}^{\ell_{max}}
\left( \frac{1}{\Delta D^{XX'}_{\ell}}
\frac{\partial C^{XX'}_{\ell}}{\partial \ln R} \right)^2} ~.
\end{eqnarray}

The further simplifications of Eq.~(\ref{analyticalsnr}) rely on further
assumptions. It is reasonable to assume that for sufficiently small $R$ one
can use the Taylor expansion
\begin{eqnarray}
C^{XX'}_{\ell}(R) = \left.C^{XX'}_{\ell}\right|_{R=0}  +
R\left( \left.\frac{\partial C^{XX'}_{\ell}}{\partial
R}\right|_{R=0}\right).
\label{lindepR}
\end{eqnarray}
Then, for $XX'=TE$ one derives
\begin{eqnarray}
\label{finial_te2}
S/N = \sqrt{
\sum_{\ell=2}^{\ell_{max}}
\left( \frac{C_{\ell}^{TE}(R) - \left.C_{\ell}^{TE}\right|_{R=0}}
{\Delta D_{\ell}^{TE}}\right)^2
},
\end{eqnarray}
and for $XX'=BB$ one derives
\begin{eqnarray}
S/N = \sqrt{ \sum_{\ell=2}^{\ell_{max}}
\left( \frac{C_{\ell}^{BB}(R) }{\Delta D_{\ell}^{BB}} \right)^2 }.
\label{semifinalBB}
\end{eqnarray}

These expressions could be expected on physical grounds. In the $TE$ case,
the quantity $\left.C_{\ell}^{TE}\right|_{R=0}$ is not zero,
and it is entirely determined by density perturbations. Therefore, it is
reasonable that the $R$-signal, allowing to detect gravitational waves,
is given (at each $\ell$) by the difference
$C_{\ell}^{TE}(R) - \left. C_{\ell}^{TE}\right|_{R=0}$, whereas the $R$-noise is
given by the total standard deviation $\Delta D_{\ell}^{TE}$. In the
$BB$ case, the quantity $\left.C^{BB}_{\ell}\right|_{R=0}$ is zero (no
gravitational waves - no $BB$ correlation), and therefore it is reasonable
that the $R$-signal is given by $C_{\ell}^{BB}(R)$, whereas the $R$-noise
is given by $\Delta D_{\ell}^{BB}$.

As one could anticipate, the linear approximation (\ref{lindepR}) is not quite
good if $R$ is not small. Some deviations from exact numerical results could
be expected. However, in Fig.~\ref{section6_fig3}, we plot the $TE$ curve
without any corrections, that is, as it follows from the analytical formula
(\ref{finial_te2}). As for the $BB$ case, the noticable deviations exist, so
a better fit to the exact numerical results requires a small correction to
Eq.~(\ref{semifinalBB}). The actually plotted $BB$ curves in
Fig.~\ref{section6_fig3} are given by the corrected analytical formula
\begin{eqnarray}
\label{finial_bb2}
S/N = \sqrt{ \sum_{\ell=2}^{\ell_{max}}
\left( \frac{1}{\left(1+0.5R\right)} \frac{C_{\ell}^{BB}(R) }{\Delta
D_{\ell}^{BB}} \right)^2 }.
\end{eqnarray}
The analytical results shown in Fig.~\ref{section6_fig3} are based on
$\ell_{max} = 100$.

\begin{figure}
\begin{center}
\includegraphics[width=16cm,height=10cm]{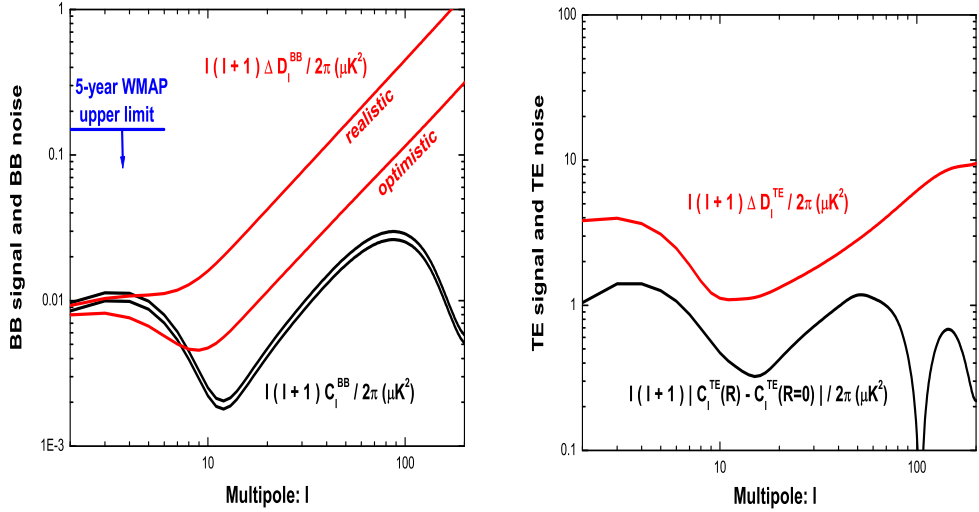}
\end{center}
\caption{The $BB$ and $TE$ gravitational wave signals (power spectra) in
comparison with noises (standard deviations) in the benchmark model $R=0.24$.
The signals are in black, the noises are in red.}
\label{BB-figure}
\end{figure}

It is natural that the total $(S/N)^2$ consists of individual contributions
$(S/N)^2_{\ell}$ at each $\ell$. Both, Eq.~(\ref{finial_te2}) and
Eq.~(\ref{finial_bb2}), have the structure
\begin{eqnarray}
 \nonumber %\label{snrR}
(S/N)^2 = \sum_{\ell=2}^{\ell_{max}}(S/N)^2_{\ell},
\end{eqnarray}
where the individual signal to noise ratios $(S/N)_{\ell}$ are given by
\begin{eqnarray}
\label{snrl}
(S/N)_{\ell}^{TE} =
\frac{C_{\ell}^{TE}(R) - C_{\ell}^{TE}(R=0)}{\Delta D_{\ell}^{TE}}, ~~~~~
(S/N)_{\ell}^{BB} =
\frac{(1+0.5R)^{-1} C_{\ell}^{BB}(R) }{\Delta D_{\ell}^{BB}}.
\end{eqnarray}
One can say that the numerators in Eq.~(\ref{snrl}) are signals depending
on $R$, while the denominators are noises. Signals and noises are shown, as
functions of $\ell$, in Fig.~\ref{BB-figure} for our benchmark model with
$R=0.24$. The lower solid black line for the $BB$ power spectrum includes the
correcting factor $(1+0.5R)^{-1}$, whereas the upper solid black line does not.

%%%%%%%%%%%%%%%%%%%%%%%%%%%%%%%%%%%%
%%%%%%%%%%%%%%%%%%%%%%%%%%%%%%%%%%%%

Fig.~\ref{BB-figure} provides a graphical illustration of what the
signal to noise ratio $S/N$ in Fig.~\ref{section6_fig3} consists of
and how it is being accumulated from the individual, predominantly
lower than the noise, contributions at each $\ell$. It is seen
from Fig.~\ref{BB-figure} that the success of the $BB$ method
strongly relies on the very low multipoles, $\ell \lesssim 10$,
and hence on the era of reionization with the optical depth $\tau$
from Eq.~(\ref{WMAPBestFit}). Most of $S/N$ in the $BB$ method
comes from these multipoles \cite{deepak,ana5,planck}.
Specifically, in the $R=0.24$ model we find
\begin{eqnarray}
\nonumber
\left(
\frac{\sum_{\ell=2}^{10}(S/N)^2_{\ell}}{\sum_{\ell=2}^{100}(S/N)^2_{\ell}}
\right)_{\rm optimistic}\simeq
0.47, ~~~  \left(
\frac{\sum_{\ell=2}^{10}(S/N)^2_{\ell}}  {\sum_{\ell=2}^{100}(S/N)^2_{\ell}}
\right)_{\rm realistic}\simeq
0.83.
\end{eqnarray}
The cosmic reionization remains to be fully understood and
quantified \cite{fan}. If it happens that the optical depth is
actually smaller than the currently accepted
$\tau_{reion}\approx0.08$, the sensitivity of $BB$ method to relic
gravitational waves will be reduced. On the other hand, if the
$BB$ detection does take place, it will be an argument for the
currently accepted or higher $\tau$.

In contrast to $BB$, the $TE$ method does not rely on very low multipoles.
In fact, the most of $S/N$ comes from the multipoles
$30 \lesssim \ell \lesssim 70$. Specifically, in the $R=0.24$ model we have
\begin{eqnarray}
\nonumber
\frac{\sum_{\ell=2}^{10}(S/N)^2_{\ell}}{\sum_{\ell=2}^{100}(S/N)^2_{\ell}}
\simeq 0.15.
\end{eqnarray}
One can say that that the $TE$ correlation function directly probes the
relic gravitational waves at recombination.

It emerges from this comparative analysis and Fig.~\ref{section6_fig3} that
both, $TE$ and $BB$, methods need to be used, with the former one being more
likely to bring the first identification of relic gravitational waves in the
Planck data.

%%%%%%%%%%%%%%%%%%%%%%%%%%%%%%%%%%%%%%%%%%%%%%%%%%%%%%%%%%%%%%%%%%%%%%%%%%%%%%%%%%%
%%%%%%%%%%%%%%%%%%%%%%%%%%%%%%%%%%  SECTION 9   %%%%%%%%%%%%%%%%%%%%%%%%%%%%%%%%%%%%%%%%%%
%%%%%%%%%%%%%%%%%%%%%%%%%%%%%%%%%%%%%%%%%%%%%%%%%%%%%%%%%%%%%%%%%%%%%%%%%%%%%%%%%%%

\section{Conclusions}

A correct theoretical picture of the physical phenomenon is important for
a proper setting of search experiment, as well as for the analysis of data
and its interpretation. The relic gravitational waves `must be out there', and
we take this understanding as our starting point. Having further developed
statistical theory of CMB anisotropies, we concentrated on the $TE$ correlation
function. We have shown that the WMAP5 $TE$ data contain a hint on the presence
of gravitational waves in the data. Our best-fit model includes a substantial
amount of relic gravitational waves, $R=0.24$. In simple terms, this means that
20\% of the temperature quadrupole is caused by gravitational waves, and 80\%
by density perturbations. Because of large WMAP5 noises, the confidence of this
conclusion is not high. The maximum likelihood result includes the WMAP's
best-fit model with no gravitational waves, $R=0$, almost within 1$\sigma$
interval.

We projected our conclusion $R=0.24$ on the forthcoming Planck mission. We
numerically simulated Planck data and compared the $TE$ and $BB$ channels
of observations. In the $BB$ channel, much will depend on contamination by
systematic effects. We distinguish the `optimistic' $BB$ case, when only the
advertised instrumental noises are taken into account, and the `realistic' $BB$
case, when the effective noises are increased as a reflection of many possible
residual effects. It is shown that the $TE$ method will see the relic gravitational
waves at a better than 3$\sigma$ level, whereas the `realistic' $BB$ method
will see them at a better than 2$\sigma$ level.

Although most of our discussion focused on the Planck experiment, some
balloon-borne \cite{EBEX,SPIDER} and ground-based experiments
\cite{BICEP,quad,Clover,QUITE}, currently in preparation, will be
sensitive to part of the lower-$\ell$ spectrum of CMB. The contributions of various
intervals of $\ell$ to the total signal to noise ratio $S/N$ can be read off
from the evaluations developed here. These experiments will be complementary
to the Planck mission and can provide a healthy competition in the race for
discovery of relic graviational waves.

%%%%%%%%%%%%%%%%%%%%%%%%%%%%%%%%%%%%%%%%%%%%%%%%%%%%%%%%%%%%%%%%%%%%%%%%%%%%%%%%%%%
%%%%%%%%%%%%%%%%%%%%%%%%%%%%%%%%%%  Acknowledgments   %%%%%%%%%%%%%%%%%%%%%%%%%%%%%%%%%%%%%%%
%%%%%%%%%%%%%%%%%%%%%%%%%%%%%%%%%%%%%%%%%%%%%%%%%%%%%%%%%%%%%%%%%%%%%%%%%%%%%%%%%%%

\section*{Acknowledgements}
We appreciate useful mathematical discussions with N. N. Leonenko.
WZ is partially supported by Chinese NSF grants No.10703005 and No.10775119.

%%%%%%%%%%%%%%%%%%%%%%%%%%%
%%%%%%%%%%% APPENDIX
%%%%%%%%%%%%%%%%%%%%%%%%%%%

\appendix

\section{Probability density function for $v_{\ell m}^{(c)}$
\label{appendixA}}

In Sec~\ref{s2.3.3} it was shown that the joint pdf's
$f(\frac{a_{\ell 0}^{T(r)}}{\sqrt{2}},\frac{a_{\ell 0}^{E(r)}} {\sqrt{2}})$,
$f(a_{\ell m}^{T(c)},a_{\ell m}^{E(c)})$ (where $m\geq1$)
are bivariate zero-mean normal distributions with standard deviations
$\frac{\sigma_{\ell}^T}{\sqrt{2}}$, $\frac{\sigma_{\ell}^E}{\sqrt{2}}$
and correlation coefficient $\rho_{\ell}$. The indices $\ell, m, c$ are
considered fixed, so the two variables are distinguished by
the labels $T$, $E$. From such a bivariate distribution one can derive
a pdf for the product variable $v_{\ell m}^{(c)}= a_{\ell m}^{T(c)}a_{\ell m}^{E(c)}$
(see Eq.~\ref{definev}). For each
$v_{\ell m}^{(c)}$, the corresponding pdf is given by \cite{grishchuk}
\begin{eqnarray}
\label{pdfvl0}
f(v_{\ell m}^{(c)})=
\frac{1}{\pi\sigma_{\ell}^T\sigma_{\ell}^E\sqrt{1-\rho_{\ell}^2}}
\exp\left[{\frac{~v^{(c)}_{\ell
m}\rho_{\ell}}
{(1-\rho_{\ell}^2)\sigma_{\ell}^T\sigma_{\ell}^E}}\right]K_0
\left(\frac{|v^{(c)}_{\ell
m}|}{(1-\rho_{\ell}^2)\sigma_{\ell}^T\sigma_{\ell}^E}\right),
\end{eqnarray}
where $K_0$ is the zero-order modified Bessel function.

Let us analyze the pdf (\ref{pdfvl0}) in more detail. Firstly,
Eq.~(\ref{pdfvl0}) shows that the pdf's with positive and negative
$\rho_{\ell}$ are related by
\begin{eqnarray}
\label{symmetry}
 f(v_{\ell m}^{(c)})\left|_{-\rho_{\ell}}\frac{}{}\right.=f(-v_{\ell
 m}^{(c)})\left|_{\rho_{\ell}}\frac{}{}\right. .
\end{eqnarray}
If $\rho_{\ell}=0$ (no correlations), the pdf is symmetric with respect
to the axis $v_{\ell m}^{(c)}=0$. For any $\rho_{\ell}$, the pdf is
divergent at $v_{\ell m}^{(c)}=0$, because the modified
Bessel function $K_0(\alpha)$ is divergent at $\alpha=0$.

%%%%%%%%%%%%%%%%%%%%%%%%%%%%%%%%%%%
\begin{figure}
\begin{center}
\includegraphics[width=12cm,height=10cm]{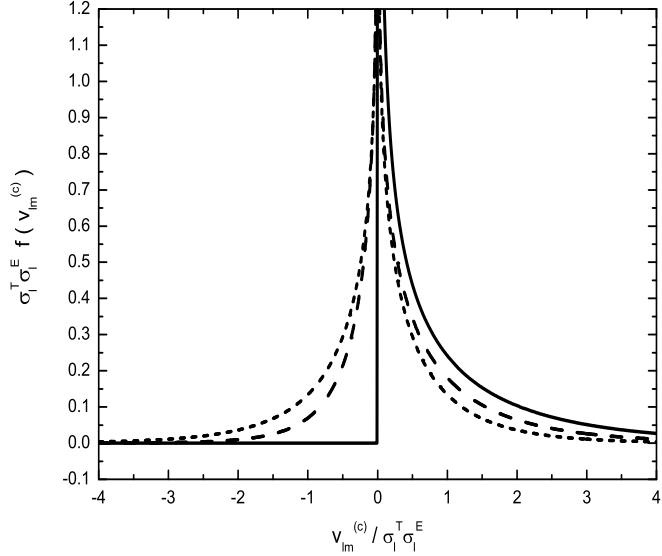}
\end{center}
\caption{Probability density functions $f(v_{\ell m}^{(c)})$ for
various positive $\rho_\ell$'s. The solid, dashed and dotted lines
depict $\rho_\ell=1.0,~0.4,~{\rm and}~0.0$.}
\label{section3_fig2.1}
\end{figure}
%%%%%%%%%%%%%%%%%%%%%%%%%%%%%%%%%%%

In Fig.~\ref{section3_fig2.1} we show the pdf's for different values
of positive $\rho_{\ell}$. When $\rho_{\ell}$ increases, the
distribution shifts to the right reflecting the fact that the mean value
$\langle v_{\ell m}^{(c)}\rangle$ is proportional to $\rho_\ell$.
In the limiting case $\rho_\ell=1$ the pdf develops a step, making
$f(v_{\ell m}^{(c)})=0$ for $v_{\ell m}^{(c)}<0$.
In general, using (6.621.3) from Ref.~\cite{book1}, one can show
that the probabilities of negative, or positive, values of $v_{\ell m}^{(c)}$
are given by
 \begin{equation}
\label{Pv10_negative}
 P\left(v_{\ell m}^{(c)} < 0\right) \equiv \int_{-\infty}^{0}
 f(v_{\ell m}^{(c)}) dv_{\ell m}^{(c)}
 =
 \frac{2}{\pi}\frac{1-\rho_{\ell}}{(1-\rho_{\ell}^2)^{1/2}}
 F\left[1,\frac{1}{2},\frac{3}{2},\frac{\rho_{\ell}-1}{\rho_{\ell}+1}\right],
 \end{equation}
or
 \begin{equation}\label{Pv10_positive}
 P\left(v_{\ell m}^{(c)} > 0\right) \equiv \int_{0}^{\infty}
 f(v_{\ell m}^{(c)})dv_{\ell m}^{(c)} =
  \frac{2}{\pi}\frac{1+\rho_{\ell}}{(1-\rho_{\ell}^2)^{1/2}}
 F\left[1,\frac{1}{2},\frac{3}{2},\frac{\rho_{\ell}+1}{\rho_{\ell}-1}\right],
 \end{equation}
where $F$ is the $hypergeometric$-function. These probabilities depend only
on $\rho_{\ell}$, but not on $\sigma_{\ell}^T$ and $\sigma_{\ell}^{E}$.

Similarly to what has been done in \cite{grishchuk}, one can show,
using (9.131.1), (9.121.7), (1.624.9), (1.623.2) from
\cite{book1}, that $P(v_{\ell m}^{(c)}< 0)+P(v_{\ell m}^{(c)}>
0)=1$, which confirms the correct normalization of $f(v_{\ell
m}^{(c)})$. As expected, when $\rho_{\ell}=1$, $P(v_{\ell
m}^{(c)}<0)=0$ and $P(v_{\ell m}^{(c)}>0)=1$. And when
$\rho_{\ell}=0$, $P(v_{\ell m}^{(c)}<0)=P(v_{\ell
m}^{(c)}>0)=0.5$.

The probability density function (\ref{pdfvl0}) allows one to define the
confidence intervals of finding specific values of $v_{\ell m}^{(c)}$.
As an example, we start from the $68.3\%$ confidence interval. We denote
by $(v_{\ell m}^{(c)})_U$ and $(v_{\ell m}^{(c)})_L$ the
upper and lower boundaries of the shortest interval, i.e. the interval
having the minimum difference $(v_{\ell m}^{(c)})_U-(v_{\ell m}^{(c)})_L$,
such that \cite{book6}
\begin{eqnarray}
 \nonumber %\label{vUL}
 \int_{(v_{\ell m}^{(c)})_L}^{(v_{\ell m}^{(c)})_U}f(v_{\ell m}^{(c)})
 ~d~v_{\ell m}^{(c)}=0.683.
\end{eqnarray}
The surface area under the pdf curve between $(v_{\ell
m}^{(c)})_U$ and $(v_{\ell m}^{(c)})_L$ gives the probability
($68.3\%$) of finding $v_{\ell m}^{(c)}$ in this interval. This is
a sensible measure \cite{book6} of uncertainty of our estimator.
For the pdf's that we are dealing with, the $68.3\%$ confidence
interval always includes points of the maximum and the mean value
of the pdf.

\begin{figure}
\begin{center}
\includegraphics[width=12cm,height=10cm]{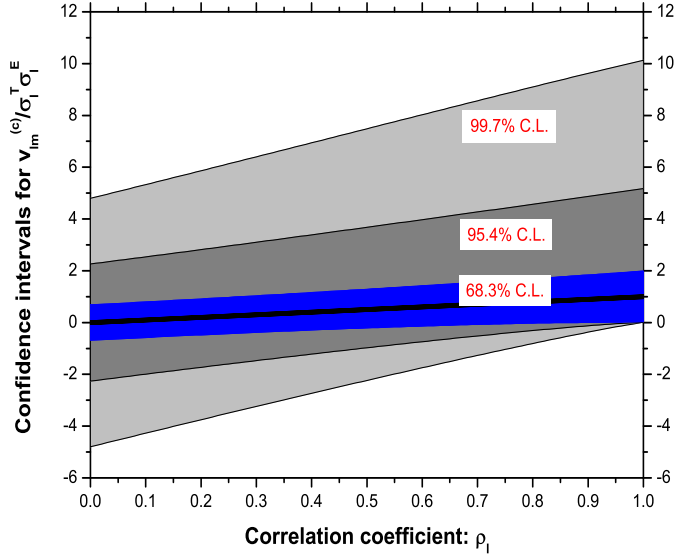}
\end{center}
\caption{The $68.3\%, 95.4\%$ and $99.7\%$ confidence intervals of
the variable $v_{\ell m}^{(c)}/\sigma_{\ell}^T\sigma_{\ell}^E$ for
different $\rho_{\ell}$.  The solid (black) line denotes the mean value of
$v_{\ell m}^{(c)}/\sigma_{\ell}^T\sigma_{\ell}^E$.} \label{section3_fig3a}
\end{figure}

In Fig.~\ref{section3_fig3a}, we plot the $68.3\%$, $95.4\%$, $99.7\%$
confidence intervals for different $\rho_\ell$'s. When
$\rho_{\ell}=0$, the confidence intervals are symmetric with respect
to the mean value (zero) of $v_{\ell m}^{(c)}$. When $\rho_{\ell}$ increases,
the intervals move toward larger values $v_{\ell m}^{(c)}$. In the limit
$\rho_{\ell}\rightarrow1$, the lower boundary $(v_{\ell
m}^{(c)})_L$ approaches $0$ for all levels of confidence. For any confidence
level, the confidence interval is not symmetric with respect to the mean
value of $v_{\ell m}^{(c)}$. It can also be seen that, unless
$\rho_{\ell}\approx1$, there is a significant probability of finding
$v_{\ell m}^{(c)}<0$, despite the fact that the mean value is strictly
non-negative.

%%%%%%%%%%%%%%%%%%%%%%%%%%%%%%%%%%%%
%%%%%%%%%%%%%%%%%%%%%%%%%%%%%%%%%%%%

\section{Some properties of $f(D_{\ell}^{TE})$ \label{appendixB}}

It is insightful to discuss properties of $f(D_{\ell}^{TE})$ in
comparison with $f(v_{\ell m}^{(c)})$ in Eq.~(\ref{pdfvl0}). Surely,
both functions go to zero when the argument $\rightarrow \pm
\infty$. But the behaviors at zero argument are different,
$f(D_{\ell}^{TE})$ does not diverge at $D_{\ell}^{TE}=0$ if $n >
1$. Although the modified Bessel functions do diverge at zero
argument \cite{book1},
\begin{eqnarray}
 K_{\nu}(\alpha)\sim\frac{(\nu-1)!}{2}\left(\frac{\alpha}{2}\right)^{-\nu},~~
{\rm where}~~(\nu\ge1,~~\alpha\rightarrow0),
\nonumber
\end{eqnarray}
the first factor in the right hand side of Eq.~(\ref{pdfwl}) compensates for
this divergency, so that the pdf (\ref{pdfwl}) remains finite.

In the limit $\rho_{\ell}\rightarrow1$, keeping other parameters fixed
and using the asymptotic formula \cite{book1} $K_{\nu}(\alpha)\sim\sqrt{\frac{\pi}{2\alpha}}e^{-\alpha}$ for $\alpha\rightarrow\infty$,
%\begin{eqnarray}
% K_{\nu}(\alpha)\sim\sqrt{\frac{\pi}{2\alpha}}e^{-\alpha},  ~~~~
%{\rm for}~~\alpha\rightarrow\infty,
%\nonumber
%\end{eqnarray}
one obtains
\begin{eqnarray}
 f(D^{TE}_{\ell})\rightarrow \frac{n{\tilde{V}}^{n/2-1}
 e^{-{\tilde{V}}/2}}{2^{n/2}\Gamma(n/2)\sigma_{\ell}^T\sigma_{\ell}^E}
 ~~{\rm with}~~
 \tilde{V}\equiv\frac{nD^{TE}_{\ell}}{\sigma_{\ell}^T\sigma_{\ell}^E}.
\nonumber
\end{eqnarray}
As expected, the asymptotic distribution is a $\chi^2$
distribution with $n$ degrees of freedom.

Similarly to Eq.~(\ref{symmetry}) one has 
\begin{eqnarray}
 \nonumber %
 f(D^{TE}_\ell)\left|_{-\rho_{\ell}}\frac{}{}\right.=f(-D^{TE}_\ell)
 \left|_{\rho_{\ell}}\frac{}{}\right.
\end{eqnarray}
The probability of negative, or positive, values of $D^{TE}_\ell$ is given by
\begin{eqnarray}
\nonumber
P(D^{TE}_\ell<0)&\equiv&\int_{-\infty}^{0} f(D^{TE}_{\ell})dD^{TE}_{\ell}
=\frac{(1-\rho_{\ell})(1-\rho_{\ell}^2)^{\frac{n}{2}-1} }{2^{n-1}} \\
\label{pwl} &\times&
\frac{\Gamma(n)}{\Gamma(\frac{n}{2}+1)\Gamma(\frac{n}{2})}
F\left[1,1-\frac{n}{2},1+\frac{n}{2},\frac{\rho_{\ell}-1}{\rho_{\ell}+1}\right],
\end{eqnarray}
or
\begin{eqnarray} \nonumber
P(D^{TE}_\ell>0)&\equiv&\int_{0}^{+\infty}f(D^{TE}_{\ell}) dD^{TE}_{\ell}
=\frac{(1+\rho_{\ell})(1-\rho_{\ell}^2)^{\frac{n}{2}-1} }{2^{n-1}} \\
\label{pwll}&\times&
\frac{\Gamma(n)}{\Gamma(\frac{n}{2}+1)\Gamma(\frac{n}{2})}
F\left[1,1-\frac{n}{2},1+\frac{n}{2},\frac{\rho_{\ell}+1}{\rho_{\ell}-1}\right].
\end{eqnarray}
These probabilities depend on $\rho_{\ell}$ but not on $\sigma_{\ell}^T,
\sigma_{\ell}^E$. As it should be, Eqs.~(\ref{pwl},~\ref{pwll}) reduce to
Eqs.~(\ref{Pv10_negative},~\ref{Pv10_positive}) when $n=1$.

We have checked numerically, for a set of $\ell$'s and
$\rho_{\ell}$'s, the normalization condition
$P(D^{TE}_\ell<0)+P(D^{TE}_\ell>0)=1$. A useful equality relating
the probabilities for positive and negative $\rho_\ell$'s is
\begin{eqnarray}
 P(D^{TE}_\ell<0)\left|_{-\rho_l}\frac{}{}\right.=1-P(D^{TE}_\ell<0)
 \left|_{\rho_l}\frac{}{}\right..
\nonumber
\end{eqnarray}
A similar equality is valid for  probabilities
$P(D^{TE}_\ell>0)|_{\pm\rho_{\ell}}$. When $\rho_\ell=1$, we find, using
formula $F[1,1-\frac{n}{2},1+\frac{n}{2},0]=1$, that $P(D^{TE}_\ell<0)=0$
for any $n$. On the other hand, when $\rho_\ell=0$, we find, using formula
$F\left[1,1-\frac{n}{2},1+\frac{n}{2},-1\right] =
\frac{\sqrt{\pi}}{2}\frac{\Gamma(\frac{n}{2}+1)}{\Gamma(\frac{n+1}{2})}$,
that $P(D^{TE}_\ell<0)= P(D^{TE}_\ell>0)=\frac{1}{2}$ for any $n$.

\begin{figure}
\begin{center}
\includegraphics[width=14cm,height=10cm]{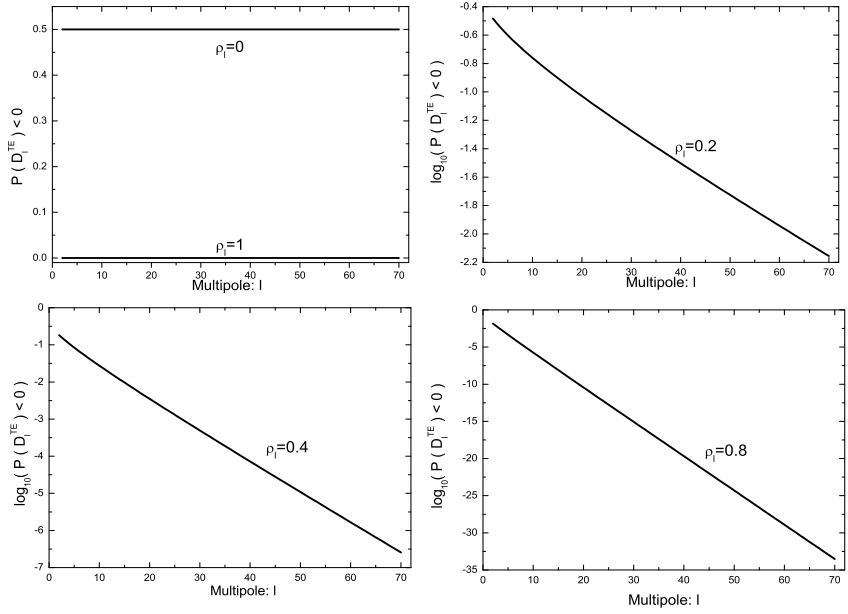}
\end{center}
\caption{$P(D^{TE}_{\ell}<0)$ as a function of $\ell$ for
$\rho_\ell=0.0, 0.2, 0.4, 0.8, 1.0$. In the upper left panel the vertical
axis is $P(D_{\ell}^{TE}<0)$, whereas in other three panels it is
$\log_{10}{P(D_{\ell}^{TE}<0)}$.}
\label{section3_fig4a}
\end{figure}

In Fig.~\ref{section3_fig4a}, we plot the probability
$P(D^{TE}_\ell<0)$ in the interval $2\leq\ell\leq70$ for selected
values of $\rho_\ell$, $\rho_\ell=0.0, 0.2, 0.4, 0.8$ and $1.0$.
It is seen that for a fixed $\rho_\ell$ the value of $P(D^{TE}_\ell<0)$
rapidly decreases as $\ell$ becomes larger, excepting the limiting
cases $\rho_\ell=0$ or $1$. For a fixed $\ell$ and growing
$\rho_{\ell}$, the $P(D^{TE}_\ell<0)$ changes from 0.5 to 1.

%%%%%%%%%%%%%%%%%%%%%%%%%%%%%%%%%%%%
%%%%%%%%%%%%%%%%%%%%%%%%%%%%%%%%%%%%

%%%%%%%%%%%%%%%%%%%%%%%%%%%%%%%%%%%%%%%%%%%%%%%%%%%%%%%%%%%%%%%%%%%%%%%%%%%%%%%%%%%
%%%%%%%%%%%%%%%%%%%%%%%%%%%%%%%%%%  BIBLIOGRAPHY  %%%%%%%%%%%%%%%%%%%%%%%%%%%%%%%%%%%%%%%%
%%%%%%%%%%%%%%%%%%%%%%%%%%%%%%%%%%%%%%%%%%%%%%%%%%%%%%%%%%%%%%%%%%%%%%%%%%%%%%%%%%%

\end{document}